\documentclass[12pt,preprint]{aastex}
\usepackage{natbib}

\newcommand{\kms}{\ifmmode {\rm km\ s}^{-1} \else km s$^{-1}$\fi}
\newcommand{\Msun}{\ifmmode {\rm M}_{\odot} \else M$_{\odot}$\fi}
\newcommand{\Lsun}{\ifmmode {\rm L}_{\odot} \else L$_{\odot}$\fi}
\newcommand{\qo}{\ifmmode q_{\rm o} \else $q_{\rm o}$\fi}
\newcommand{\Ho}{\ifmmode H_{\rm o} \else $H_{\rm o}$\fi}
\newcommand{\ho}{\ifmmode h_{\rm o} \else $h_{\rm o}$\fi}
\newcommand{\ltsim}{\raisebox{-.5ex}{$\;\stackrel{<}{\sim}\;$}}
\newcommand{\gtsim}{\raisebox{-.5ex}{$\;\stackrel{>}{\sim}\;$}}
\newcommand{\vFWHM}{\ifmmode v_{\mbox{\tiny FWHM}} \else
                    $v_{\mbox{\tiny FWHM}}$\fi}
\newcommand{\CCF}{\ifmmode F_{\it CCF} \else $F_{\it CCF}$\fi}
\newcommand{\ACF}{\ifmmode F_{\it ACF} \else $F_{\it ACF}$\fi}
\newcommand{\Halpha}{\ifmmode {\rm H}\alpha \else H$\alpha$\fi}
\newcommand{\Hbeta}{\ifmmode {\rm H}\beta \else H$\beta$\fi}
\newcommand{\Hgamma}{\ifmmode {\rm H}\gamma \else H$\gamma$\fi}
\newcommand{\Hdelta}{\ifmmode {\rm H}\delta \else H$\delta$\fi}
\newcommand{\Lya}{\ifmmode {\rm Ly}\alpha \else Ly$\alpha$\fi}
\newcommand{\Lyb}{\ifmmode {\rm Ly}\beta \else Ly$\beta$\fi}
\newcommand{\HeI}{\ifmmode {\rm He}\,{\sc i}\,\lambda5876 \else 
	          He\,{\sc i}\,$\lambda5876$\fi}
\newcommand{\HeII}{\ifmmode {\rm He}\,{\sc ii}\,\lambda4686 \else 
	           He\,{\sc ii}\,$\lambda4686$\fi}

\newcommand{\hei}{He\,{\sc i}}
\newcommand{\heii}{He\,{\sc ii}}

\newcommand{\feii}{Fe\,{\sc ii}}

\newcommand{\fevii}{Fe\,{\sc vii}}
\newcommand{\fex}{Fe\,{\sc x}}

\newcommand{\neiii}{Ne\,{\sc iii}}

\newcommand{\ciii}{\ifmmode {\rm C}\,{\sc iii} \else C\,{\sc iii}\fi}
\newcommand{\civ}{\ifmmode {\rm C}\,{\sc iv} \else C\,{\sc iv}\fi}

\newcommand{\nii}{N\,{\sc ii}}

\newcommand{\oi}{O\,{\sc i}}
\newcommand{\oii}{O\,{\sc ii}}
\newcommand{\oiii}{O\,{\sc iii}}

\newcommand{\sii}{S\,{\sc ii}}

\newcommand{\caii}{Ca\,{\sc ii}}

\slugcomment{To be published in Astrophysical Journal}

\shorttitle{3C390.3 variability}
\shortauthors{Dietrich et al.}

\begin{document}

%\received{December 02, 2002}
%\revised{REVISION DATE}
%\accepted{September 11, 2001}
%\cpright{type}{year}
%
%\journalid{VOL}{JOURNAL DATE}
%\articleid{START PAGE}{END PAGE}
%\paperid{MANUSCRIPT ID} 
%
%\cpright{TYPE}{YEAR}
%\ccc{CODE} 

\title{Optical Monitoring of the Broad-Line Radio Galaxy 3C\,390.3 $^\ast$}

\author{
Matthias Dietrich
 \altaffilmark{1},
Bradley M. Peterson
 \altaffilmark{1},
Catherine J. Grier
 \altaffilmark{1},
Misty C. Bentz
\altaffilmark{1,2},
Jason Eastman
 \altaffilmark{1,3},
Stephan Frank
 \altaffilmark{1,4},
Raymond Gonzalez
 \altaffilmark{1},
Jennifer L. Marshall
 \altaffilmark{1,5},
Darren L. DePoy
 \altaffilmark{1,5,6},
and
Jose L. Prieto 
 \altaffilmark{1,7},
}
\altaffiltext{1}
{Department of Astronomy, The Ohio State University, 140 West 18th Av.,
 Columbus, OH 43210, USA}
\altaffiltext{2}
{Department of Physics and Astronomy, Georgia State University,
 One Park Place South, Suite 700, Atlanta, GA 30303, USA}
\altaffiltext{3}
{Department of Physics, University of California, Broida Hall, Santa Barbara, 
 CA 93\,106, USA}
\altaffiltext{4}
{Laboratoire d'Astrophysique de Marseille/LAM
 Observatoire Astronomique de Marseille-Provence, Pole de l'Etoile Site de 
 Chateau-Gombert 38, rue Frederic Joliot-Curie, 13388 Marseille cedex 13, 
 France}
\altaffiltext{5}
{Department of Physics and Astronomy, Texas A\&M University, 4242 TAMU, 
 College Station, TX 77843, USA}
\altaffiltext{6}
{George P. and Cynthia Woods Mitchell Institute for Fundamental Physics 
 and Astronomy, Texas A\&M University, 4242 TAMU, College Station, 
 TX 77843, USA}
\altaffiltext{7}
{Observatories of the Carnegie Institute of Washington, 813 Santa Barbara 
 Street, Pasadena, CA 91101, USA\\
$^\ast$ Based on observations collected at the MDM Observatory}
\email{dietrich@astronomy.ohio-state.edu}
%}
% the } above closes the list of footnotes

\begin{abstract}
\noindent
We have undertaken a new ground-based monitoring campaign on the broad-line
radio galaxy 3C\,390.3 to improve the measurement of the size of the broad 
emission-line region and to estimate the black hole mass. 
Optical spectra and {\it g}-band images were observed in late 2005 for three 
months using the 2.4-m telescope at MDM Observatory. 
Integrated emission-line flux variations were measured for the hydrogen Balmer 
lines H$\alpha$, H$\beta $, H$\gamma$, and for the helium line 
\heii $\lambda 4686$, as well as
{\it g}-band fluxes and the optical AGN continuum at $\lambda =$\,5100\,\AA .
The {\it g}-band fluxes and the optical AGN continuum vary simultaneously 
within the uncertainties, $\tau _{cent} = (0.2\pm1.1)$\,days.
We find that the emission-line variations are delayed with respect to the 
variable {\it g}-band continuum by 
$\tau (H\alpha)        = 56.3^{+2.4}_{-6.6}$\,days,
$\tau (H\beta )         = 44.3^{+3.0}_{-3.3}$\,days,
$\tau (H\gamma)        = 58.1^{+4.3}_{-6.1}$\,days, and
$\tau ($\heii \,4686$) = 22.3^{+6.5}_{-3.8}$\,days.
The blue and red peak in the double peaked line profiles, as well as the blue 
and red outer profile wings, vary simultaneously within $\pm3$ days. This
provides strong support for gravitationally bound orbital motion of the
dominant part of the line emitting gas.
Combining the time delay of the strong Balmer emission lines of H$\alpha$ and
H$\beta$ and the separation of the blue and red peak in the broad double-peaked
profiles in their rms spectra, 
we determine 
M$_{bh}^{vir} = 1.77^{+0.29}_{-0.31}\,\times\,10^8 M_\odot$ 
and using $\sigma_{line}$ of the rms spectra
M$_{bh}^{vir} = 2.60^{+0.23}_{-0.31}\,\times\,10^8 M_\odot$ 
for the central black hole of 3C\,390.3, respectively. 
Using the inclination angle of the line emitting region which is measured from
superluminal motion detected in the radio range, accretion disk models to
fit the optical double-peaked emission line profiles, and X-ray observations, 
the mass of the black hole amounts to 
M$_{bh} = 0.86^{+0.19}_{-0.18}\,\times10^9 M_\odot$ (peak-separation) and
M$_{bh} = 1.26^{+0.21}_{-0.16}\,\times10^9 M_\odot$ ($\sigma_{line}$),
respectively.
This result is consistent with the black hole masses indicated by simple 
accretion disk models to describe the observed double-peaked profiles, 
derived from the stellar dynamics of 3C\,390.3, and with the AGN 
radius\,--\,luminosity relation. Thus, 3C\,390.3 as a radio-loud AGN with a low
Eddington ratio, L$_{edd}$/L$_{bol}$ = 0.02, follows the same AGN 
radius\,--\,luminosity relation as radio-quiet AGN.
\end{abstract}

\section{Introduction}
The mass of super-massive black holes (SMBHs) is of fundamental importance for
understanding active galactic nuclei (AGN) which are powered by mass accretion 
onto a SMBH.
Analysis of AGN variability by applying the technique of reverberation 
mapping (RM) has been established as a powerful tool to determine the size of 
the broad emission-line region (BLR) of AGN (Blandford \& McKee 1982; 
Horne et al.\,2004; Netzer \& Peterson 1997; Peterson 2003) and, under the 
assumption of virial gas motion, consequently the SMBH mass 
(e.g., Peterson et al.\,2004). 

Large observational efforts are required to obtain data of sufficient quality
and temporal coverage to study AGN variability and to employ RM methods. 
Currently, about 50 AGN have been monitored for periods of at least a few 
months, usually as part of international collaborations. 
The entire available database of reverberation observations obtained through
2003 was uniformly re-analyzed with improved methods of time series analysis to
minimize systematic errors (Peterson et al.\,2004). 
Some of these results have been superseded by more recent experiments
(Bentz et al.\,2006b,\,2007a; Denney et al.\,2006,\,2009a; Grier et al.\,2008; 
Woo et al.\,2010) and
additional reverberation results steadily increase the size and quality of the
database (e.g., Bentz et al.\,2009b,\,2010a; Denney et al.\,2009b,\,2010).

It has been shown that AGN and non-active galaxies follow the same relation 
between the black hole mass, M$_{bh}$, and stellar velocity dispersion, 
$\sigma_\ast$, of the central spheroidal component of the host galaxy 
(Ferrarese et al.\,2001; Gebhardt et al.\,2000; G\"{u}ltekin et al.\,2009;
Merritt \& Ferrarese 2001; Tremaine et al.\,2002) and that the masses based on 
RM studies are consistent with the masses derived from the 
M$_{bh}$ - $\sigma_\ast$ relation (Nelson et al.\,2004; Onken et al.\,2004; 
Woo et al.\,2010).
The close relations of the black hole mass with physical properties of the host
galaxy, e.g. stellar velocity dispersion, bulge mass and bulge luminosity, 
indicate a coupled growth of the black hole and the formation and evolution
of the host galaxy (e.g., Silk \& Rees 1998; Haiman \& Loeb 1998,2001; 
Bromm \& Loeb 2003; Di\,Matteo et al.\,2004; Yoo \& Miralda-Escud\'e 2004; 
Volonteri \& Rees 2006; Volonteri \& Begelman 2010).

The broad-line radio galaxy (BLRG) 3C\,390.3 is a bright and nearby
($m_v = 15.0$, $z=0.056$; Osterbrock, Koski, \& Phillips 1976) FR\,II radio 
galaxy with extended double-lobed radio emission (Leahy \& Perley 1995). 
After the identification as optical counterpart of the radio source 3C\,390.3
(Wyndham 1966), it was classified as a N-type galaxy by Sandage (1966).
Soon, it was discovered that 3C\,390.3 shows very broad Balmer emission-lines
(Lynds 1968) which were later recognized as prominent double-peaked 
emission-line profiles (Burbidge \& Burbidge 1971). These double-peaked 
profiles are generally regarded as a characteristic signature of accretion 
disk emission (e.g., Eracleous \& Halpern 1994,2003; Gezari, Halpern, \&
Eracleous 2007).

So far almost all AGN variability studies have focused on radio-quiet AGN.
However, 3C\,390.3 has a well-known variability history (e.g.,
Cannon, Penston, \& Penston 1968; Selmes, Tritton, \& Wordsworth 1975; 
Barr et al.\,1980; Penston \& Perez 1984; Veilleux \& Zheng 1991; Zheng 1996; 
Wamsteker et al.\,1997; O'Brien et al.\,1998; Dietrich et al.\,1998; 
Sergeev et al.\,2002; Tao et al.\,2008), with photometric measurements 
going back to 1968 (Yee \& Oke 1981). Using the Harvard plate collection Shen, 
Usher, \& Barrett (1972) traced back brightness variations to 1895.
Thus, 3C\,390.3 was a prime target of a multiwavelength monitoring campaign
in 1994/95 especially because of
previous reports of dramatic changes in the Balmer line profile shape and 
strength and to perform coordinated X-ray/UV/optical monitoring of a radio-loud
AGN for the first time. 
Currently, 3C\,120 and 3C\,390.3 are the only radio-loud AGN which have been 
monitored in detail.
The multiwavelength study of 3C\,390.3 in 1994/95 (Leighly et al.\,1997; 
O'Brien et al.\,1998; Dietrich et al.\,1998) shows significant variations in 
the X-ray, UV, and optical domain over a period of one year.
However, the light curves of the UV and optical variations show a nearly 
monotonic increase of the continuum and emission-line flux with only moderately
strong substructures.  
The measured delays of several broad emission-lines relative to the observed
continuum variations are about $\tau \simeq 20\pm6$\,days for H$\beta $ 
and H$\alpha$ (Dietrich et al.\,1998) and $\tau \simeq 36$ to 60\,days 
($\pm 18$\,days) for \civ\ and Ly$\alpha$ (O'Brien et al.\,1998).
More recently, Sergeev et al.\,(2002) studied the variability characteristics
of 3C\,390.3 for a period of nearly one decade. They found a longer delay of 
the response of the  H$\beta $ emission with respect to continuum variations 
compared to the result of the 1994/95 study. They concluded that the difference
might be caused by different continuum variability characteristics which 
manifest in different continuum auto-correlation functions. 
Recently, Shapovalova et al.\,(2010), Jovanovic et al.\,(2010), 
Popovic et al.\,(2011), and Sergeev et al.\,(2011) discussed profile variations
over periods from a few years up to about 15 years.

In the following work, we present the results of a monitoring campaign 
undertaken at MDM Observatory in late 2005. 
Based on imaging and spectroscopic data, we find that 3C\,390.3 clearly showed 
broad band variations, delayed variations of the broad Balmer 
emission-lines H$\alpha$, H$\beta $, and H$\gamma $, and of the helium 
line \heii $\lambda 4686$. 
Using the time delays and the peak separation in the rms spectra of these
emission lines, we estimate a virial black hole mass of
M$_{bh}^{vir} = 1.77^{+0.29}_{-0.31}\,\times\,10^8 M_\odot$ for 3C\,390.3
and using $\sigma_{line}$ of the rms spectra
M$_{bh}^{vir} = 2.60^{+0.23}_{-0.31}\,\times\,10^8 M_\odot$, respectively. 
The detection of superluminal motion for 3C\,390.3 indicates an inclination 
angle of $i = 27^o \pm 2^o$ which allows to correct the observed velocity for 
the tilt of the line emitting region relative to the observer.
Taking this into account, we derive a black hole mass of 
M$_{bh} = 0.86^{+0.19}_{-0.18}\times 10^9 M_\odot$ (peak separation) and
M$_{bh} = 1.26^{+0.21}_{-0.16}\times 10^9 M_\odot$ ($\sigma_{line}$) for 
3C\,390.3, respectively,
which is, within the uncertainties, consistent with recent results based on the
\caii $\lambda \lambda 8494, 8542,8662$ stellar absorption triplet and the
$M_{bh} - \sigma_\ast$ relation
(Lewis \& Eracleous 2006; Nelson et al.\,2004). 

\section{Observations}
\noindent
We revisited 3C\,390.3 to determine the size of the BLR with improved accuracy.
In 2005 from September until December we used the 2.4-m Hiltner telescope at 
MDM Observatory to obtain spectroscopic and imaging data (Table\,1).

The instrumental setup of the Boller \& Chivens CCD Spectrograph (CCDS) and 
imaging camera RetroCam (Morgan et al.\,2005) was kept unchanged for the 
entire observing campaign to achieve a homogeneous data set, i.e., avoiding the
impact of different instrumental settings.
The spectrograph was equipped with a Loral 1200\,x\,800 pixel CCD, with a
projected pixel size that corresponds to $0\farcs4$/pixel.
To cover the entire optical wavelength range, a 150 grooves/mm grating was used
with a fixed grating tilt for the entire observing season. The slit 
width was set to a uniform value of $3\arcsec$ and a position angle of 
PA\,$= 90^\circ$.
To allow for correction of cosmic ray event contamination, at each epoch two 
spectra were recorded for 3C\,390.3 with an integration time of 
t$_{\rm int}$\,=\,900\,s each.
In contrast to previous campaigns we observed HD\,217086 (BD$+61^\circ 2373$, 
O7Vn) for the entire campaign for flux calibration
to minimize the uncertainty which is introduced by flux calibrations that are
based on different standard stars.
In addition, HD\,217086 was employed to correct for strong atmospheric 
absorption bands, i.e., the O$_2$ C-band ($\sim 6275 - 6300$\,\AA ), 
                 O$_2$ B-band ($\sim 6850 - 6950$\,\AA ), and
                 broad H$_2$O absorption ($\sim 7160 - 7320$\,\AA ).
This is especially important for the broad H$\alpha$ emission line profile
which is severely affected by the B-band absorption.
In total, optical spectra were recorded for 28 epochs (Table\,1).

In addition, 3C\,390.3 was simultaneously observed for 25 epochs in SDSS 
{\it g}-band (Fu\-ku\-gi\-ta et al.\,1996) using RetroCam at the 2.4-m 
telescope (Table\,1).
RetroCam was equipped with a E2V CCD\,55-20 (1152\,x\,770 pixel) with a pixel
size of $22.5\mu$m, which corresponds to $0\farcs 259$/pixel.
The optical design of RetroCam provides the valuable option to switch between 
spectroscopic and imaging mode in less than 2 minutes.
Thus, the {\it g}-band measurements were taken no more than a few minutes 
before the spectroscopic observations. The exposure time was uniformly set to 
t$_{\rm int}$\,=\,120 sec and generally one exposure was recorded.

The spectra of 3C\,390.3 and the standard star HD\,217086, as well as the
{\it g}-band imaging data were processed using MIDAS\footnote{Munich Image 
Data Analysis System, trade-mark of the European Southern Observatory} 
software. Standard dark and flat-field corrections were applied to the spectra 
and {\it g}-band imaging data of 3C\,390.3 and the HD\,217086 spectra.
We subtracted the night sky intensity for each spectrum individually.
The night sky fit was based on two regions, $\sim 100\arcsec$ to $130\arcsec$
wide on average, which were separated by $\sim 60\arcsec$ relative to the 
spectrum of 3C\,390.3 and HD\,217086, respectively. A $3^{rd}$ order 
polynomial fit was used for each wavelength element to fit the spatial 
intensity distribution of the night sky emission. 

Argon\,--\,xenon comparison spectra were recorded for each spectrum of
3C\,390.3 for wavelength calibration, which is based on $\sim 15$ to 20
individual lines. The wavelength calibration yields a sampling of 
$(3.07\pm 0.05)$\,\AA /pxl. The spectral resolution, measuring the full width 
at half maximum (FWHM) of isolated lines in the argon\,--\,xenon spectra, 
amounts to $R \simeq 280$ at $\lambda \simeq 5600$\,\AA .
The spectra of 3C\,390.3 were corrected for galactic reddening using
$E_{B-V} = 0.072$ (Schlegel et al.\,1998).

To transform the observed spectra into the rest-frame of 3C\,390.3, we measured
a redshift of z =$\,0.056 \pm 0.001$ which is based on the narrow emission 
lines of H$\alpha$, H$\beta $, [\oi ]$\lambda 6300$, 
[\oiii ]$\lambda \lambda 4959,5007$, and [\oiii ]$\lambda 4363$.
In the following, we present the analysis of the rest-frame spectra of 
3C\,390.3 and discuss the results based on this spectra.

\section{Data Analysis}
\subsection{{\it g}-Band Imaging}
We used the flat-field corrected images to measure the {\it g}-band flux of 
3C\,390.3
relative to comparison stars in the field. The approach of relative photometry 
has the advantage of being quite insensitive to weather conditions, i.e.,  
photometric conditions are not required.
The BLRG 3C\,390.3 is among a sample of 34 AGN which were carefully studied by
Penston, Penston, \& Sandage (1971). 
They provide Johnson U,\,B,\,V magnitudes for three non-variable stars in the 
field of 3C\,390.3. RetroCam which has a field of view of 
$5\farcm0 \times 3\farcm3$,
allows us to observe stars A and B together with 3C\,390.3 (Fig.\,1). However, 
with an apparent brightness of $m_V = 11.^m71$, star A is usually saturated in 
the observed {\it g}-band frames. 
Star B has a magnitude ($m_V = 14.^m28$) comparable to that of 3C\,390.3. This 
star was already successfully employed in a prior monitoring campaign of 
3C\,390.3 as a flux standard (Dietrich et al.\,1998). Hence, we used star B to 
determine the relative broad-band flux changes of 3C\,390.3.

To measure the {\it g}-band flux of 3C\,390.3 and of star B, we applied a 
square aperture with a size of of $13\arcsec \times 13\arcsec $ (Fig.\,1). By 
using such a large aperture for the comparison star and particularly for 
3C\,390.3 we ensured that the total flux is recorded independently of the 
seeing, which varied from $1\farcs1$ to $2\farcs7$ with an average of 
$(1\farcs7 \pm 0\farcs4)$ during this monitoring campaign.
To correct for the night sky flux, the object aperture was separated by a 
2\arcsec\ wide region (so-called no-man's-land) from a 2\arcsec\ wide region 
that enclosed the square aperture (Fig.\,1). This outer region provided an 
average sky brightness which was used to correct the observed flux of the 
object for the night sky contribution.

\subsection{Optical Spectroscopy}
\subsubsection{Intercalibration of the 3C\,390.3 Spectra}
To study broad emission-line flux variations, it is necessary to calibrate the
spectra to a uniform flux level. This can be achieved using the flux of 
forbidden narrow emission-lines from the narrow-line region (NLR). 
Due to the large spatial extension of the NLR (of the order of 100\,pc in 
Seyfert galaxies) and the long recombination time scales which are due to the
low gas density of the NLR
($\tau_{\rm rec} \approx 100$\,years for $n_e \approx 10^3$\,cm$^{-3}$), 
light travel-time effects and long recombination time scales 
will damp out short time-scale variability.
Thus, it can be assumed that the flux of narrow emission lines can be treated 
as constant, at least on time scales of years to decades 
(e.g., Peterson 1993,\,2000).

However, 3C\,390.3 needs careful additional attention. The NLR of 3C\,390.3 
seems to be very compact (Baum et al.\,1988; Sergeev et al.\,2002) and the 
small emission-line flux ratio of F([\oiii ]$\lambda\lambda 4959,5007$) to 
F([\oiii ]$\lambda 4363$) indicates the presence of high density gas 
($n_e \simeq 10^5$\,cm$^{-3}$).	
Furthermore, narrow-line variability has been reported for this object. 
Clavel \& Wamsteker (1987) analyzed the strength of the narrow Ly$\alpha$ 
and \civ $\lambda 1549$ emission lines. They found that the strength of these
two narrow lines decreased by a factor of about 2 to 3 from 1978 to 1986.
On the basis of optical spectra obtained between 1974 and 1990, 
Zheng et al.\,(1995) presented evidence that the fluxes of 
[\oiii]\,$\lambda\lambda 4959,5007$ follow continuum variations, although on a
longer time scale than the broad emission-lines. 
During the period of decreasing or increasing [\oiii] flux there might also 
be periods of several months or years of nearly constant [\oiii] flux.
Hence, we examined the data closely to test for variability of the 
[\oiii] emission lines before we used them for flux calibration.

To test the reliability of the narrow [\oiii ] emission lines as an internal
calibrator, the {\it g}-band imaging data are of great value. 
As noted earlier, these broad band flux measurements were taken nearly 
simultaneously with the spectra, i.e., within minutes of the spectral 
observations (Table\,1) and the {\it g}-band brightness of 3C\,390.3 was 
measured with high precision (see \S 4.2, $\sim 0.1$\% ).
To compare the {\it g}-band variations to those which can be derived from the
spectral data, the spectra were convolved with the transmission curve of the 
SDSS {\it g}-band filter. The fluxes of the convolved spectra were measured for
the transmission wavelength range of the {\it g}-band filter curve 
($\lambda = 4000$\,\AA\ to 6000\,\AA ).
To adjust the flux level of the spectra to the {\it g}-band measurements, the 
spectra were rescaled. These scaling factors are on average $(1.10 \pm 0.07)$.

We used these adjusted spectra, based on the observed {\it g}-band flux 
variations, to test whether the [\oiii]$\lambda 5007$ emission line flux was
constant during this monitoring campaign.
Although the broad-band flux measurements include the broad and narrow H$\beta$
emission-line flux, as well as the [\oiii ]$\lambda \lambda 4959,5007$ line 
emission, the contribution of emission line variability to photometric 
variations can be neglected (Walsh et al.\,2009) if it amounts to less than
$\sim10$\,\%\ of the broad band flux. For 3C\,390.3 we find that the H$\beta$ 
flux amounts to less than 5\,\%\ of the g-band flux.
To test the assumption that the [\oiii ]$\lambda 5007$ emission line flux was 
constant for the duration of this monitoring campaign, these adjusted spectra 
were internally flux calibrated by scaling each individual spectrum to a 
uniform flux of the [\oiii ]$\lambda 5007$ emission line. 
First, we calculated a mean spectrum which was used as a reference spectrum. 
Next, we applied a modified version of the scaling algorithm of van Groningen 
\& Wanders (1992) to adjust the flux of the [\oiii ]$\lambda 5007$ in the 
individual spectra to the reference spectrum.
This routine minimizes the residuals of the [\oiii ]$\lambda 5007$ in a
difference spectrum which is calculated for each spectrum with the reference 
spectrum.
The intercalibration to a uniform flux level was achieved by allowing different
flux scales, small wavelength shifts, and different spectral resolutions. The 
average scaling factor amounts to $(0.99 \pm 0.02)$. This result indicates that
the [\oiii ]$\lambda 5007$ flux was constant in the adjusted spectra with a 
flux of 
F([\oiii ]$\lambda 5007$) = $(189\pm4)\times10^{-15}$\,erg\,s$^{-1}$\,cm$^{-2}$
(Table 2).
It is interesting to note that during the monitoring campaign of 3C\,390.3 in
1995 (Dietrich et al.\,1998) the [\oiii ]$\lambda 5007$ emission line flux was
measured to be 
F([\oiii ]$\lambda 5007$) = $(161\pm4)\times10^{-15}$\,erg\,s$^{-1}$\,cm$^{-2}$
(rest frame flux) which is about $\sim 15$\,\%\ lower. This indicates that the 
[\oiii ]$\lambda 5007$ emission line flux was lower in 1995 compared to 2005
when 3C\,390.3 was in a less luminous state.
Based on the average scaling factor and the small scatter, it is justified to 
assume that the F([\oiii ]$\lambda\lambda 4959,5007$) emission-line flux was 
constant during this monitoring campaign.

While the H$\beta $ -- [\oiii ]$\lambda \lambda 4959,5007$ region is scaled 
with respect to the narrow [\oiii ]$\lambda 5007$ line, we investigated whether
this internal calibration holds for the H$\alpha$ region at longer wavelengths
and the H$\gamma $ -- [\oiii ]$\lambda 4363$ region at shorter wavelengths.
We intercalibrated the spectra for the H$\alpha$ region with respect to the 
fluxes of the relatively strong [\oi ]$\lambda 6300$ emission line and the 
H$\gamma $ -- [\oiii ]$\lambda 4363$ region using the narrow
[\oiii ]$\lambda 4363$ emission line. 
We found that the H$\alpha$ region required small adjustments to achieve a
constant [\oi ]$\lambda 6300$ flux. The average scaling factor for the 
H$\alpha$ region amounts to $(1.014 \pm 0.021)$.
For the H$\gamma $ -- [\oiii ]$\lambda 4363$ region, minor rescaling of the
spectra was also necessary. These scaling factors are on average 
$(1.014\pm0.004)$.

\subsubsection{Accounting for Host Galaxy and Optical \feii\ Emission}
A spectrum of an active galactic nucleus is the superposition of various 
sources, including the host galaxy, thermal and non-thermal continuum emission 
from the AGN, which can be described by a power-law continuum, Balmer continuum
emission, \feii\ line emission, and emission from other metal line transitions,
e.g., carbon, nitrogen, oxygen, magnesium, neon, sulfur. These contributions 
need to be accounted for in order to obtain reliable line measurements and the 
strength of the AGN continuum that most of our study relies on. 
We adopted a multi-component fit approach (e.g., Wills et al.\,1985) that we
have successfully applied in prior studies of high-$z$ quasars and Narrow Line
Seyfert\,1 (NLS1) 
galaxies (for more details see Dietrich et al.\,2002a,\,2003b,\,2005,\,2009).
We assume that the spectrum of 3C\,390.3 can be described as a superposition of
three components, 
      (i) a power law continuum (F$_\nu \sim \nu ^{\alpha}$),
     (ii) a host galaxy spectrum (Kinney et al.\,1996),
and (iii) a pseudo-continuum due to merging \feii\ emission blends.
Neither a Balmer nor a Paschen continuum emission were taken into account 
because the Balmer edge at $\lambda = 3645$\,\AA\ is close to the short 
wavelength end of our spectra of 3C\,390.3 and the strength of the Paschen 
continuum is only weakly constrained (Korista \& Goad 2001).

As demonstrated by Bentz et al.\,(2006a,\,2009a) it is crucial to correct for 
the host galaxy contribution to measure the strength of the AGN continuum flux 
density. 
Based on ACS/HST images which were obtained for 14 AGN of the AGN-watch sample 
(Peterson et al.\,2004), Bentz et al.\,(2009a) estimated the host galaxy 
contribution to the continuum flux density at $\lambda =$\,5100\,\AA\ for those
AGN, including 3C\,390.3. For the large aperture of 
($3\arcsec \times 9\farcs4$) which we used, the galaxy contribution amounts to 
$F_{gal}(5100{\rm \AA}) = (0.87\pm0.09)\,\times\,10^{-15}$ 
 erg\,s$^{-1}$\,cm$^{-2}$\,\AA $^{-1}$, based on ACS/HST imaging data
(Bentz priv.comm.). 
The average host galaxy contribution which we 
determined, based on the multi-component fits, is $F_{gal}^{obs}(5100{\rm \AA})
= (0.77\pm0.10)\,\times 10^{-15}$ erg\,s$^{-1}$\,cm$^{-2}$\,\AA $^{-1}$, i.e., 
within the errors consistent with the measurements based on the ACS image.
To construct representative host galaxy template spectra we combined several 
individual spectra of various Hubble types. These galaxy spectra were 
retrieved from the publicly available sample published by Kinney 
et al.\,(1996). The spectral resolution of these galaxy spectra amounts to 
$R\simeq 8$ \AA\ which is comparable to the spectral resolution of our spectra.

To account for \feii\ emission, we used the rest-frame optical \feii\ template 
that is based on observations of I\,Zw1 by Boroson \& Green (1992) which covers
the wavelength range from 4250\,\AA\ to 7000\,\AA . 
An alternative optical \feii\ emission line template was presented by 
V\'eron-Cetty et al.\,(2004) which in addition takes into account
NLR contributions to the \feii\ emission. In general, these two templates are
similar but in detail they differ at around $\lambda \simeq 5000$\,\AA\ and
$\lambda \gtsim 6400$\,\AA . 
The optical \feii\ emission in the spectrum of 3C\,390.3 is weak, however, and 
the width of the \feii\ emission is expected to be quite broad, as is seen for
other permitted emission lines. Therefore, the choice of the \feii\ emission 
template has no impact on the results of this study.

The components (i), (ii), and (iii) were simultaneously fitted to the 
spectra of 3C\,390.3 that were intercalibrated with respect to the 
[\oiii ]$\lambda\lambda 4959,5007$ emission-line flux, to determine the minimum
$\chi^2$ of the fit. 
Various host galaxy templates were employed and the best results were obtained
using an appropriately scaled spectrum of the E0 galaxy NGC\,1407, which is
consistent with the morphological type of the host galaxy of 3C\,390.3. The 
width of the \feii\ emission template was allowed to vary from 
FWHM\,=\,3000 km\,s$^{-1}$ up to 9000 km\,s$^{-1}$ in steps of 
$\Delta$FWHM\,=\,500 km\,s$^{-1}$. 
The best fit results were achieved for \feii\ emission templates whose width 
was on average FWHM\,=\,$(6000 \pm 1000)$ km\,s$^{-1}$, ranging from 
FWHM\,=\,5000 km\,s$^{-1}$ to FWHM\,=\,8500 km\,s$^{-1}$. 
In Fig.\,2 we show the fit for a typical spectrum of 3C\,390.3 from this study,
to illustrate the method to determine the different components.

The uncertainty of the fit was estimated based on the distribution of $\chi^2$
around the minimum. The spectral slope $\alpha$ of the continuum fit has an
average error of $\Delta \alpha \simeq 0.01$, while the uncertainties in the
continuum flux density at $\lambda = 5100$ and the \feii\ emission flux are of
the order of $\sim 5$\,\% .
The best fit of the components (i) to (iii) were subtracted before we continued
to analyze the emission-line profiles of interest.

\subsection{Narrow-Line Region Emission Lines}
The spectrum of 3C\,390.3 shows strong narrow-line emission (Fig.\,2). 
Hence, it is necessary to correct the broad emission-lines for NLR 
contributions before line flux and profile parameters are measured. 
In addition, the study of the narrow emission-lines provide information about
the ionizing continuum strength. Under the assumption that the NLR emission is
constant during the observing period in 2005 (September to December), we 
measured the line strength of the narrow emission-lines in the individual 
spectra of the entire campaign. 
This is an additional test for the quality of the intercalibration of the 
spectral region of the broad emission-lines, studied for variability.

We used a strong NLR emission line as a template profile to fit the narrow-line
spectrum and determine its flux (e.g., Whittle 1985c; Dietrich et al.\,2005).
Using high-quality, high-resolution spectra, Whittle (1985a,b) 
showed that NLR emission-line profiles are similar and in particular that the 
Balmer emission-line profiles are identical to the [\oiii]\,$\lambda 5007$ 
line profile within the uncertainties, even in the case of strong profile
asymmetries. Therefore, we used the strong emission-line profile of 
[\oiii]\,$\lambda 5007$ to obtain a representative NLR line profile template.
To obtain such a profile template we made use of a spectrum of 3C\,390.3 which 
we observed with the 2.2\,m telescope at Calar Alto Observatory/Spain
(Aug.\,30, 1994).
At this time 3C\,390.3 was at a lower intensity level than in 2005, i.e., the
narrow emission-lines are more pronounced (Fig.\,3). 
This spectrum which covers a wavelength range comparable to this study, has
the advantage of a higher spectral resolution (R\,$\simeq 700$ at $\lambda 
\simeq 6500$\,\AA ).

To isolate the [\oiii ] emission lines which are located in the extended red
wing of the double-peaked H$\beta $ line profile we used an appropriately
scaled H$\alpha $ profile to recover the red wing of H$\beta  $. This scaled
H$\alpha$ profile was subtracted and the residual was used to fit the
[\oiii ]$\lambda \lambda 4959,5007$ line profiles (Fig.\,4).
The [\oiii ] line profiles are well described using a dominant narrow 
Gaussian component (FWHM\,=\,$11.0\,{\rm \AA}\pm0.1\,{\rm \AA}$) and a 
broader Gaussian component (FWHM\,=\,$26.0\,{\rm \AA}\pm1.5\,{\rm \AA}$) 
which is slightly blue-shifted by $v_{shift}\simeq100$\,km\,s$^{-1}$, to 
account for the 
base of the line profile (Fig.\,5). We find that the narrow component carries 
$\sim 75$\,\%\ of the [\oiii ]$\lambda 5007$ emission-line flux. The individual
Gaussian components have a priori no physical meaning by themselves, but are 
just used to obtain a NLR line profile template.

\section{Results}
\subsection{The Narrow-Line Region Spectrum}
The [\oiii ]$\lambda 5007$ emission-line profile fit (\S\,3.3) was used to 
measure the NLR emission-line flux of the narrow lines in the spectra of 
3C\,390.3.
While the NLR emission-line profile was kept fixed, we allowed the strength of 
the profile template to vary and also allowed variations of the profile width 
and the location in wavelength space. 
For measurable emission-line doublets, 
we assumed that the line ratios are 
F([\oiii ]$\lambda 5007$)/F([\oiii ]$\lambda 4959$) = 2.94,
F([\oi ]$\lambda 6300$)/F([\oiii ]$\lambda 6364$) = 3.05, and
F([\nii ]$\lambda 6583$)/F([\nii ]$\lambda 6548$) = 3.06
(Osterbrock \& Ferland 2005) in order to fit the weaker line of each doublet.

Applying this approach we measured the strength of the emission lines and 
listed them in Table 2.
The measurement uncertainty for an individual spectrum is of the order of 
$\sim$\,5\,\%\ for strong narrow emission-lines like 
[\oiii ]$\lambda\lambda 4959,5007$. In the case of moderately strong lines like
the [\oi ]$\lambda 6300$, [\oii ]$\lambda 3727$, [\neiii ]$\lambda 3869$, and 
blended lines like [\nii]$\lambda \lambda 6548,6583$ the error is of the order 
of $\sim$\,10\,\%\ or less, while for the weakest lines it is $\sim$\,25\,\% , 
e.g., [\fex ]$\lambda 6374$, [\sii ]$\lambda \lambda 6716,6731$. In Table 2 we 
present the mean narrow emission-line fluxes as measured for all epochs and the
scatter of the distribution.

We corrected the observed line fluxes for internal reddening before the 
emission-line flux ratios were analyzed. 
Generally, it is assumed that the observed Balmer decrement, i.e. 
F(H$\alpha$)/F(H$\beta$)$_{obs}$, can be used to estimate the internal 
reddening (e.g., Davidson \& Netzer 1979; Netzer 1982).
However, there are strong indications that the intrinsic Balmer decrement of 
the emission-line gas of AGN is about $F(H\alpha)/F(H\beta) = 3.1$ instead of
the classical case B value for the BLR
due to collisional enhancement in the higher density partially ionized zone 
(e.g., Gaskell \& Ferland 1984; Halpern \& Steiner 1983; Netzer 1982).

For the NLR gas the strength of the [\oii ]\,$\lambda 3727$ emission 
indicates the presence of a large fraction of low density gas 
($n_e^{crit}(O^+) = 3.1 \times 10^3$ cm$^{-3}$ for $2D_{5/2}$ and 
$n_e^{crit}(O^+) = 1.6 \times 10^4$ cm$^{-3}$ for $2D_{3/2}$; 
Osterbrock \& Ferland 2005). Thus, additional collisional enhancement, 
particularly of H$\alpha$ emission, is expected to be weak.
Therefore, we used the classical value of the Balmer decrement of 
$F(H\alpha)/F(H\beta) _{int} = 2.85$, for pure recombination case\,B 
(e.g., Osterbrock \& Ferland 2005) to correct
the narrow emission-lines for reddening (e.g., Crenshaw et al.\,2001). 
We determine an E$_{B-V} = 0.205\pm 0.004$ for the NLR of 3C\,390.3.
Together with a Galactic reddening curve (Seaton 1979), we corrected the 
observed emission line ratios relative to $F(H\beta)$, applying

\begin{equation}
{F(line) \over F(H\beta)}_{int} = 
  {F(line) \over F(H\beta)}_{obs} \times 
   10^{\,0.4\,\,E_{B-V}\,\,(R_\lambda - 3.68)}
\end{equation}

\noindent
with $R_\lambda$ as the Galactic extinction coefficient at the wavelength of 
the corresponding emission line (Seaton 1979).
We also investigated the impact of a larger Balmer decrement, 
$F(H\alpha)/F(H\beta) = 3.1$ which has been suggested to be more typical for 
emission line ratios of the BLR, i.e., it takes into account collisional 
excitation and radiative transfer effects - (Osterbrock \& Ferland 2005).
We find that it results in reddening-corrected 
line ratios that are less than $\sim 8$\,\%\ larger for line ratios shortward 
of H$\beta$ and less $\sim 8$\,\%\ smaller for line ratios longward of 
H$\beta$. The reddening-corrected narrow emission-line measurements 
(flux$_{cor}$), are given in Table 2.

\subsection{Continuum Variability}

We measured the relative broad-band flux variations of 3C\,390.3 relative to 
the calibrated comparison star\,B in the field of 3C\,390.3 (Fig.\,1). The 
Johnson $V$ magnitude of this star is given as $m_V = 14.28$ (Penston, Penston,
\& Sandage 1971). During the monitoring campaign, star\,B and 3C\,390.3 showed 
comparable {\it g}-band brightness. This already indicates that 3C\,390.3 was 
in a brighter state during this monitoring campaign in late 2005 than in 1995
(Dietrich et al.\,1998).

Based on the measured {\it g}-band flux ratios, we calculate the {\it g}-band 
magnitude variations of 3C\,390.3. To do this we utilized the known $V$-band 
magnitude of the comparison star\,B. 
Fukugita et al.\,(1996) and Smith et al.\,(2002) give the transformation of a 
$g'$\,-band into $V$-band magnitude, which involves in addition the ($B-V$) 
color. 
Applying these relations we find a $g'$\,-band magnitude of $m_{g'} = 14.59$ 
(Fukugita et al.\,1996) and $m_{g'} = 14.62$ (Smith et al.\,2002) for star\,B,
respectively. 
However, we used the {\it g}-band filter of RetroCam (Morgan et al.\,2005) at 
the MDM 2.4\,m telescope to measure the broad-band flux of 3C\,390.3. It has 
been pointed out that there are small differences between the $g'$\,band 
magnitudes (used at the 0.5\,m telescope of the SDSS) and the {\it g}-band 
filter (used at the 2.5\,m telescope of the SDSS; Tucker et al.\,2006; 
Davenport et al.\,2007; Smith et al.\,2007). 

Thus, we had to transform the $g'$\,-band magnitude of the comparison star\,B
into a {\it g}-band magnitude.
We employed the transformation given by Tucker et al.\,(2006) with
$g = g' + 0.060 \times ((g' -r') - 0.53)$. To get an estimate of the color
$g' - r'$ we used the color of the star L107-S97 in Bilir et at.\,(2008) which
has a very similar ($B-V$) color to comparison star\,B. This results in a 
correction of $\Delta (g-g') = -0.0014$, which can be neglected. Therefore,
for comparison star\,B the {\it g}-band and the {\it g'}-,band magnitudes can 
be treated as identical.
In Figure 6 (top panel), we show the broad {\it g}-band flux density variations
of 3C\,390.3.

The errors of the {\it g}-band magnitudes of 3C\,390.3 and comparison star\,B
are estimates using the uncertainties of the flux measurements. These 
uncertainties are combined in quadrature to obtain the error of the flux ratio
and hence of the flux density of 3C\,390.3. Due to the high integrated count 
rates of comparison star\,B and of 3C\,390.3, in the range of several $10^6$ 
counts, the errors of the flux ratios are of the order of $\sim 0.1$\,\% .

Next, we measured the continuum flux density variations at 
$\lambda = 5100$\,\AA\ directly from the power law continuum fit for each 
spectrum (Table 3).
In Figure 6, the variations of $F_\lambda$(5100 \AA ) are displayed. It can be
seen that the AGN continuum and the {\it g}-band variations follow the same 
pattern and have comparable amplitudes. 

\subsection{Broad Emission-Line Flux Variations}
\subsubsection{H$\alpha \,\lambda $\,6563}
In the spectral region of the broad and double-peaked H$\alpha$ emission line
some of the corrected spectra still exhibit a weak residual continuum flux 
level. To adjust for the residual continua, we fit a linear local 
pseudo-continuum. The H$\alpha$ emission-line flux was measured within 
the wavelength range of 6200\,\AA\ to 6800\,\AA.
This spectral range contains contributions of several narrow emission lines, 
including H$\alpha _{nar}$. To correct for narrow line flux contamination the 
average intensities of the narrow emission-lines of 
[\oi ]$\lambda \lambda 6300,6364$, [\fex ]$\lambda 6374$, 
[\nii ]$\lambda \lambda 6548,6583$, [\sii ]$\lambda \lambda 6716,6731$, and 
H$\alpha _{nar}$ were subtracted (Table\,2) and the resulting broad H$\alpha$
emission line flux is given in Table 3. 

To estimate the uncertainties of the broad H$\alpha$ flux measurements, first 
the uncertainties of the multi-component fit were taken into account. 
As noted earlier, we assumed
that the strength of the power law continuum fit, the host galaxy
contribution, and the optical \feii \,emission can be determined within 1\,\%.
The next source of error is given by the uncertainties of the narrow 
emission-line intensities (Table\,2). The largest source of error is the level 
of the pseudo-continuum. The strength of the linear pseudo-continuum fit is 
based on two windows which are located at 6135\,\AA\ - 6160\,\AA\ 
and 6875\,\AA\ - 6895\,\AA , respectively. Using the scatter of the
continuum level of these two windows, the error introduced by the 
pseudo-continuum fit was determined.
The individual errors were combined in quadrature to obtain the uncertainty of
the individual broad H$\alpha$ emission-line flux measurements (Table\,3).
In Figure 6, we present the light curve of the broad H$\alpha$ emission line
flux.
Although the scatter is larger than for the {\it g}-band and 
$F_\lambda$(5100 \AA ) variations, the broad H$\alpha$ flux shows a similar 
variation pattern which is shifted to later epochs.

\subsubsection{H$\beta \,\lambda$\,4861 and \heii \,$\lambda 4686$}
As in the case of the H$\alpha $ emission line, it was necessary to correct for
a residual continuum level of the H$\beta \,\lambda 4861$ -- 
\heii $\lambda 4686$ wavelength range. The linear local pseudo-continuum fits 
are based on two 20\,\AA\ wide continuum regions centered at 
$\lambda = 4520$\,\AA\ and $\lambda = 5240$\,\AA , respectively. The corrected 
spectra were used to measure the broad H$\beta \,\lambda 4861$ and 
\heii $\lambda 4686$ emission-line fluxes. 
Because the emission-line profiles of 3C\,390.3 are so broad (Fig.\,2) it is 
necessary to evaluate and to correct for the mutual blending of the 
H$\beta \,\lambda 4861$ and \heii $\lambda 4686$ emission, i.e., to identify 
the contributions of the H$\beta \,\lambda 4861$ and \heii $\lambda 4686$ 
emission in the overlapping region.

We used the broad H$\alpha \,\lambda 6563$ emission-line profile as a 
template to estimate the shape and strength of the broad \heii $\lambda 4686$ 
emission. To extract the broad H$\alpha \,\lambda 6563$ emission-line profile, 
the mean spectrum of the campaign was used (\S 4.4) and the mean H$\alpha$ 
profile was corrected for the narrow emission-line contributions 
([\oi ]$\lambda \lambda 6300,6364$, [\fex]$\lambda 6374$, 
[\nii]$\lambda \lambda 6548,6583$, H$\alpha _{nar}$, and
[\sii ]$\lambda \lambda 6716,6731$). The width of the double-peaked broad 
emission-line profile of H$\alpha \,\lambda 6563$ was scaled to have the same 
width in velocity space at the location of the H$\beta \,\lambda 4861$ emission
line and the strength was rescaled to fit the central part and red wing of 
the H$\beta \,\lambda 4861$ profile. In Figure 7 we show the 
H$\beta \,\lambda 4861$ -- \heii $\lambda 4686$ range corrected for narrow 
emission of \heii $\lambda 4686$, H$\beta \,\lambda 4861$, and 
[\oiii]$\lambda\lambda 4959,5007$. The scaled double-peaked H$\alpha$ 
emission-line profile describes the broad H$\beta \,\lambda 4861$ emission
profile well, except the blue hump which is less prominent in the H$\beta$
profile compared to the H$\alpha \,\lambda 6563$ emission-line profile. This 
results in an absorption line-like feature in the residual spectrum (Fig.\,7, 
bottom panel) which indicates a steeper Balmer decrement for the blue peak
in the double-peaked profiles.
To measure the \heii $\lambda 4686$ emission-line flux, the residual emission 
at the location of the \heii $\lambda 4686$ line was fit with a rescaled broad 
H$\alpha \,\lambda 6563$ double-peaked profile and also with a single broad 
Gaussian profile. 
The profile in the residual spectrum of the mean spectrum is better represented
by a Gaussian profile than using an appropriately scaled double-peaked 
H$\alpha$ profile (Fig.\,7). 
To determine an appropriate Gaussian profile for the broad \heii $\lambda 4686$
emission, the blue wing of the \heii\ line in the residuum was extracted and 
assuming that the broad \heii\ profile is symmetric the blue wing was mirrored 
to represent the red wing of the \heii $\lambda 4686$ line profile (Fig.\,8). 
The resulting profile can be described with a single Gaussian profile at 
$\lambda _c = 4686.14\,{\rm \AA}\pm0.04$\,\AA\ and a profile width of 
FWHM\,=\,$242.3\,{\rm \AA}\pm1.4$\,\AA\ 
($15500\,{\rm km\,s}^{-1}\pm100$ km\,s$^{-1}$).
To measure the \heii $\lambda 4686$ emission-line light curve (Fig.\,6) a 
Gaussian profile of fixed width (FWHM = 15500 km\,s$^{-1}$) was fit to the 
blue wing of the \heii $\lambda 4686$ profile in the spectra of 3C\,390.3
(Fig.\,8). 
The flux of the \heii $\lambda 4686$ emission was measured from this Gaussian 
fit for a wavelength range of $\lambda = 4472$\,\AA\ to $\lambda = 4900$\,\AA.
This Gaussian profile fit was subtracted from the spectrum to measure the broad
H$\beta \,\lambda 4861$ emission flux in the range from $\lambda = 4680$\,\AA\ 
to $\lambda = 5000$\,\AA\ (Table\,3). The resulting light curves for the broad
H$\beta \,\lambda 4861$ emission line flux, corrected for narrow H$\beta$ 
emission, and the \heii $\lambda 4686$ emission line, are displayed in Fig.\,6.

To estimate the total error of these flux measurements, we assumed that the 
uncertainties of the multi-component fit are of the order of 1\,\% .
The errors introduced by the correction of the narrow emission lines are taken 
from Table 2. The uncertainty due to the broad \heii $\lambda 4686$ emission is
of the same order as the total error that is caused by the narrow emission 
lines.
As in the case of H$\alpha \,\lambda 6563$, the dominant source of error is the
level of the pseudo-continuum fit. The level of the pseudo continuum fit was 
varied within the scatter given by the two ranges centered at 
$\lambda_c = 4520$\,\AA\ and $\lambda _c = 5240$\,\AA\ that are 20\,\AA\ wide.
The individual contributions to the uncertainty were combined in quadrature to 
obtain the total uncertainties of the individual broad H$\beta$ and 
\heii $\lambda 4686$ emission-line flux measurements (Table\,3).

\subsubsection{H$\gamma \,\lambda $\,4340}
In the case of the H$\gamma $ emission line profile region, we correct for a 
residual continuum level as well. The linear local pseudo-continuum fits are 
based on two 20\,\AA\ wide continuum regions centered at 
$\lambda _c = 4160$\,\AA\ and $\lambda _c = 4530$\,\AA , respectively. The 
corrected spectra were used to measure the H$\gamma \,\lambda 4340$ 
emission-line flux (4220\,\AA\ to 4500\,\AA ).
These flux measurements were corrected for the emission line contributions of 
the narrow H$\gamma \,\lambda 4340$ emission and of the [\oiii ]$\lambda 4363$ 
emission, using the average narrow line fluxes as given in Table 2. The 
resulting light curve of the variable broad H$\gamma \,\lambda 4340$ emission 
line is shown in Fig.\,6. The variability pattern is very similar to the broad 
H$\beta$ variations.

To estimate the total error of the broad H$\gamma$ flux measurements 
(Table\,3), we assumed that the uncertainties of the multi-component fit are
of the order of 1\,\%. The errors introduced by the correction of the narrow 
emission lines are taken from Table 2. 
The dominant source of error is the level of the pseudo-continuum fit. The 
level of the pseudo-continuum fit was varied within the scatter given by the 
two 20\,\AA\ wide ranges ($\lambda_c = 4160$\,\AA\ and 
$\lambda _c = 4530$\,\AA ).
The contributions of the individual error sources were combined in quadrature
to estimate the total uncertainty of the broad H$\gamma \,\lambda 4340$ 
emission line flux.

\subsection{Mean and rms Spectra}
\subsubsection{H$\alpha \,\lambda 6563$, H$\beta \,\lambda 4861$, and 
               H$\gamma \,\lambda 4340$}
Based on the calibrated spectra of 3C\,390.3, we calculated mean and residual
(hereafter simply {\it rms}) spectra
of the broad emission profiles of Balmer lines H$\alpha$, H$\beta$, and 
H$\gamma$ which are shown in Figure 9. 
These mean and rms spectra are based on the entire set of MDM spectra, except
the spectrum that was recorded at the epoch JD\,=\,244\,3696. The 
signal-to-noise (S/N) ratio for this spectrum is significantly lower than for 
all the other spectra of 3C\,390.3 on account of bad weather conditions
(S/N$_{2443696} \simeq 10$, compared to an average signal-to-noise ratio for 
the sample of S/N\,=\,$35\pm7$).

The mean spectra of the Balmer emission lines show a characteristic 
double-peaked emission-line profile with a strong blue hump while the red
hump is very weak. 
The rms spectra of the Balmer lines show a similar shape.
The blue hump is the dominant feature for the H$\alpha $ and H$\beta$ 
rms spectra while it is less pronounced for H$\gamma$.
Only weak residuals of narrow emission-lines can be seen in the Balmer 
rms spectra of H$\alpha$, H$\beta$, and H$\gamma$. 
The rms spectra of the Balmer emission lines also display extended red wings.

The line width can be characterized by its full width at half maximum 
(FWHM) and by the line dispersion $\sigma _{line}$ (the second moment of the 
line profile; e.g., Peterson et al.\,2004). In order to
characterize the uncertainties in each of these parameters we employed a 
procedure described by Peterson et al.\,(2004). For a set of {\it N} spectra, 
{\it N} spectra are randomly selected, in particular without regard whether the
spectrum has been previously selected or not. These {\it N} randomly selected 
spectra are used to construct mean and rms spectra from which the line width 
measurements are made. This process yields one Monte Carlo realization, and a 
large number ($N\simeq 10\,000$) of these realizations yield a mean and 
standard deviation for each of the measurements of the line width.
The results of the line profile measurements and the corresponding errors are
presented in Table 4.

We find that the mean spectra of the Balmer emission lines show a similar 
profile width of FWHM = 11900 km\,s$^{-1}$ to 13200 km\,s$^{-1}$ and the
second moment of the line profiles $\sigma_{line}$ ranges from 
4000 km\,s$^{-1}$ to 5400 km\,s$^{-1}$, respectively. 
The comparison of the rms spectrum profile widths show basically the same
result. The Balmer emission line rms spectra have an average profile width
of FWHM = $(11430 \pm 790)$\,km\,s$^{-1}$ and an average second moment of
the line profiles of 
$\sigma_{line}$\,= $(5160 \pm 310)$\,km\,s$^{-1}$, respectively.

\subsubsection{\heii $\lambda 4686$}
The mean and rms profiles are a Gaussian due to our approach to measure the 
\heii $\lambda 4686$ emission line flux and hence also the rms spectrum
(\S 4.3.2).
However, the rms spectrum of the \heii $\lambda 4686$ variations is contained 
in the rms spectrum of the broad H$\beta$ emission line (Fig.\,9). To recover
the \heii $\lambda 4686$ rms spectrum we examined two approaches.

First, the fit of the \heii $\lambda 4686$ emission-line was subtracted from 
each spectrum of the H$\beta $-[\oiii ]$\lambda \lambda 4959,5007$ emission
line region. This results in an uncontaminated broad H$\beta$ profile. These 
spectra are used to calculate a mean and rms spectrum which represent the 
H$\beta$ variation alone. In Figure 10 the mean spectrum of the 
H$\beta $-[\oiii ]$\lambda \lambda 4959,5007$ emission-line region with and 
without correction for the broad \heii $\lambda 4686$ emission is displayed,
as well as the comparison of the rms spectra. The difference of these rms
spectra can be associated with the \heii $\lambda 4686$ rms spectrum 
(Fig.\,10).

As an alternative approach to recover the \heii $\lambda 4686$ rms spectrum, we
assumed that the rms spectrum of the broad  H$\alpha $ emission-line can be 
used as a template for the rms spectrum of the H$\beta$ emission line. 
The H$\alpha$ rms spectrum was re-binned that the width is unchanged in 
velocity space at the location of the H$\beta$ emission-line. In Figure 11 the
rms spectra of the broad H$\beta$ and H$\alpha$ emission lines are shown.
Next, the re-binned H$\alpha $ rms spectrum was rescaled by dividing by 3.50 
to fit the red wing of the H$\beta$ rms spectrum. The difference between the
H$\beta $ rms spectrum and the scaled, re-binned H$\alpha$ rms spectrum can be 
associated with the \heii $\lambda 4686$ rms spectrum as well. This difference 
is displayed in the bottom panel of Fig.\,11. It can be seen that this 
representation of the \heii $\lambda 4686$ rms spectrum is nearly identical 
to the rms spectrum which is obtained by comparison of the rms spectra of the
H$\beta$-[\oiii ]$\lambda \lambda 4959,5007$ spectral region, corrected and
uncorrected for \heii $\lambda 4686$ emission, as described above.

The comparison of the \heii$\lambda 4686$ rms spectrum with the mean Gaussian 
profile fits to the \heii $\lambda 4686$ line profile indicates that the 
recovered rms spectrum of the broad \heii $\lambda 4686$ emission line tends to
show less variability in the far blue wing than the mean spectrum and thus the
FWHM is smaller. 
This result is consistent with the properties of the mean and rms spectra of 
the Balmer emission lines which also show narrower profiles in the rms spectra 
compared with the corresponding mean spectra (Table\,4).
We used both \heii $\lambda 4686$ rms spectra to measure their FWHM. While the
overall shapes of the rms spectra are very similar, the signal-to-noise ratio 
is different. Hence, we used the average of the FWHM, as well as the second 
moment $\sigma _{line}$ of the profile. The results are given in Table 4.

\subsection{Variability Characteristics}
We calculated basic variability parameters to characterize the statistical 
properties of the observed emission-line and continuum variations (Fig.\,6).
The results are presented in Table 5. Column (1) gives the spectral feature.
The mean flux and the standard deviation are presented in column (2) and (3). 
The excess variance, $F_{var}$, in column (4) is computed as

\begin{equation}
F_{var} = {1 \over {\overline{F}}} \, \sqrt{\sigma_F^2 - \Delta^2}
\end{equation}

\noindent
with
$\overline{F}$ as the mean flux (column 2), $\sigma_F$ as the rms of the flux
variations and $\Delta^2$ as the mean square value of the measurement errors 
(Rodriguez-Pascual et al.\,1997). Finally, in columns (5) and (6) the minimum 
and maximum fluxes are given for each light curve.
The relative variations of the AGN continua (F$_\lambda$(5100\AA) and 
{\it g}-band) and of the emission lines of H$\alpha$, H$\beta$, and H$\gamma $ 
are of the order of $\sim 5$\,\% . The weak \heii $\lambda 4686$ emission line,
however, shows a relative variation of about $\sim 15$\,\% . 
In spite of the larger uncertainties of the emission-line flux measurements of 
\heii $\lambda 4686$ these uncertainties are accounted for by the parameter
$\Delta $ in eq.\,2 to calculate the relative variability. 
There is a trend for stronger relative variations of emission lines with a 
higher ionization potential of the involved atom. The larger relative
variability is also consistent with the results of former variability studies
which find that F$_{var}$ of high-ionization lines like \heii $\lambda 1640$ or
\heii $\lambda 4686$ are always larger by a factor of about $\sim 2$ to 3 times
than F$_{var}$ for low ionization lines like H$\beta$ or H$\alpha$ (e.g.
Clavel et al.\,1991; Collier et al.\,1998; Dietrich et al.\,1993,\,1998;
Korista et al.\,1995; O'Brien et al.\,1998; Peterson et al.\,1991; 
Reichert et al.\,1994; Rodriguez-Pascual et al.\,1997;
Santos-Lle\'o et al.\,1997; Stirpe et al.\,1994; Wanders et al.\,1997).

\subsection{Time Series Analysis}
Generally, AGN monitoring campaigns have focused on the H$\beta$ -- 
[\oiii ]$\lambda \lambda 4959,5007$ spectral range
because the strong forbidden [\oiii ]$\lambda \lambda 4959,5007$ emission
lines can be used to establish a robust calibration of the individual spectra. 
With the large wavelength coverage of this campaign for 3C\,390.3 we are able 
to study the delayed response of not only H$\beta$ but H$\alpha$, H$\gamma$, 
and \heii $\lambda 4686$ as well. 

To measure the time delay between AGN continuum variations and variable broad 
line emission, the continuum light curve is used as the driving light curve 
with the line emission in response to these variations. To calculate the time 
delay between the observed variations, we use the interpolation correlation 
function (ICCF) method (Gaskell \& Peterson 1987; White \& Peterson 1994; 
Peterson et al.\,1998,\,2004). 
For this monitoring campaign we derived an AGN continuum light curve from the 
power-law continuum fits at $\lambda =5100$\,\AA , as well as the {\it g}-band 
light curve. 

First, we selected an appropriate continuum light curve. Since the {\it g}-band
flux contains variable broad H$\beta $ emission in addition to the AGN 
continuum, we studied the impact of the broad H$\beta $ line emission on the 
{\it g}-band variations. 
The contribution of the host galaxy can be assumed to be constant.
The {\it g}-band light curve and the F$_\lambda$(5100\AA) light curve (Fig.\,6)
are cross correlated using the ICCF method with F$_\lambda$(5100\AA) as the 
driving light curve. The resulting CCF is shown in Figure 12. 
Close inspection of this CCF shows that the {\it g}-band light curve is delayed
with respect to the F$_\lambda$(5100\AA) variations by 
$\tau _{cent} = (0.2 \pm 1.1$) days and 
$\tau _{peak} = (0.6 \pm 1.2$) days, respectively (Table\,6). 
Within the uncertainties, the F$_\lambda$(5100\AA) variations and the 
{\it g}-band variations are simultaneous and the detected small delay can be 
neglected. It is reasonable to attribute this hint of a delay to the variable 
broad H$\beta$ emission which is part of the {\it g}-band flux. 
Since the scatter in the {\it g}-band fluxes is much smaller than in the 
F$_\lambda$(5100\AA) continuum light curve, we selected the {\it g}-band 
variations as the driving continuum in the subsequent cross-correlation 
analysis. 
For comparison, we also provide the cross-correlation results using 
F$_\lambda$(5100\AA) as the driving continuum light curve (Table\,6).
The cross-correlation functions (ICCFs) for the broad emission lines are 
displayed in Figure 13. The ICCFs show some asymmetric shape and the region 
around the peak has some structure as well. 

To quantify the uncertainties in the time delay measurements, we employ 
the model-independent Monte Carlo FR/RSS method (Peterson et al.\,1998), 
including modifications described by Peterson et al.\,(2004). For each single
realization of the method, random subset sampling (RSS) is applied in which a 
light curve with N data points is randomly sampled N times without regard to 
the previous selection of each point. 
Flux randomization (FR), in which a random Gaussian deviation, based on the 
associated error, is then applied to each of the selected N points.
This FR/RSS-altered subset of data points is then cross-correlated as though it
were real data. The peak of the cross-correlation function, r$_{max}$, which 
occurs at a time lag $\tau _{peak}$, is determined, as is the centroid,
$\tau _{cent}$.
A cross-correlation peak distribution (CCPD) for $\tau _{peak}$ and 
a cross-correlation centroid distribution (CCCD) for $\tau _{cent}$ are built 
with a large number (N\,$\simeq$\,10,000) of Monte Carlo realizations. 
We take the average value of the CCPD to be $\tau _{peak}$ 
and the average of the CCCD to be $\tau _{cent}$.
To calculate the uncertainties $\Delta \tau _{up}$ and $\Delta \tau _{low}$ 
we assume that 15.87\,\%\ of the cross-correlation centroid realizations
have values $\tau > \tau _{cent} + \Delta \tau _{up}$ and 
15.87\,\%\ have values $\tau < \tau _{cent} - \Delta \tau _{low}$, with analog 
estimated uncertainties for $\tau _{peak}$.
These definitions of the uncertainty correspond to $\pm1\,\sigma$ for a 
Gaussian distribution. 

\subsubsection{Velocity-Dependent Broad Emission-Line Variations}
While the variability characteristics of the integrated broad-line flux can be 
used to infer the size of the BLR, the analysis of the response of different 
parts of the broad emission-line profile, e.g., the line core, and the profile
wings, yield information on the dominant gas motion, i.e., the kinematics of
the broad emission-line gas (e.g., Blandford \& McKee 1982; 
Horne et al.\,2004). 
Previous investigations of 3C\,390.3 have found that the profile wings vary in 
phase (Dietrich et al.\,1998; Popovic et al.\,2011; Sergeev et al.\,2002; 
Shapovalova et al.\,2010) which strongly favors orbital motion, i.e., virial
motion dominates the gas kinematics. This is also supported by accretion disk 
models to describe the double-peaked  Balmer emission line profiles 
(e.g., Eracleous \& Halpern 1994; Flohic \& Eracleous 2008).
Recently, results for several AGN have been presented (Denney et al.\,2009b; 
Bentz at al.\,2009b) which indicate a more complex picture with signatures for 
orbital motion, as well as for radial motions. 
In addition, for Arp\,151 the analysis of spectroscopically resolved broad 
emission-lines results into velocity-delay maps which indicate the presence of 
substructures like orbiting hot spots (Bentz et al.\,2010b).

As in former studies of 3C\,390.3, we extracted light curves for different 
parts of the broad emission-line profile. For better comparison with prior 
studies (e.g., Dietrich et al.\,1998; Gezari et al.\,2007; 
Shapovalova et al.\,2010) the width and placement of those regions are 
motivated by the location and the widths of the blue and red peak in the 
double-peaked line profile. 

In Table 7 the location of the blue peak and red peak for the H$\alpha$, 
H$\beta $, and H$\gamma $ emission-line profiles are given. Within 
$\sim 100$ km\,s$^{-1}$ both peaks are at similar velocities. Based on the mean
line profiles (Fig.\,9) the width of the peaks amounts to 
FWHM\,$\simeq$\,50\,\AA\ and FWHM\,$\simeq$\,70\,\AA\ for H$\beta$ and 
H$\alpha$, respectively. 
We investigated the profiles of the H$\alpha$ and H$\beta$ emission lines only
because the H$\gamma $ emission line is too weak and \heii $\lambda 4686$ is 
best represented by a single Gaussian profile.
We extracted the light curves of the blue and red peak for a 3000 km\,s$^{-1}$
wide region. In Figure 14 the boundaries of the blue and red wings, the blue 
and red peaks, and the line center are shown. For comparison the location of 
those profile sections are given in Table 8 which were used by 
Gezari et al.\,(2007).
The errors for the extracted light curves were determined following the same
approach as for the integrated emission line fluxes. 

Next, the H$\alpha$ and H$\beta$ line profile section variations are 
cross-correlated with the {\it g}-band light curve, as well as with the AGN 
continuum F$_\lambda$(5100\,\AA).
In Figure 15 the corresponding ICCF($\tau$) are shown. While
for H$\alpha$ the light curves of the blue wings, the blue and red peak and of
the line center result into ICCFs with well defined structures, the light curve
of the red wing of the broad H$\alpha$ line profile results into a ICCF which 
provides merely an indication for a possible delay (Fig.\,15). For the H$\beta$
line profile the ICCFs are broad and in the case of the red wing for H$\beta$ 
there is only a weak indication for a marginally defined peak (Fig.\,15).

The location of the centroid, the peak and the uncertainties in the ICCFs for 
the wings, the peaks, and the line center of the H$\alpha$ and H$\beta$ 
emission lines are determined using the technique described above (\S 4.3) and 
the results are given in Table 9. As can be seen already in Figure 15, the blue
and red peak of the H$\alpha$ line and of the H$\beta$ line vary basically with
the same time delay relative to the continuum variations, consistent with the 
model that the hydrogen Balmer line emission originates in a rotating disk 
(e.g., Eracleous \& Halpern 1994; Flohic \& Eracleous 2008). Within the 
uncertainties there is also no delay of the profile wings with respect to the 
peaks or the center of the line profile. 
To test this result we also applied the time series analysis to the variations 
of the blue and red wings and the blue and red peaks to eliminate uncertainties
which are introduced by the continuum light curves. The corresponding ICCFs for
the H$\alpha $ and H$\beta$ lines are displayed in Figures 16. 
Within the uncertainties we detect no time delay between the variations of the
blue and red peaks and wings (Table\,9).

\subsection{Stochastic Process Estimation for AGN Reverberation}

Generally, the time delay between continuum and broad emission line flux 
variations is determined by calculating cross-correlation functions. However,
for a better specification of the possible impact of gaps in the temporal 
coverage of the light curves, recently an alternative approach has been 
suggested by Zu et al.\,(2011) and applied by Grier et al.\,(2012) to the
measured variability of the NLS1 galaxy Mkn\,335. 
This method to analyze irregularly sampled light curves with a more statistical
approach was developed by Press et al.\,(1992) and Rybicki \& Press (1992).
They assumed that an irregularly sampled light curve can be treated as a damped
random walk process with a specific amplitude $\hat{\sigma}$ and an exponential
damping time scale $\tau_d$. 
In recent years it has been shown that quasar light curves can be, indeed, well
described with this approach (Kelly et al.\,2009; Zu et al.\,2012) and 
furthermore that the amplitude $\hat{\sigma}$ and damping time scale $\tau_d$ 
for quasar variability are found in a well defined region of the $\hat{\sigma}$
-- $\tau_d$ space (Koz\l owski et al.\,2010; MacLeod et al.\,2010). 
Recently, Mushotzky et al.\,(2011), however, have shown using Kepler data for
four AGN, that the variability of all AGN cannot be described with the random 
walk approach but that individual AGN have intrinsically different variability
characteristics.

To investigate also AGN variability data and in particular the impact of gaps 
in the temporal sampling in AGN light curves Zu et al.\,(2011) have modified 
this statistical approach.
We used their method of Stochastic Process Estimation for AGN Reverberation 
(SPEAR) as presented by Zu et al.\,(2011) to investigate the variations of
3C\,390.3 (Fig.\,6).
First, we used the {\it g}-band variation and the F$_\lambda$(5100\AA) 
variability to test whether the SPEAR results are consistent with our 
cross-correlation results (Table\,6). We started with an amplitude of 
$\hat{\sigma}=0.05$ and an damping time scale of $\tau_d = 60$\,days. With the 
{\it g}-band variations as a driver light curve (because of its much smaller 
errors than the F$_\lambda$(5100\AA) light curve), we find using SPEAR a delay 
of $\tau = 0.55 ^{+1.56}_{-0.45}$\,days which is consistent with the 
cross-correlation result. 

Next, we used the {\it g}-band light curve as the driver to recover the time 
delays of the H$\alpha$, H$\beta$, H$\gamma$, and \heii $\lambda 4686$ 
variations. Using SPEAR we computed time delays for the strong Balmer lines 
H$\alpha$ and  H$\beta$ which are basically identical within the errors to the
cross-correlation based delays. 
In the case of the H$\gamma$ and \heii $\lambda 4686$ emission lines, however, 
the delays determined with SPEAR and based on the cross-correlation analysis 
show some differences. The time delays of the Balmer lines H$\alpha$, H$\beta$,
and H$\gamma$ derived with {\it SPEAR} are consistent with those of the ICCF
analysis using F$_\lambda$(5100\AA ) as the driving continuum. However, for
the \heii $\lambda 4686$ emission lines the delays are just consistent within 
the 1-$\sigma$ errors. The time delay of the H$\gamma$ variations based on the
g-band light curve as the driver continuum and the {\it SPEAR} result differ
by nearly a factor of 2 (Table 6).
Furthermore, due to the larger uncertainties of the flux measurements for the 
H$\gamma$ and \heii $\lambda 4686$ emission lines, {\it SPEAR} needed to be 
restricted to a narrow range of allowed time delays to avoid spurious delays 
at $\tau \simeq 120$ days, which is longer than the period covered by the 
experiment.

SPEAR treats large gaps in the temporal coverage of a light curve in a
statistically based approach and also follows the impact on the uncertainty of
the time delay. However, in the case of the study of 3C\,390.3 with no large
gaps in the light curve, we applied SPEAR predominantly to compare the results 
with those of the ICCF analysis and the results of both methods do well agree. 
Hence, we used the results of the ICCF analysis in the following for 
consistency with former variability investigations which allows an easier 
comparison with those results.

\subsection{Black Hole Mass and Eddington Ratio}
\subsubsection{Black Hole Mass Estimates}

To determine the mass of the black hole in the center of 3C\,390.3, we assumed
that the line-emitting gas is bound, i.e., that the virial theorem can be 
applied. This approach is well-motivated by the results of the fitting of the
double-peaked emission line profiles of 3C\,390.3 with an elliptical disk model
(Flohic \& Eracleous 2008). In addition, for several AGN it has been possible 
to measure the time delay of additional broad emission line flux variations. 
These studies consistently show that emission-lines with higher ionization 
energies show shorter time delays, in keeping with the observed ionization 
stratification of the BLR and their closer locations to the central continuum 
source (Bentz et al.\,2009a; Kollatschny 2003; Onken \& Peterson 2002; 
Peterson \& Wandel 1999,\,2000).
Together with broader emission line profile widths, these results provide 
strong support that the gas is in gravitational bound motion.
The virial mass of the black hole is given by

\begin{equation}
M_{bh}^{vir} = f\,{{c\,\tau \,\,\Delta v^2} \over G} 
\end{equation}

\noindent
with $c\,\tau$ as the size of the broad emission line region, $\Delta v$ which
is associated with the line profile width, and $G$ as the gravitational 
constant. The factor $f$ depends of the geometry, kinematics, and the 
orientation of the line emitting region. It has been shown by 
Onken et al.\,(2004) that the average of $f$ amounts to $<f> = 5.5$ when the 
virial black hole mass, based on a rms spectrum profile width, is calibrated to
the $M_{bh} - \sigma_\ast$ relation for quiescent galaxies
(e.g., Tremaine et al.\,2002; G\"{u}ltekin et al.\,2009, 2011).
In the following we calculate the pure virial product, i.e., we will assume 
$f = 1.0$ because in the case of 3C\,390.3 and the successful accretion disk
models $v_{obs} = v_{intr}\times {\rm sin}\,i$ can be used to derive the
intrinsic velocity, $v_{intr}$, of the gas.

\begin{equation}
M_{bh}^{vir} = 1.951\times 10^5\,\left({\tau \over days}\right) 
               \left({\sigma_{line} \over {1000\,km/s}}\right)^2 \,\,M_\odot
\end{equation}

Using the second moment of the emission-line profiles $\sigma_{line}$ 
(cf.\,Peterson et al.\,2004) for the mean and rms spectra (Table 4) and the 
measured time delay of the broad 
emission-line flux response to continuum variations, given by $\tau _{cent}$ 
(Table 6), derived from a threshold of 80\,\%\ of the maximum of the ICCF, we 
calculated the virial black hole mass using eq.(4). The results are presented 
in Table 10. For comparison we estimated black hole masses for the individual 
emission lines using $\sigma_{line}$ from the the mean and the rms spectra 
and the time delay which we derived using the AGN continuum variations of 
F$_\lambda$(5100\AA) and of the {\it g}-band, as well (Table 6). 
We find that within the uncertainties the virial mass, M$_{bh}^{vir}$, of the 
black hole of 3C\,390.3, based on the mean emission line profiles, is in the 
range of 
$1.0\times10^8 M_\odot$ to $2.6\times10^8 M_\odot$ 
based on the properties of the mean profiles of the Balmer lines and the 
\heii $\lambda 4686$ emission line, independent of the continuum used 
(Fig.\,17, Table\,10). 
However, if only the strong Balmer emission lines H$\alpha$ and H$\beta$ are
considered because the uncertainties for the H$\gamma$ and \heii $\lambda 4686$
emission lines are larger, the virial products are in better agreement,
$2.3\times10^8 M_\odot$ to $2.6\times10^8 M_\odot$. 
The rms spectra yield results which are consistent with the mean-spectrum
with virial black hole masses of 
$1.1\times10^8 M_\odot$ to $3.0\times10^8 M_\odot$ (Table\,10) and the strong 
Balmer emission lines H$\alpha$ and H$\beta$ yield black hole
mass estimates of about $\sim 2.6\times10^8 M_\odot$.
Since, the rms spectrum represents the actually variable part of the emission 
line profile, we conclude that the virial product of the black hole of 
3C\,390.3, based on the strong Balmer emission lines H$\alpha$ and H$\beta$ is
M$_{bh}^{vir}(3C390.3) = (2.60^{+0.23}_{-0.31})\times 10^8\,M_\odot$.

To estimate the black hole mass M$_{bh}$ using reverberation mapping studies,
it is still necessary to specify the factor $f$ in eq.\,(3) which depends on
the not well-constrained geometry and kinematics of the gas, as well as on the
orientation of the line emitting region. While it is generally assumed that 
this factor $f$ is of the order of unity, it has been pointed out that a small 
inclination of the line emitting region can result in a significant 
underestimation of the black hole mass (e.g., Collin et al.\,2006)
if the gas motions are almost entirely in the disk plane with little motion in 
the polar direction.
In contrast to almost all AGN monitored so far, the BLRG 3C\,390.3 is unique 
since it is the only radio loud AGN with extended double-lobed radio emission 
(Leahy \& Perley 1995) whose variability has been studied in detail.
This allows us to estimate the impact of the inclination $i$ of the line 
emitting region on the black hole mass measurement and hence the factor $f$, 
assuming pure disk motion of the gas. 
Based on VLBI observations, Alef et al.\,(1988) reported indications for 
super-luminal motion for 3C\,390.3. Further radio observations and a detailed 
analysis provided information on the inclination $i$ of the axis of the radio
emission in 3C\,390.3 (Alef et al.\,1994,\,1998), which was found to be
$i = 28^o$\,$^{+5^o}_{-9^o}$.
The inclination of the radio axis of 3C\,390.3 has been also estimated based on
X-ray observations (Eracleous et al.\,1996). 
They successfully modeled the resolved Fe K$\alpha$ line emission with X-ray 
reprocessing in a cool, dense disk of gas at an inclination of $i \simeq 26^o$.
A third way to estimated the inclination of the disk-like line-emitting region 
is provided by fitting the double-peaked profile. Using a relativistic 
Keplerian disk Eracleous \& Halpern (1994) derived an inclination angle of 
$i = 26^o$\,$^{+4^o}_{-2^o}$. Recently, Flohic \& Eracleous (2008) analyzed
the variations of the double-peaked profile shape of H$\alpha$ for Arp\,102B
and 3C\,390.3. They found that the best disk models for 3C\,390.3 strongly
indicate an inclination angle of $i = 27^o \pm 2^o$.

In a simple model of an accretion disk as the dominant source of the observed
double-peaked emission-line profile, the observed velocity is actually
$v_{obs} = v_{intr}\times {\rm sin}\,i$.
Assuming an inclination angle of $i=27^o \pm 2^o$ this implies a correction 
factor of $f = 4.85^{+0.75}_{-0.60}$. Within the  uncertainties, this factor is
consistent with $<f> = 5.5$, determined by Onken et al.\,(2004) 
and $<f> = 5.25$ by G\"{u}ltekin et al.\,(2009) 
which were determined by comparing the reverberation-based black hole masses 
with the masses determined using the M\,--\,$\sigma$ relation, with the 
stellar velocity dispersion $\sigma$ based on stellar or gas dynamics of the 
host galaxy's bulge.
Applying the correction due to the measured inclination of the accretion disk, 
the black hole mass of 3C\,390.3 based on the H$\alpha$ and H$\beta$ rms 
spectrum profile properties and $\tau_{cent}$ yields
M$_{bh}({\rm 3C\,390.3}) = (1.26^{+0.21}_{-0.16})\times 10^9\,M_\odot$.

Accretion disk models have been shown to be successful in explaining the 
observed profiles and even profile variations over longer time scales for
3C\,390.3, Thus, the typical velocity of the line emitting gas can be 
associated with the location of the blue and red peak of the double-peaked 
emission-line profile, i.e., the location in velocity space of the dominant 
part of the line emitting gas. Using the separation of the blue and red peak 
of the Balmer emission-line profiles as $\Delta v$, we estimated the virial 
product using eq.(4) and $\tau_{cent}$ (80\,\%\ threshold level). The derived 
black hole masses are given in Table 10.
Based on the Balmer emission lines H$\alpha$, H$\beta$, and H$\gamma$ we find 
virial black hole masses in the range of 
$1.1\times10^8 \ltsim M_{bh} \ltsim 2.0\times10^8 M_\odot$ using the mean 
emission-line profiles and 
$0.9\times10^8 \ltsim M_{bh} \ltsim 1.9\times10^8 M_\odot$ based on
the rms spectra (Fig.\,17, Table\,10). 
However, this wide range is caused by the H$\gamma$ based results which have a 
larger uncertainty than the black hole mass estimates obtained from the strong 
H$\alpha$ and H$\beta$ emission lines which are consistent within the errors. 
Since the rms spectrum represents the actual variable part of the emission line
gas, we calculated the average virial black hole mass which is based on the rms
spectra of the Balmer emission lines H$\alpha$, H$\beta$, and H$\gamma$. We 
find a virial black hole mass of 
M$_{bh}^{vir} = (1.77^{+0.29}_{-0.31})\times10^8 M_\odot$.
Together with the correction factor $f = 4.85^{+0.75}_{-0.60}$ which is given 
by the inclination of the accretion disk, we find for the black hole mass for 
3C\,390.3 a value of $M_{bh} = 0.86^{+0.19}_{-0.18}\,\times\,10^9\,M_\odot$.

\subsubsection{Eddington Ratio}
Using the black hole mass estimates based on the broad Balmer and emission-line
profiles in the rms spectrum and the optical continuum luminosity of  
3C\,390.3 at $\lambda = 5100$\,\AA , we computed the Eddington ratio 
${\rm L}_{bol} / {\rm L}_{edd}$. The value of the conversion factor $f_L$ 
between the monochromatic luminosity and the bolometric luminosity 
(${\rm L}_{bol} = f_L \times \lambda \, {\rm F}_\lambda (5100)$), is still
debated in the literature (e.g., Elvis et al.\,1994; Laor 2000; Netzer 2003; 
Richards et al.\,2006). Recently, Marconi et al.\,(2008) suggested even a 
luminosity-dependent correction factor $f_L$, due to radiation pressure
effects. We assumed that the bolometric luminosity is given by 
$L_{bol} = 9.74\times \lambda \,L_\lambda (5100\,{\rm \AA })$ 
(Vestergaard 2004). We find $L_{bol} = 2.22\times10^{45}$ erg\,s$^{-1}$.
To calculate the Eddington luminosity ${\rm L}_{edd}$, we assumed that the gas 
is a mixture of hydrogen and helium ($\mu = 1.15$), i.e., 
${\rm L}_{edd} = 
             1.45\times 10^{38}\,{\rm M}_{bh}\,/\,{\rm M}_\odot$ erg\,s$^{-1}$.
The derived Eddington ratio $L_{bol} / L_{edd}$ for 3C\,390.3, based on this 
monitoring campaign, amounts to $L_{bol} / L_{edd} = 0.018^{+0.006}_{-0.005}$.
This low Eddington ratio is typical for radio-loud AGN like 3C\,390.3 (e.g.,
Boroson \& Green 1992; Boroson 2002). 

\section{Discussion}

\subsection{Times Series Analysis and the Size of the BLR}
Recently, results of studies on the long-term variability properties of 
3C\,390.3 have been presented by Sergeev et al.\,(2002,\,2011) and Shapovalova 
et al.\,(2010). In two studies Sergeev et al.\,(2002,\,2011) investigated the 
correlated variations of the optical continuum and the response of the broad 
H$\beta$ emission-line flux for the years 1992 to 2000 and 2000 to 2007, 
respectively.
In both studies they found that the H$\beta$ variations are delayed by
$\tau (H\beta) = 82^{+12}_{-10}$\,days ($1992 - 2000$) and 
$\tau (H\beta) = 94\pm6$\,days ($2000 - 2007$) and for H$\alpha$ delays of
$\tau (H\alpha) = 162^{+32}_{-15}$\,days ($1992 - 2000$) and 
$\tau (H\alpha) = 174\pm16$\,days ($2000 - 2007$) were determined. 
A comparable result was found by Shapovalova et al.\,(2010) who studied the 
optical variations for 3C\,390.3 from 1995 to 2007. They report that the 
variations of the broad H$\beta$ and H$\alpha$ emission-line flux are delayed 
by $\tau (H\beta) = 96^{+28}_{-47}$\,days and 
   $\tau (H\alpha) = 127\pm18$\,days,
both too long to be measurable from our data. 
However, due to the lower sampling rate of the H$\alpha$ light curve the ICCF 
analysis displays two possible peaks also at $\tau (H\alpha) \simeq 24$\,days 
and $\tau (H\alpha) \simeq 151$\,days, respectively.

Those results are inconsistent with the delays which were reported by Dietrich 
et al.\,(1998) using the data of the X-ray\,--\,UV\,--\,optical
monitoring campaign in 1994/95 (Leighly et al.\,1997; O'Brien et al.\,1998;
Dietrich et al.\,1998), i.e.,
$\tau (H\beta) = 23\pm4$\,days and $\tau (H\alpha) = 19\pm9$\,days
and also with the results of this study (Table 6).
However, the time delays which we determined for 3C\,390.3 in 1995 are 
consistent with the delays we find for H$\alpha$ and H$\beta$ in this study.
In 1995 the strength of the optical continuum corrected for host-galaxy
contributions was F$_\lambda$(5100\AA )\,$= 
1.16\times10^{-15}$ erg\,s$^{-1}$\,cm$^{-2}$\,\AA $^{-1}$. 
During the 2005 monitoring campaign the continuum was about $\sim 6\times$ 
stronger (Table 5).
Using the radius\,--\,luminosity relation for AGN (Bentz et al.\,2009a) it can 
be expected that the derived delays of the H$\alpha$ and H$\beta$ 
emission-lines should be about $\sim 2.5\times$ as long than those of 1995.
This is in good agreement with the delays which we have measured (Table 6).

To investigate the cause of the discrepancy between the results based on the 
analysis of the variations over about 10 years compared with those covering 
about three months up to one year, we re-analyzed the data published
by Sergeev et al.\,(2002,2011) and Shapovalova et al.\,(2010).
Using the data of the Sergeev et al. study,
we compiled H$\beta$ and F$_\lambda$(5100\AA ) continuum light curves which
cover about 15 years with 413 epochs for the continuum and 131 epochs for
H$\beta$. We applied the same ICCF analysis to these light curves as we did for
the measurements for this study. 
We found $\tau (H\beta) = 81.6^{+16.4}_{-22.5}$\,days which is consistent with
the results of Sergeev et al. and due to the large uncertainties, within 
1-$\sigma$ it is even in the range with the result of this study. Next, we 
studied parts of the light curve 
which covered approximately 1000 days to study the impact of the luminosity 
state of 3C\,390.3 on the delay. We found delays of the broad H$\beta$ 
variations ranging from $\tau (H\beta) = 13^{+34}_{-30}$\,days to
             $\tau (H\beta) = 102^{+11}_{-16}$\,days with no clear relation
between the continuum state and the delay. However, the mean spacing,
especially for the H$\beta$ emission-line flux measurements, is only 
$\sim20$\,days to $\sim160$\,days, i.e., probably not sufficient. Only for the 
last three years of the light curve the temporal sampling is about 
$\sim$$6$\,days for 
the continuum and $\sim$$20$\,days for the H$\beta$ emission-line flux. 
For this period we find a delay of $\tau (H\beta) = 102^{+6}_{-6}$\,days. 
The H$\beta$ emission line flux measurements by Sergeev et al.\,(2011) and by
Shapovalova et al.\,(2010) do overlap with our monitoring campaign. Therefore, 
we can compare those with our H$\beta$ emission line light curve (Fig.\,18).
It can be seen that the continuum light curve follows the variations we 
measured while the coverage in H$\beta$ is not as high. 

In the same way, we re-analyzed the continuum and H$\beta$ emission-line light 
curves given by Shapovalova et al.\,(2010). For the entire data set which 
covers about 12.5 years we find $\tau (H\beta) = 96.2^{+35.4}_{-35.2}$\,days,
consistent with the results of Sergeev et al.
Similar to the Sergeev et al. data the light curves were split up into 
periods of about 1000 days duration with temporal sampling of about 40 days.
We found delays of the broad H$\beta$ variations
ranging from $\tau (H\beta) = 16^{+39}_{-41}$\,days to
             $\tau (H\beta) = 210^{+55}_{-140}$\,days with no clear relation
between the continuum state and the delay.

In a study on the reliability of cross-correlation function time delay 
determinations Welsh (1999) has pointed out that variations on longer time 
scales, e.g., on the dynamical time scale of the BLR which is of the order of 
years, will shift the response time to longer delays. 
Long term variations are different from direct reverberation signals which 
measure the instantaneous response of the emission-line gas to continuum 
variations, while long term variations trace a more gradual response to an 
overall increase or decrease of the continuum strength.
To correct for this effect it is necessary to detrend light curves from those
gradual changes. To detrend the continuum and H$\beta$ emission-line 
variations, we fit a low order polynomial to the light curves.
Next, we applied the analog time series analysis to the detrended light curves 
of the Sergeev et al.\,(2002,2011) and Shapovalova et al.\,(2010) studies. We 
derived time delays of the H$\beta$ variations of 
$\tau (H\beta) = 90.4^{+8.9}_{-9.2}$\,days (Sergeev et al.) and
$\tau (H\beta) = 82.7^{+13.9}_{-12.9}$\,days (Shapovalova et al.).
Within the uncertainties (1-$\sigma$\,errors) these delays are consistent with 
those obtained with the original light curves. However, the uncertainties of 
the time delays using the detrented light curves are about a factor 2 to 3 
smaller than those for the light curves including gradual long term variations.
 
Furthermore, the properties of continuum strength variations have a significant
effect on the time delay measured from cross-correlation functions. This issue 
can be responsible for the different delays measured in our study and those 
using variations of the continuum and the broad emission lines over more than
10 years, in particular, the auto-correlation function of the continuum, 
ACF$_{cont}$. 
It has been already noted by Sergeev et al.\,(2002,2011) that the width of the 
ACF$_{cont}$ of the continuum light curve which is covering several years up to
more than a decade is much broader than the ACF$_{cont}$ of a shorter campaign 
like the one in 1994/95 and that this will result in a longer time delay
(see also the Appendix for a more detailed discussion).

In addition, our campaign covered only a little more than 80 days and hence we 
are not able to measure time delays of about 100 days and more.
We think that the different time delays are caused by the long time period of 
about a decade to cover the variation of 3C\,390.3 and the wider temporal 
sampling of the measurements.
Furthermore, the significantly different widths of the ACF$_{cont}$ and the 
fact that in addition to reverberation signals, i.e., the direct response of 
line emitting gas to continuum variations, also variations of the emission-line
flux are included. These are associated with changes of the physical conditions
and the distribution of the gas which happen on dynamical time scales and 
which are uncorrelated with continuum variability also contributes {\bf to} the
different time delays.
Therefore, a long duration campaign for 3C\,390.3 with densely sampled 
measurements will be necessary to find a definitive result.

\subsection{Black Hole Mass Estimates}
A wide range of mass estimates for the super-massive black hole of 3C\,390.3 
have been reported, ranging from 
$\sim 1.3\times10^8\,M_\odot$ up to $\sim 7\times10^9\,M_\odot$ using emission
line profile properties and estimates of the size of the BLR
(e.g., Barr et al.\,1980; Bentz et al.\,2009b; Clavel \& Wamsteker 1987;
       Gaskell 1996; Peterson et al.\,2004; Sergeev et al.\,2002,2011;
       Wamsteker et al.\,1997),
as well as using the \caii\ triplet in the near infrared employing the
M\,--\,$\sigma_\ast$ relation yielding about M$_{bh}\simeq 4$ to $5\times$
$10^8\,M_\odot$ (Nelson et al.\,2004).
Most of these studies favor a black hole mass for 3C\,390.3 of the order of 
5 to $10\times$ $10^8\,M_\odot$. Our measurement of the mass of the black hole
of 3C\,390.3 with 
$M_{bh} = 0.86^{+0.19}_{-0.18}\,\times\,10^9\,M_\odot$ 
(based on the separation of the blue and red peak in the rms spectrum)
and
$M_{bh} = 1.26^{+0.21}_{-0.16}\,\times\,10^9\,M_\odot$ 
(using $\sigma_{line}$) is consistent with the black hole mass based on the 
$M - \sigma_\ast $ relation.
Using the M-$\sigma_\ast$ relation from G\"{u}ltekin et al.\,(2009), we 
calculated the stellar velocity dispersion $\sigma_\ast$ which is expected for 
the host galaxy of 3C\,390.3 based on our estimated black hole mass. We find 
$\sigma_\ast = 257^{+90}_{-70}$\,km\,s$^{-1}$ for 3C\,390.3 which is consistent
with $\sigma_\ast = 273\pm16 $\,km\,s$^{-1}$ as measured by 
Nelson et al.\,(2004).
Furthermore, simple disk models which we applied to describe the overall 
structure of the broad double-peaked hydrogen Balmer emission lines yield a 
mass estimate of $M_{bh}\simeq 10^9\,M_\odot$.

An additional test for the reliability of the derived black hole mass for 
3C\,390.3 is given by the comparison of our measured values for the size and
continuum luminosity of 3C\,390.3 with the expected values derived from the
radius\,--\,luminosity relation.
Guided by simple photoionization models, a relation between the continuum
luminosity of an AGN and the radius of the BLR is expected and 
Kaspi et al.\,(2000) provided the first convincing evidence for such a 
relation. Careful re-analysis and additional observations 
(e.g., Peterson et al.\,2004; Bentz et al.\,2006b,2007a,2009b; 
Denney et al.\,2006,2009a,2010; Grier et al.\,2008; Onken et al.\,2003)
have reduced the uncertainty of the slope of the relation. It turned out that 
the correction for host galaxy contamination has a profound impact on the 
R\,--\,L relation (Bentz et al.\,2006a,2009a). With a slope of 
$\alpha =0.52\pm0.04$ (Bentz et al.\,2009a), we estimated the black hole mass 
of 3C\,390.3.
For the continuum flux at $\lambda = 5100$\,\AA\ we used the average continuum 
flux of the AGN continuum (Table 5), corrected for host galaxy contributions, 
with
$\lambda\,L_\lambda(5100{\rm \AA}) = 2.28\times10^{44}$ erg\,s$^{-1}$ 
and the measured delay for the H$\beta$ emission line is
$\tau _{cent} \simeq 44$ days. Using the radius\,--\,luminosity relation as 
given in Bentz et al.\,(2009a) 

\begin{equation}
\log\,R_{BLR} = K + \alpha \,\log\,(\lambda\,F_\lambda (5100{\rm \AA}))
\end{equation}

\noindent
with K\,=\,$-21.3^{+2.9}_{-2.8}$ and $\alpha = 0.519^{+0.063}_{-0.066}$ the 
H$\beta$ BLR radius amounts to $\tau = 52.7^{+2.9}_{-2.8}$ days,
which given the intrinsic scatter in the relationship, is consistent with the
measured $\tau _{cent}(H\beta) = 44.3^{+3.0}_{-3.3}$\,days or the 
$\tau _{cent}(H\beta) = 47.9^{+2.4}_{-4.2}$\,days using SPEAR.

\section{Summary}
We present results of a ground-based monitoring campaign on the broad-line
radio galaxy 3C\,390.3. 
Optical spectra and {\it g}-band imaging were obtained in late 2005 for three 
months using the 2.4-m telescope at MDM Observatory. 
Integrated emission-line flux variations were measured for the Balmer lines 
H$\alpha$, H$\beta$, H$\gamma$, and for the helium line \heii $\lambda 4686$, 
as well as {\it g}-band fluxes and the optical AGN continuum at 5100\,\AA.
The {\it g}-band fluxes and the optical AGN continuum are varying 
simultaneously within the uncertainties ($\tau _{cent} = -0.2\pm1.1$ days).
We measure time delays for the emission-line variations with respect to the 
variable {\it g}-band continuum of
$\tau (H\alpha)    = 56.3^{+2.4}_{-6.6}$\,days,
$\tau (H\beta)     = 44.3^{+3.0}_{-3.3}$\,days,
$\tau (H\gamma)    = 58.1^{+4.3}_{-6.1}$\,days, and
$\tau ($\heii \,4686$) = 22.3^{+6.5}_{-3.8}$\,days.
The blue and red peak in the double peaked line profiles, as well as the
blue and red outer profile wings vary simultaneously within $\pm3$ days. This
provides strong support for gravitationally bound orbital motion for the
dominant part of the line emitting gas.
Using the separation of the blue and red peak in the broad double-peaked 
profiles in the rms spectra of the Balmer emission lines 
and the corresponding time delays we determine a virial black hole mass of 
M$_{bh}^{vir} = 1.77^{+0.29}_{-0.31}\,\times\,10^8 M_\odot$ for the black hole
of 3C\,390.3. Using the inclination angle $i = 27^o\pm2^o$ of the line emitting
region the intrinsic velocity, $v_{intr}$, can be recovered from the measured 
$v_{obs} = v_{intr}\sin\,i$. This results in a black hole mass of 
M$_{bh} = (0.86^{+0.19}_{-0.18})\,\times10^9 M_\odot$ for 3C\,390.3
and M$_{bh} = (1.26^{+0.21}_{-0.16})\,\times10^9 M_\odot$ based on 
$\sigma_{line}$ of the rms-spectrum.
This mass estimate is consistent with the mass indicated by simple accretion 
disk models to describe the observed double-peaked profiles, as well as with
black hole masses derived from studies on the stellar dynamics of 3C\,390.3.
Furthermore, the mean continuum luminosity and the measured time delay for the 
broad emission-line flux variations of H$\beta$ is consistent with the most
recent AGN radius\,--\,luminosity relation. Thus 3C\,390.3 as a radio-loud AGN
with a low Eddington ratio of only L$_{edd}$/L$_{bol}$ = 0.02 follows the same
AGN radius\,--\,luminosity relation as radio-quiet AGN.

\begin{acknowledgements}
      We thank J.\,Halpern, S.\,Tyagi, and all the observers at MDM Observatory
      who conducted the observations in Fall 2005, for the first time in 
      service mode at MDM.
      We also acknowledge financial support from NSF grants AST-0604066 and
      AST-1008882 to OSU.
\end{acknowledgements}

\appendix
\section{Auto-Correlation Function and Time Delay}

Under the assumption that the observed emission-line flux $L(t)$ is the 
superposition of the response of the emission-line gas to continuum variations 
$C(t)$ of gas with the same time-delay, which are related by the transfer
function $\psi(t)$, the measured emission-line flux can be written as

\begin{equation}
  L(t) = \int ^{\infty}_{-\infty} \psi(\tau)\,C(t - \tau)\,d\tau
\end{equation}

\noindent
This equation can be also written using the CCF(t), ACF$_{cont}$(t) and the
transfer function $\psi(t)$ in the form

\begin{equation}
 CCF(t) =  \int ^{\infty}_{-\infty} \psi(\tau)\,ACF_{cont}(t - \tau)\,d\tau
\end{equation}

\noindent
as has been shown by Penston (1991), Koratkar \& Gaskell (1991), and 
Peterson (1993,\,2000). Hence, the measured time delay depends on both 
the ACF$_{cont}$ of the AGN continuum and the transfer function $\psi(t)$.
While the width of the ACF$_{cont}$ at full-width-at-half maximum (FWHM)
of the continuum variations studied by Sergeev et al.\,(2002,2011) and 
Shapovalova et al.\,(2010) amounts to about FWHM$\,\simeq 1600$\,days to 
2000\,days, the FWHM of the continuum ACF in our 2005 monitoring campaign is 
FWHM(ACF$_{cont}$)\,$\simeq 27$\,days, i.e., about $\sim60$ to $\sim75$ times 
narrower.
In the context of investigating the impact of variations on longer time scales 
and the necessity of detrending light curves for those trends to recover the
time delay, Welsh (1999) mentioned that the points in the ACF and CCF are 
highly correlated (e.g., Jenkins \& Watts 1969) and that those correlations 
result in spurious large values of the CCF, especially for light curves
which are characterized by intrinsically broad peaks. We suspect that the
different time delays are predominantly caused by the significantly different 
widths of the ACF$_{cont}$ (cf., \S 5.1). 

In addition, using the variations of the broad emission-line flux over times 
scales of the order of the dynamical time will include not only variations 
which are directly associated with continuum variations, i.e., 
pure reverberation events, but will also include variations caused by changes 
in the distribution and conditions of the line emitting gas.
Those variations are manifested, for example, in emission-line profile changes
which are uncorrelated with continuum variability, as shown by Wanders \& 
Peterson (1996). Hence, including emission-line profile variability will 
dilute a reverberation signal, i.e., the direct response of the gas on 
continuum variations. This aspect needs to be addressed when studying the 
relation of continuum and emission-line variability over time scales which are
comparable or even longer than the dynamical time scale for an AGN.

Based on the variability of NGC\,5548 which has currently the best covered
spectroscopically variability history for the optical continuum and the broad 
H$\beta$ emission-line flux, it has been shown that the measured time delay is 
strongly correlated with the strength of the continuum, i.e., whether the AGN 
is in a high or low state. Depending on the continuum luminosity, the delay of 
the H$\beta$ flux response to continuum variations varies between $\tau(H\beta)
= 4.2$\,days and $\tau(H\beta) = 26.4$\,days (Bentz et al.\,2009b; 
Peterson et al.\,2002). 
The continuum strength in the 13 years which were investigated in these studies
varied by a factor of about $\sim10$. The light curves which were studied by 
Sergeev et al.\,(2002,\,2011) and Shapovalova et al.\,(2010) cover more than 10
years. During this time the strength of the continuum of 3C\,390.3 varied by a 
factor of nearly $\sim 6$.

Finally, Horne et al.\,(2004) have studied the impact of the duration of a 
monitoring campaign on the measured time delay. They provide a guide line for 
the optimal length of monitoring campaign depending on the brightness of an 
AGN, to optimize also the use of telescope time. They found that a campaign 
should last at least about 3 times the light crossing time of the BLR to
recover the velocity resolved transfer function $\psi(\tau ,v)$. 
In the case of cross-correlation functions, shorter campaigns can still yield
reliable time delays, in particular if the light curve displays features of
increasing and decreasing continuum and emission-line flux and not only a
monotonic increase or decrease of the continuum and emission-line flux 
strength.


%\clearpage 

\begin{figure}
\epsscale{0.80}
\plotone{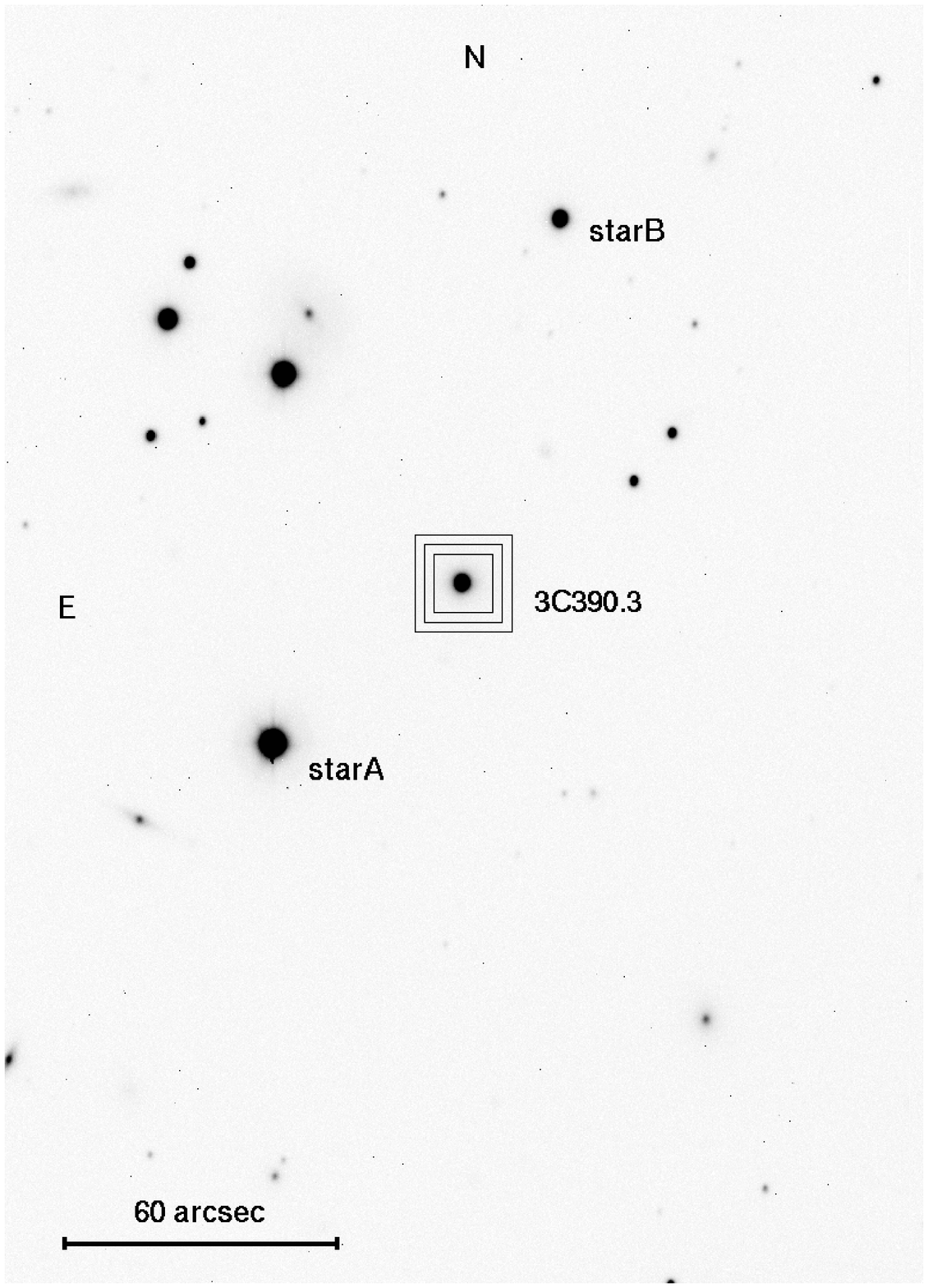}
\label{fig1}

\noindent
Figure 1 -- A typical {\it g}-band image of the field of 3C\,390.3, taken with 
            RetroCam on September 25, 2005 (t$_{\rm int}$\,=\,120\,sec). 
            3C\,390.3 and the comparison stars are marked. In addition, the 
            aperture and the location and size of the region which 
            we used to correct for the sky background, is shown.
\end{figure}

%\clearpage

\begin{figure}
\epsscale{1.00}
\plotone{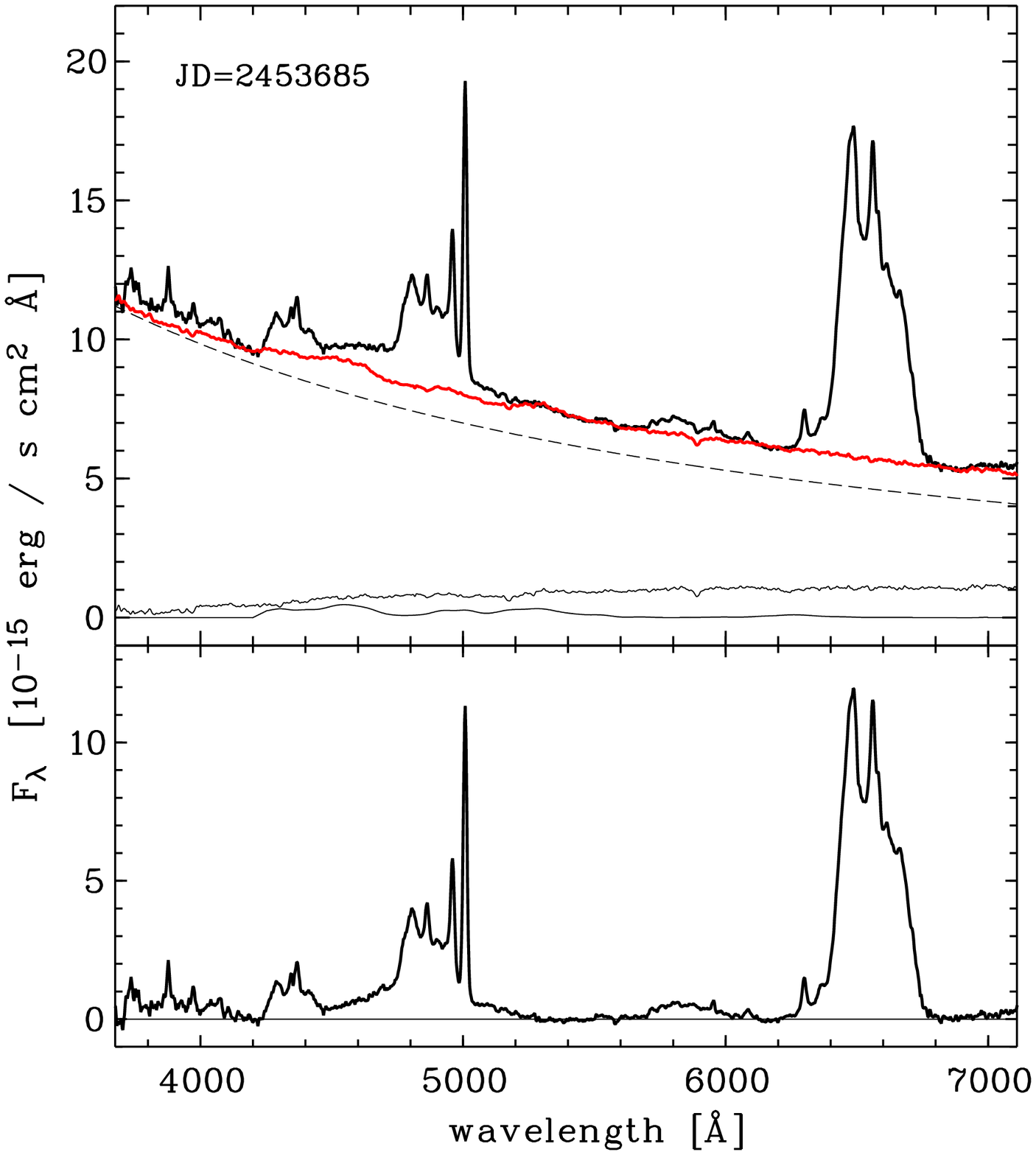}
\label{fig2}

\noindent
Figure 2 -- Decomposition of a spectrum of 3C\,390.3, observed on November 11, 
            2005. In the top panel the rest-frame spectrum is shown together 
            with the power law continuum fit (dashed line), the host galaxy 
            fit (upper thin solid line), the optical \feii\ emission fit (lower
            thin solid line), and the resulting combined fit (red line). 
            In the bottom panel the residual spectrum is displayed which shows 
            the pure emission-line spectrum of 3C\,390.3.
\end{figure}

%\clearpage

\begin{figure}
\epsscale{1.00}
\plotone{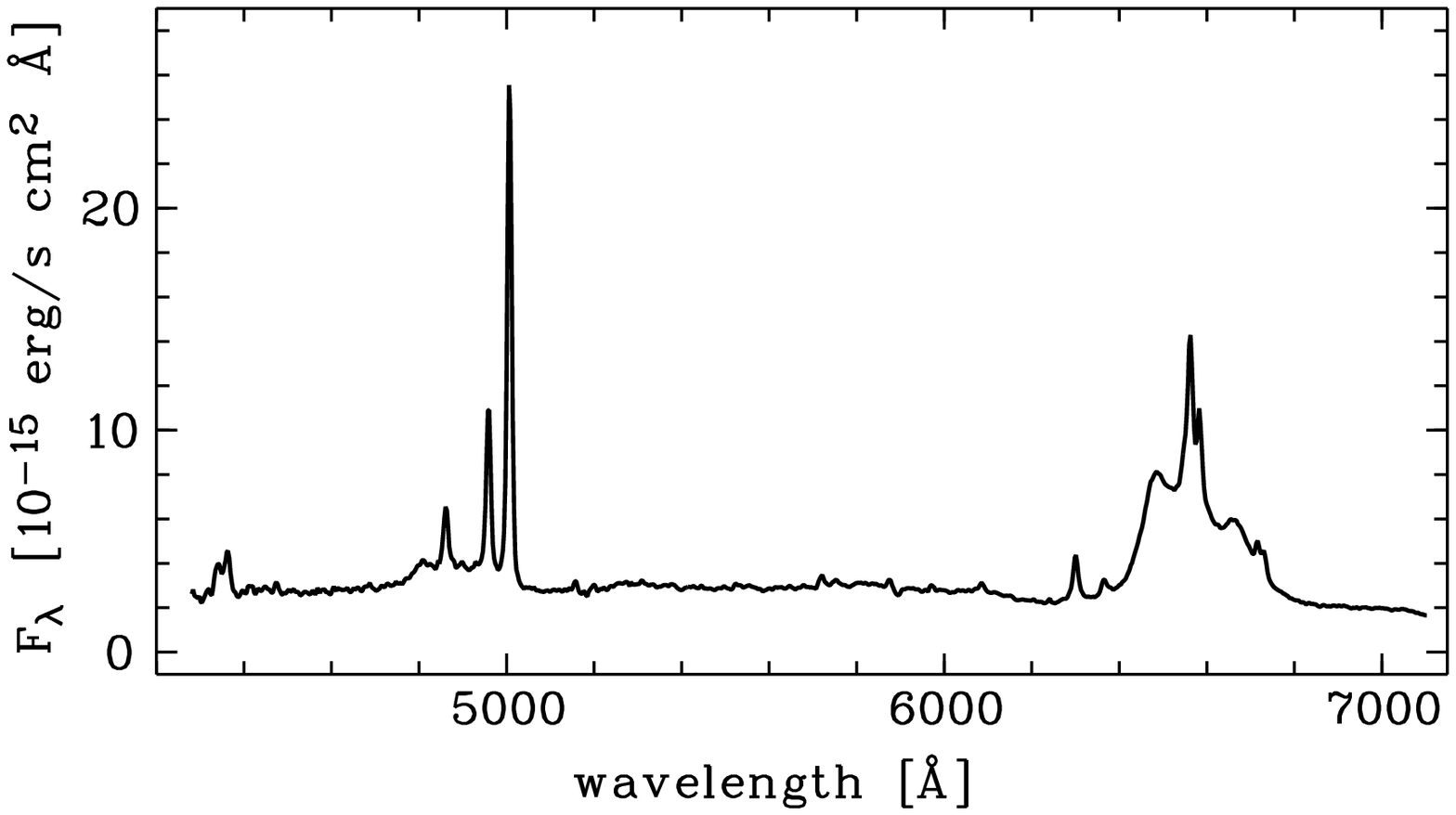}
\label{fig3}

\noindent
Figure 3 -- The spectrum of 3C\,390.3 during a low state
            (August 30, 1994, 2.2\,m at Calar Alto Observatory/Spain).
\end{figure}

%\clearpage

\begin{figure}
\epsscale{1.00}
\plotone{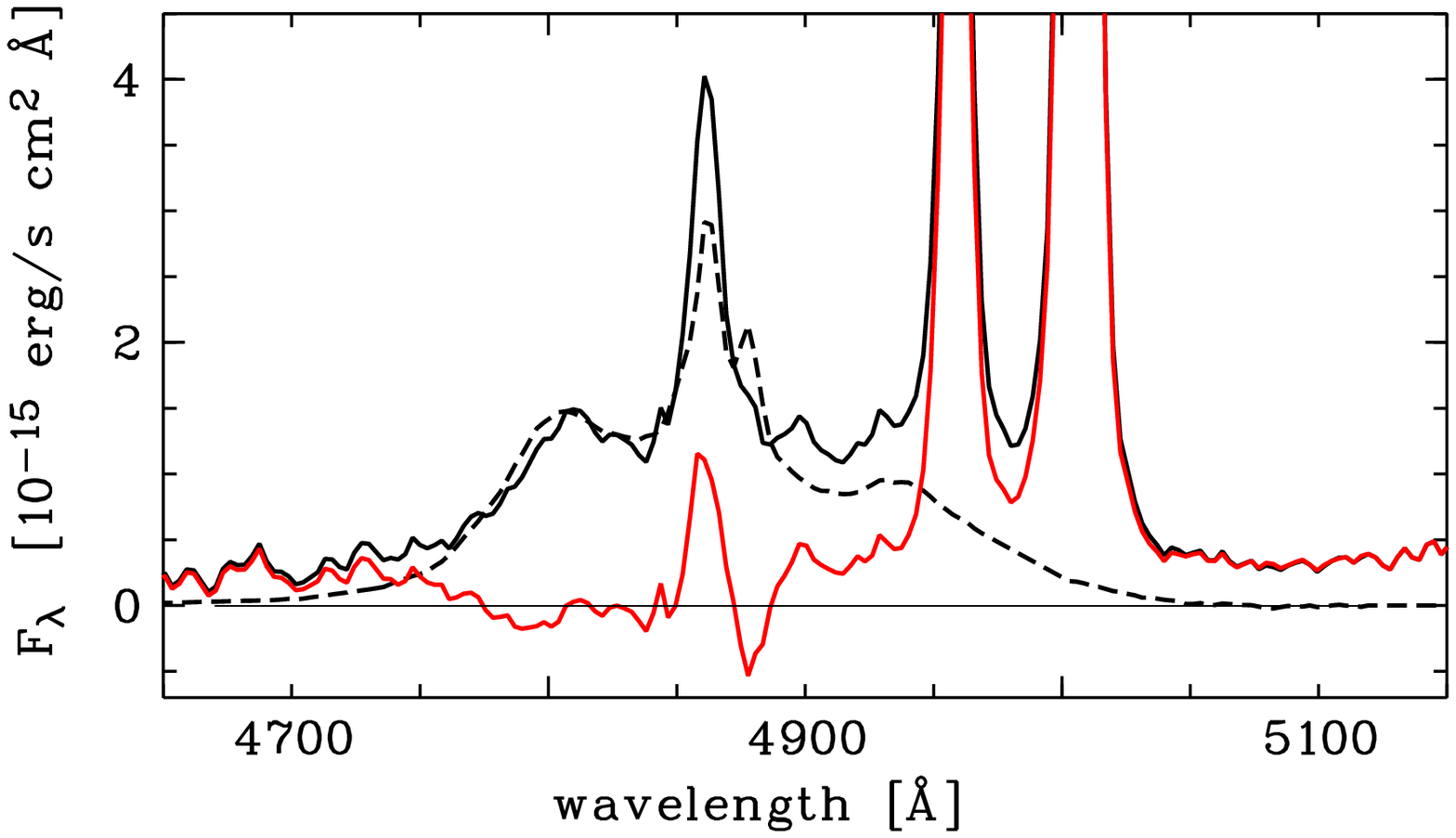}
\label{fig4}

\noindent
Figure 4 -- The H$\beta$ -- [\oiii ]$\lambda \lambda 4959,5007$ emission-line
            complex of the spectrum shown in Fig.\,3 (thick line). The scaled
            H$\alpha $ profile is shown as a long-dashed line. The residual
            spectrum, which is used to fit the [\oiii ] emission-line profiles,
            is shown as the red line.
\end{figure}

%\clearpage

\begin{figure}
\epsscale{1.00}
\plotone{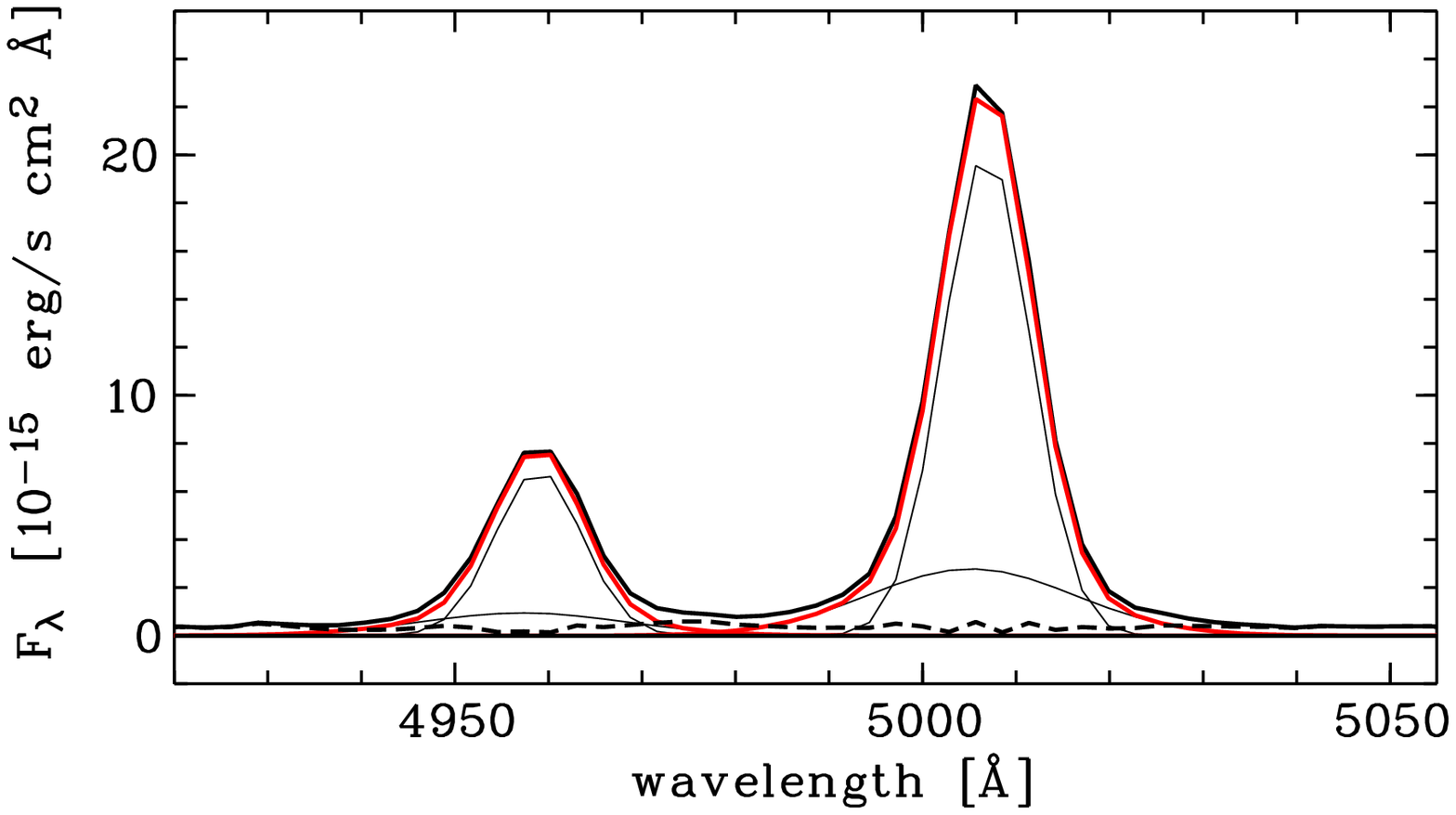}
\label{fig5}

\noindent
Figure 5 -- The fit of the [\oiii ]$\lambda 4959$ and [\oiii ]$\lambda 5007$
            emission-line profiles employing a strong narrow and a weaker,
            slightly blue-shifted broader Gaussian component for each line 
            (thin black lines). The resulting fit is shown as a thick red line 
            which recovers the observed spectrum well, as can be seen by the 
            residuum (dashed line).
\end{figure}

%\clearpage

\begin{figure}
\epsscale{0.55}
\plotone{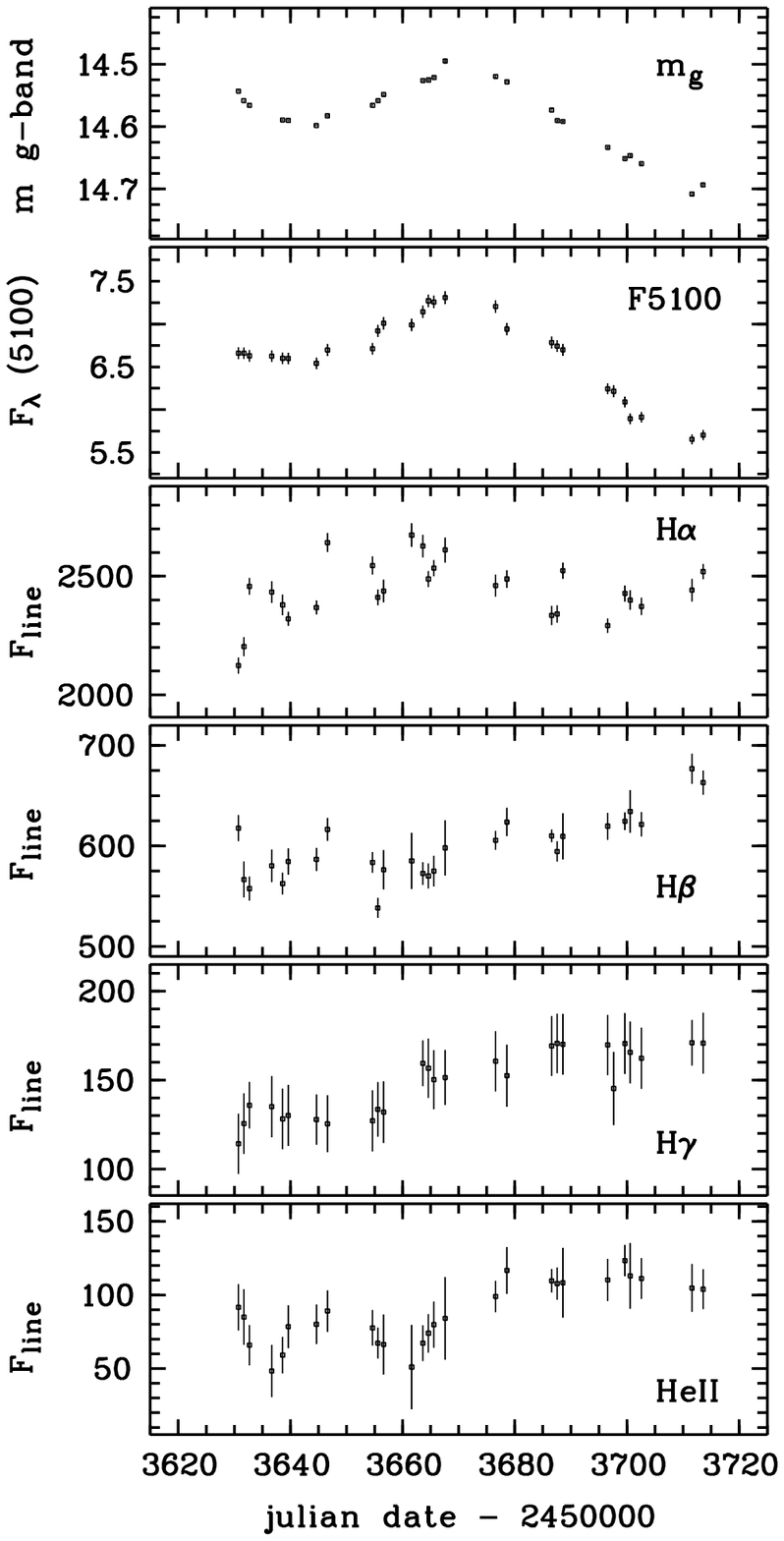}
\label{fig6}

\noindent
Figure 6 -- Light curves of 3C\,390.3 for the {\it g}-band (in {\it g}-band
            magnitudes), the F$_ \lambda$(5100 \AA ) AGN continuum (in units 
            of $10^{-15}$\,erg\,s$^{-1}$\,cm$^{-2}$\,\AA$^{-1}$) and
            the variations of the broad emission-line fluxes of H$\alpha$, 
            H$\beta$, H$\gamma$, and \heii\,$\lambda 4686$ (displayed in units 
            of $10^{-15}$\,erg\,s$^{-1}$\,cm$^{-2}$). The uncertainties of the 
            {\it g}-band magnitudes are smaller than the plot symbol.
\end{figure}

%\clearpage

\begin{figure}
\epsscale{1.00}
\plotone{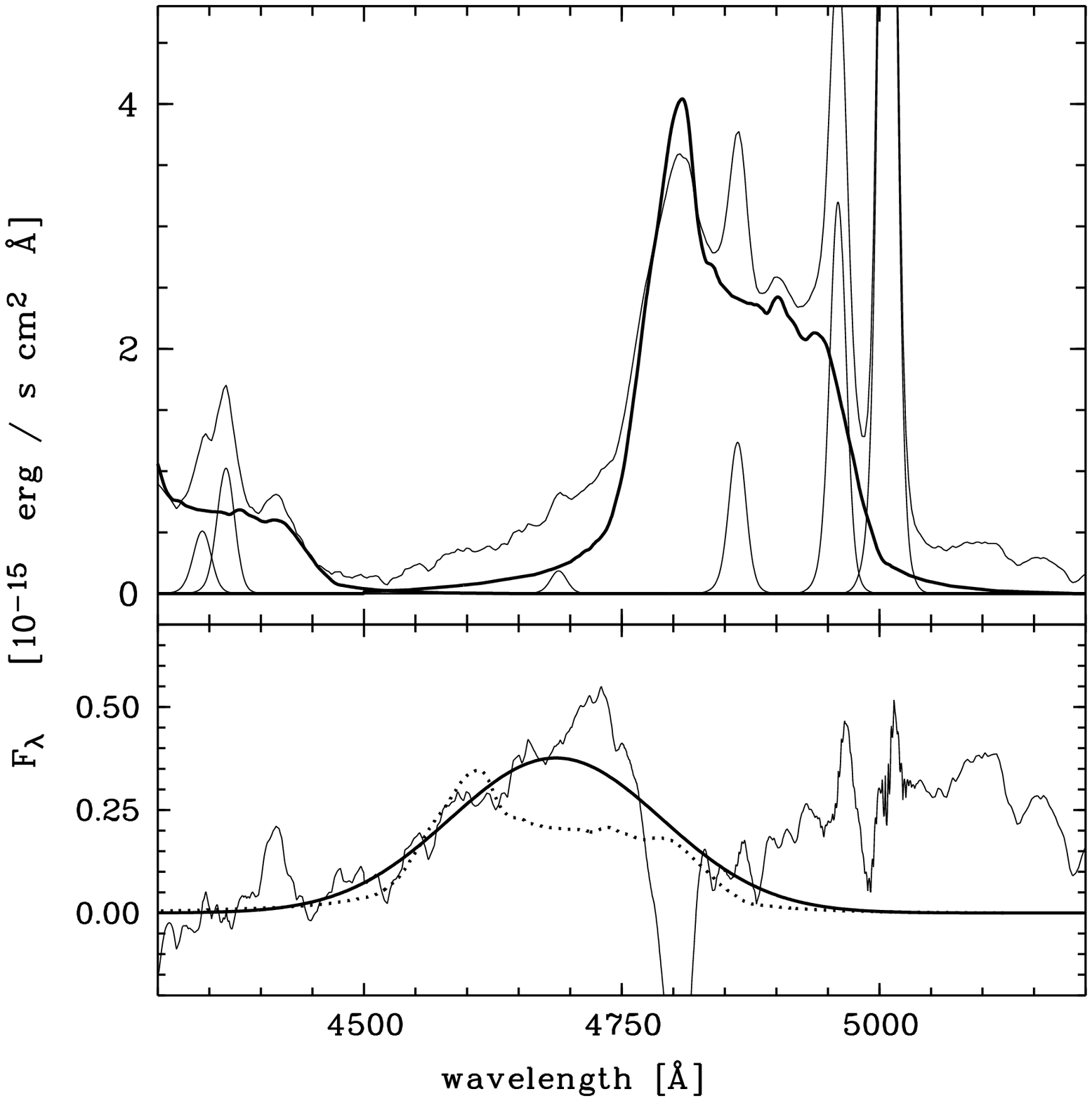}
\label{fig7}

\noindent
Figure 7 -- In the top panel the H$\beta $ -- [\oiii ]$\lambda \lambda 
            4959,5007$ complex is shown together with the scaled broad, 
            double-peaked profile fit of H$\alpha$ for the H$\beta $ and 
            H$\gamma $ emission lines (thick line). In addition, the profile 
            fits of the narrow emission lines for this wavelength range are 
            displayed. In the bottom panel, the residual spectrum is presented 
            after subtracting the profile fits of the narrow emission lines and
            the broad profile fits for H$\beta$ and H$\gamma $. The residual 
            profile at the location of the \heii\,$\lambda 4686$ emission line 
            is reconstructed with a scaled double-peaked profile (dotted line)
            and with a Gaussian profile (thick line).
\end{figure}

%\clearpage

\begin{figure}
\epsscale{1.00}
\plotone{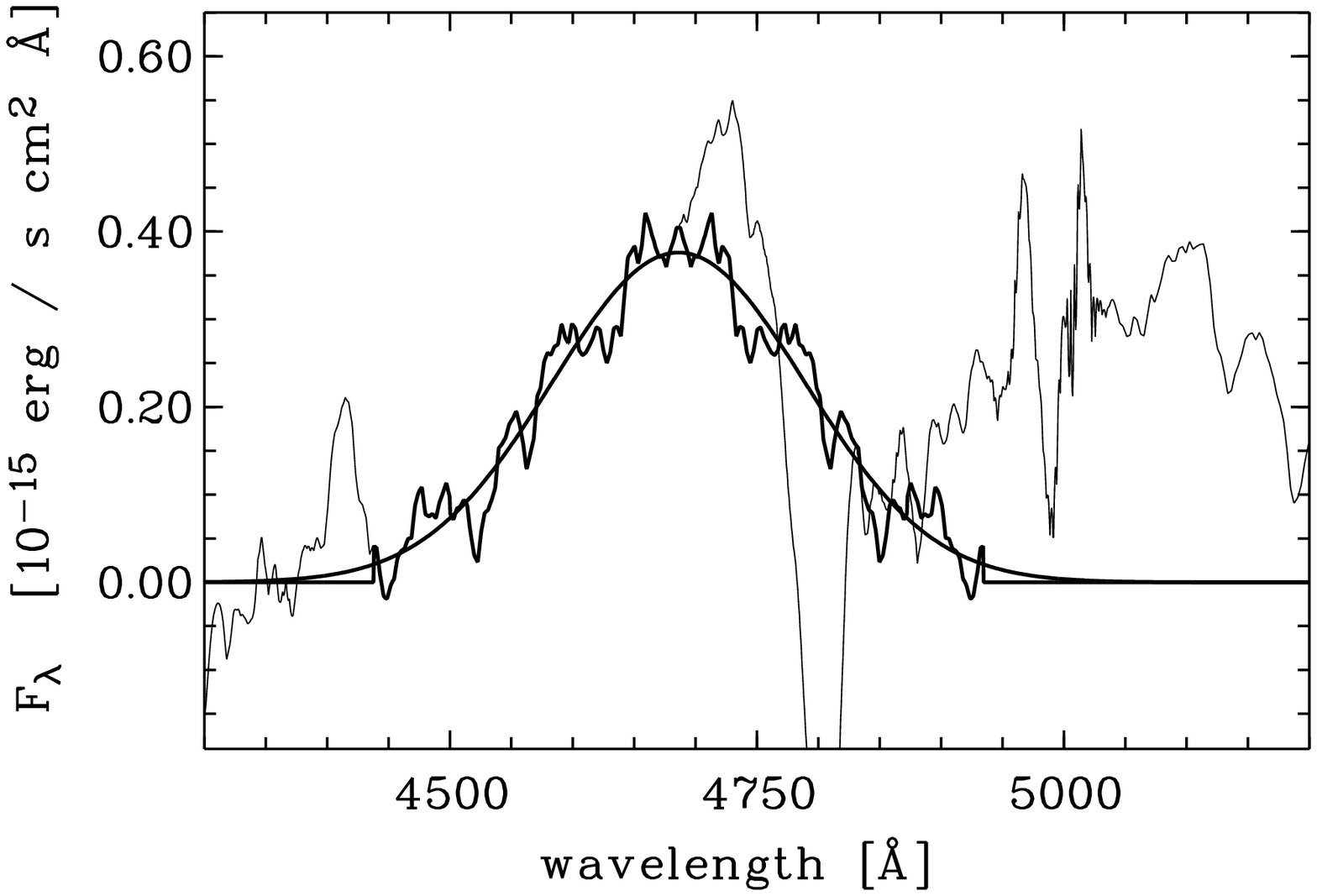}
\label{fig8}

\noindent
Figure 8 -- The Gaussian profile fit for the \heii $\lambda 4686$ emission-line
            profile, under the assumption that the blue-wing can be used to 
            represent the red wing (thick solid lines). 
\end{figure}

%\clearpage

\begin{figure}
\epsscale{0.50}
\plotone{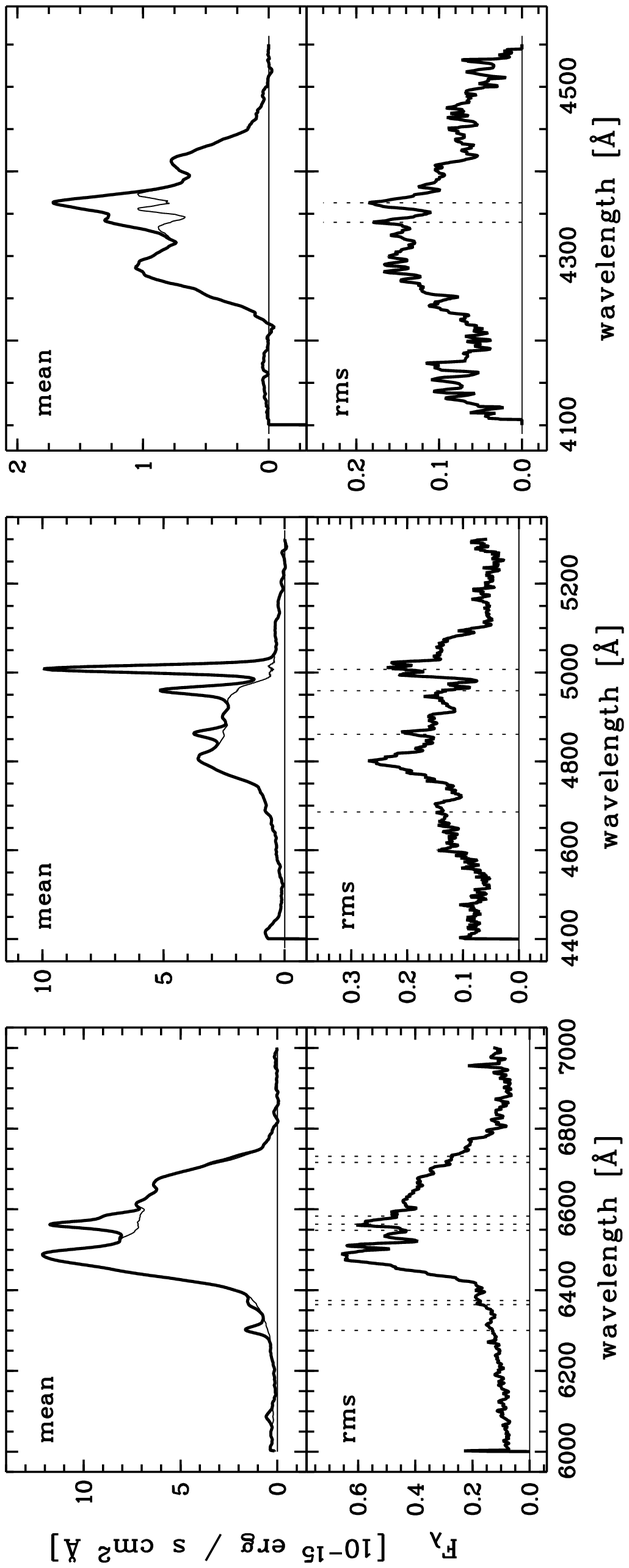}
\label{fig9}

\noindent
Figure 9 -- The mean and rms spectra of the hydrogen Balmer lines H$\alpha $, 
            H$\beta$, and H$\gamma$.
            The top panel shows the mean spectra (thick lines) and the 
            narrow-line subtracted mean spectra (thin lines).
            In the bottom panel the rms spectrum is presented (solid line) 
            together with the location of narrow emission lines (dotted lines).
\end{figure}

%\clearpage

\begin{figure}
\epsscale{1.00}
\plotone{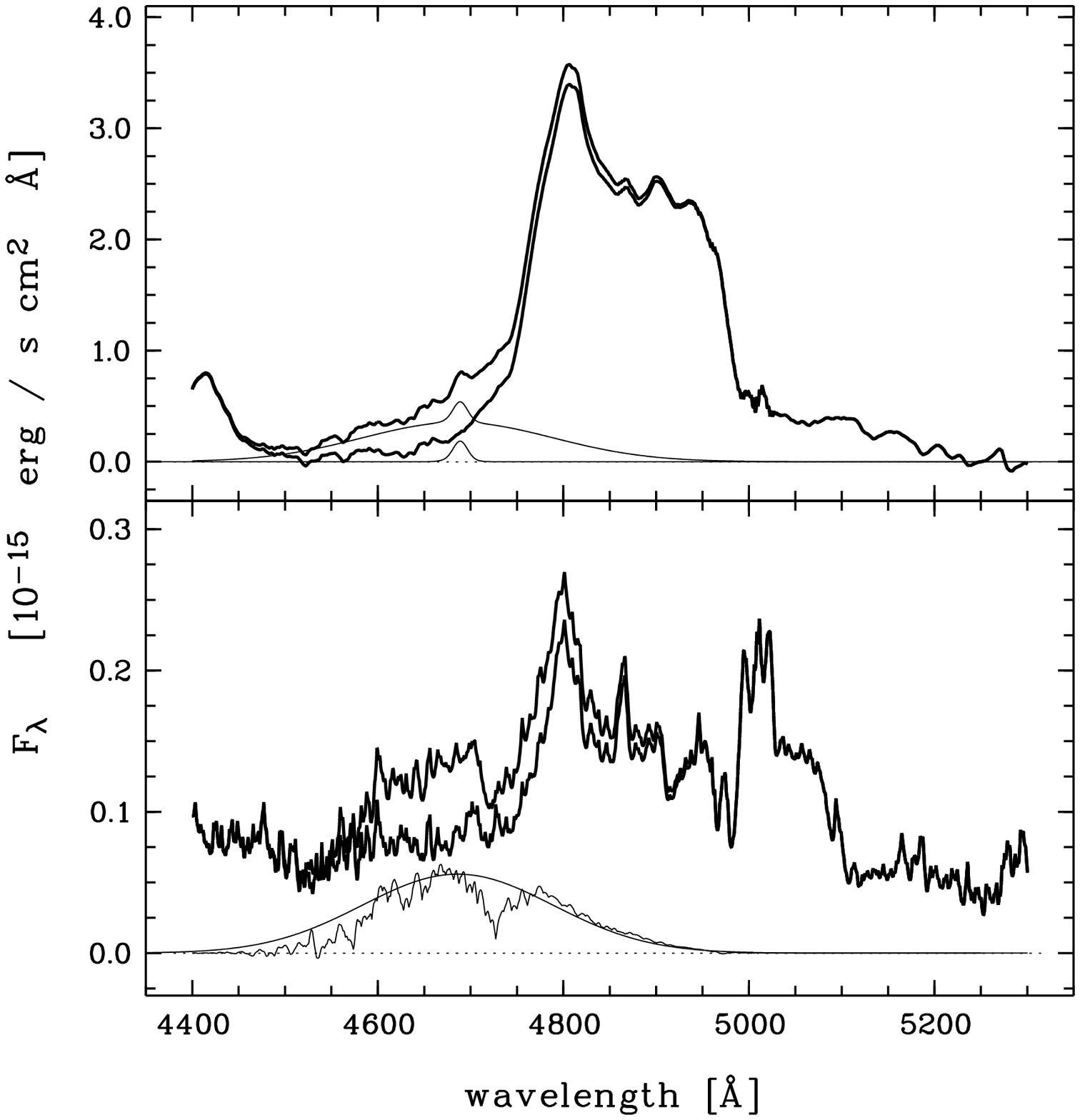}
\label{fig10}

\noindent
Figure 10 -- The mean spectrum of the H$\beta $\,4861 emission line is shown
             (top panel), uncorrected and corrected for contamination by 
             \heii $\lambda 4686$ emission.
             In addition, the mean narrow and broad \heii $\lambda 4686$ line 
             profile is plotted (thin solid line).
             In the bottom panel the corresponding rms spectra of the 
             H$\beta \,\lambda$\,4861 line are shown (thick solid lines). 
             The difference between these rms spectra can be associated with 
             the \heii $\lambda 4686$ rms spectrum (thin line). The Gaussian
             profile shaped rms spectrum for \heii $\lambda 4686$ is shown
             for comparison (thin solid line).
\end{figure}

%\clearpage

\begin{figure}
\epsscale{1.00}
\plotone{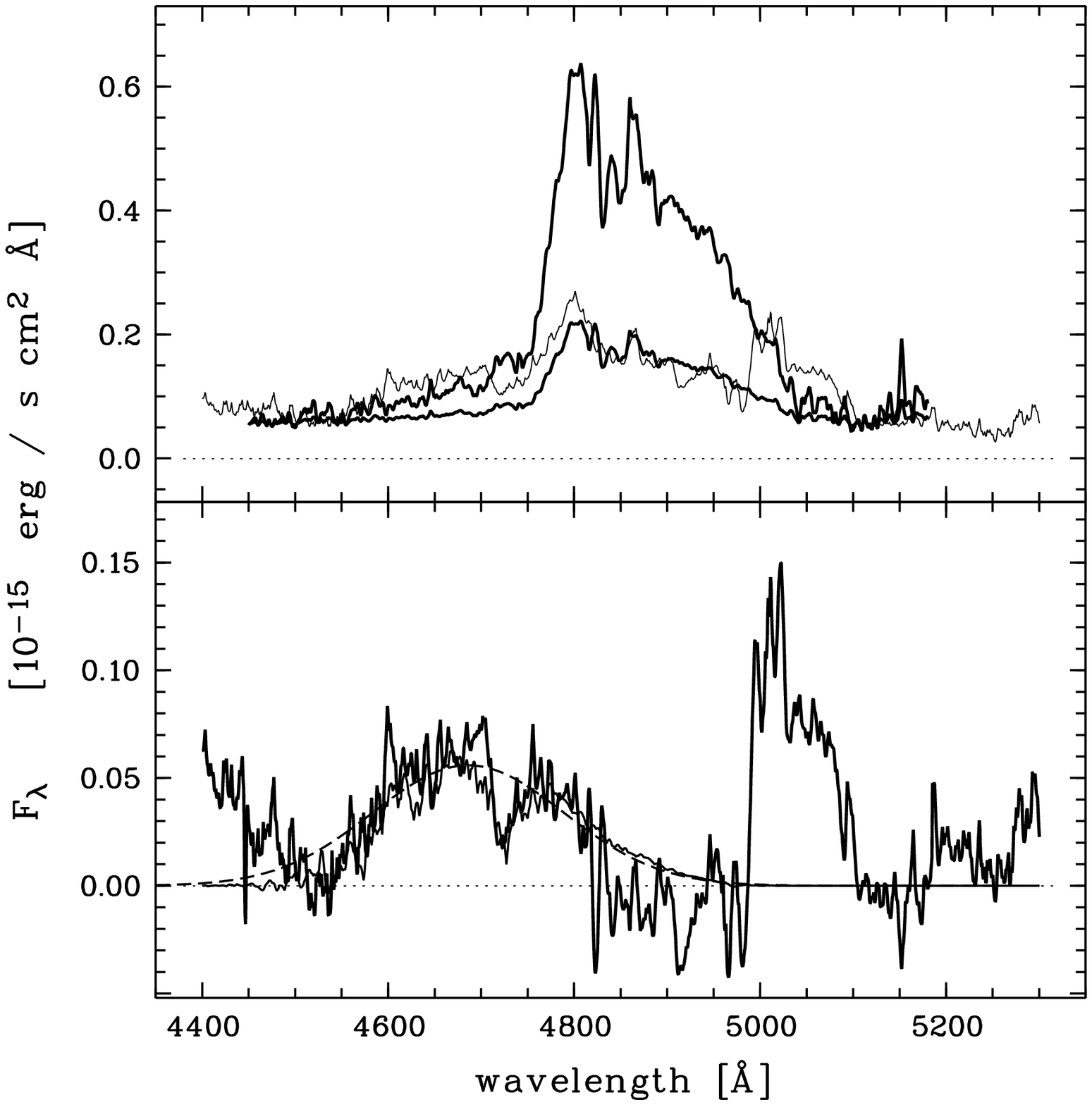}
\label{fig11}

\noindent
Figure 11 -- The H$\beta \,\lambda 4861$ rms spectrum (thin solid line) and the
             re-binned, as well as the re-binned and scaled
              H$\alpha \,\lambda 6563$ rms spectrum (thick solid line) are 
             shown (top panel). In the bottom panel the difference of the 
             H$\beta$ and the
             re-binned and scaled H$\alpha$ rms spectra is displayed (thick 
             solid line). For comparison, the \heii $\lambda 4686$ rms spectrum
             as shown in Fig.\,10 is shown as thin solid line, as well as the 
             Gaussian profile shaped rms spectrum for \heii $\lambda 4686$
             (dashed line).
\end{figure}

%\clearpage

\begin{figure}
\epsscale{1.00}
\plotone{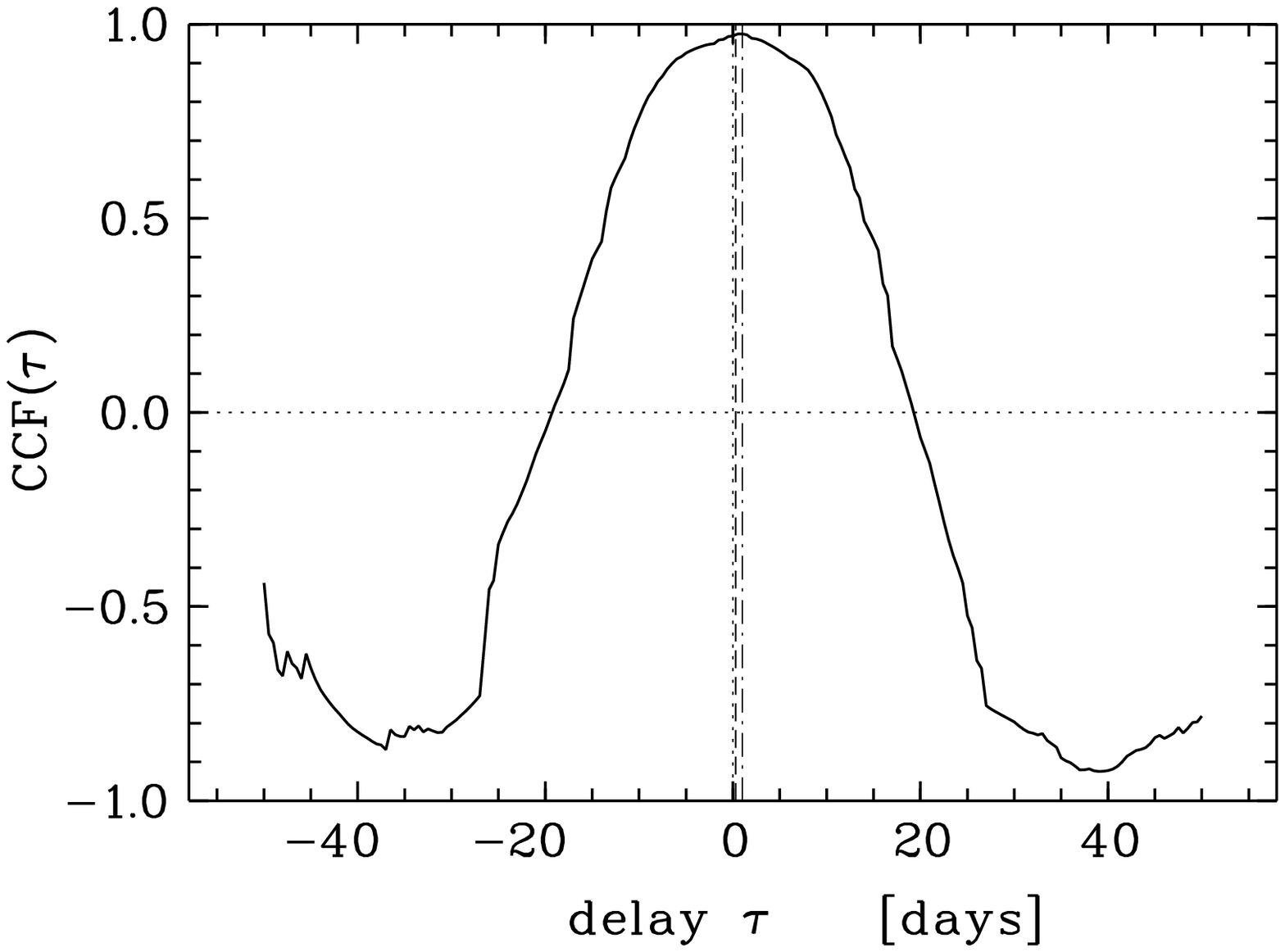}
\label{fig12}

\noindent
Figure 12 -- The cross-correlation function of the {\it g}-band flux variations
             with the variable AGN continuum F$_c$(5100\,\AA) as the driving 
             light curve. The location of the centroid 
             $(r \geq 0.8\,{\rm CCF}_{max})$ is shown as dashed line, while the
             peak of the ICCF is marked by the dashed-dotted line.
\end{figure}

%\clearpage

\begin{figure}
\epsscale{1.00}
\plotone{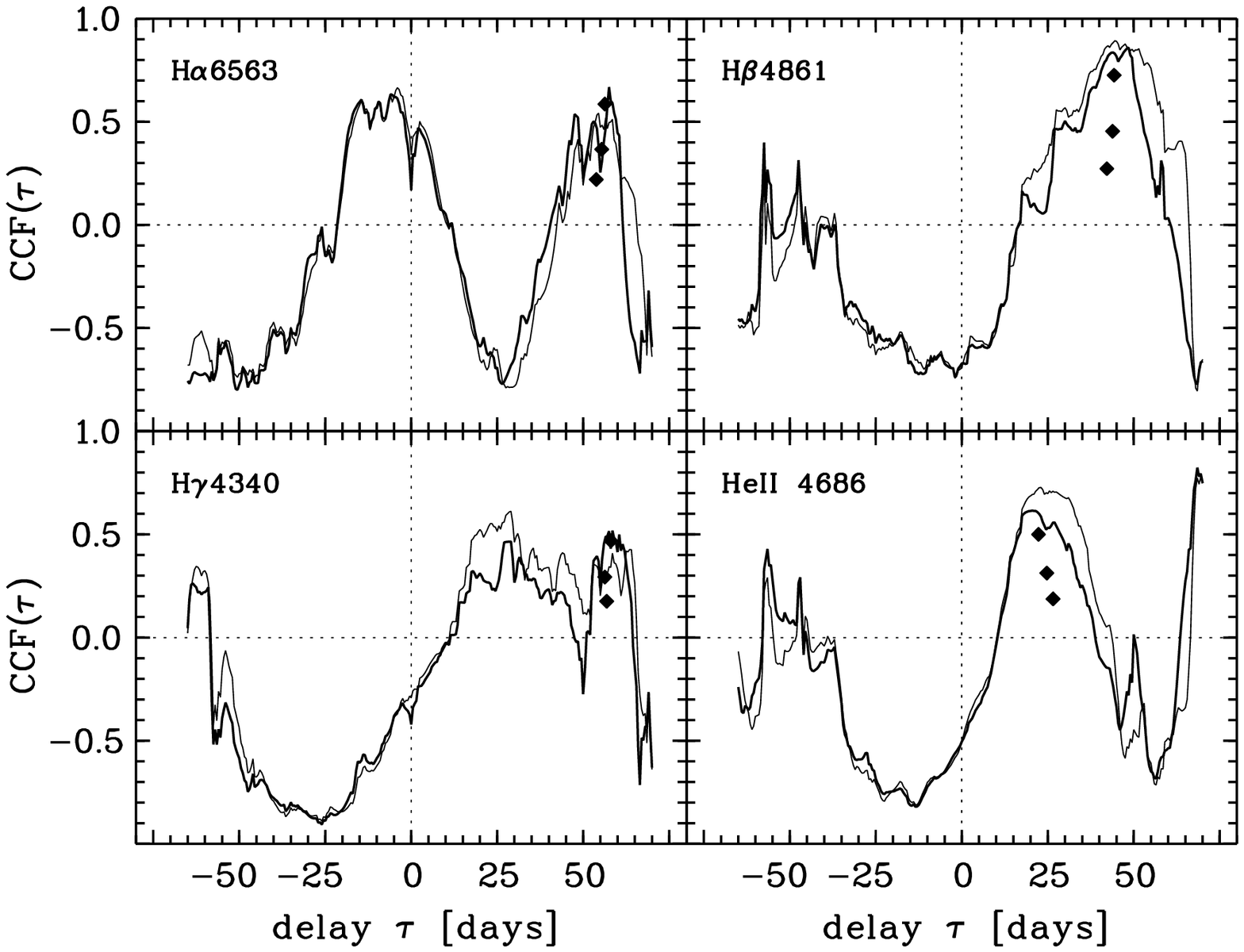}
\label{fig13}

\noindent
Figure 13 -- The cross-correlation functions of the broad emission lines of 
             H$\alpha$, H$\beta$, H$\gamma$, and \heii $\lambda 4686$
             (thick line) with the {\it g}-band variations as the driving light
             curve. For comparison, the thin lines show the cross-correlation
             functions using the continuum variations at $\lambda = 5100$\,\AA\
             as driving light curve.            
             The locations of the centroid for a threshold of 30\%\, 50\%\, and
             80\%\ are shown, respectively, for the ICCF based on the 
             {\it g}-band variation (filled diamonds).
\end{figure}

%\clearpage

\begin{figure}
\epsscale{0.50}
\plotone{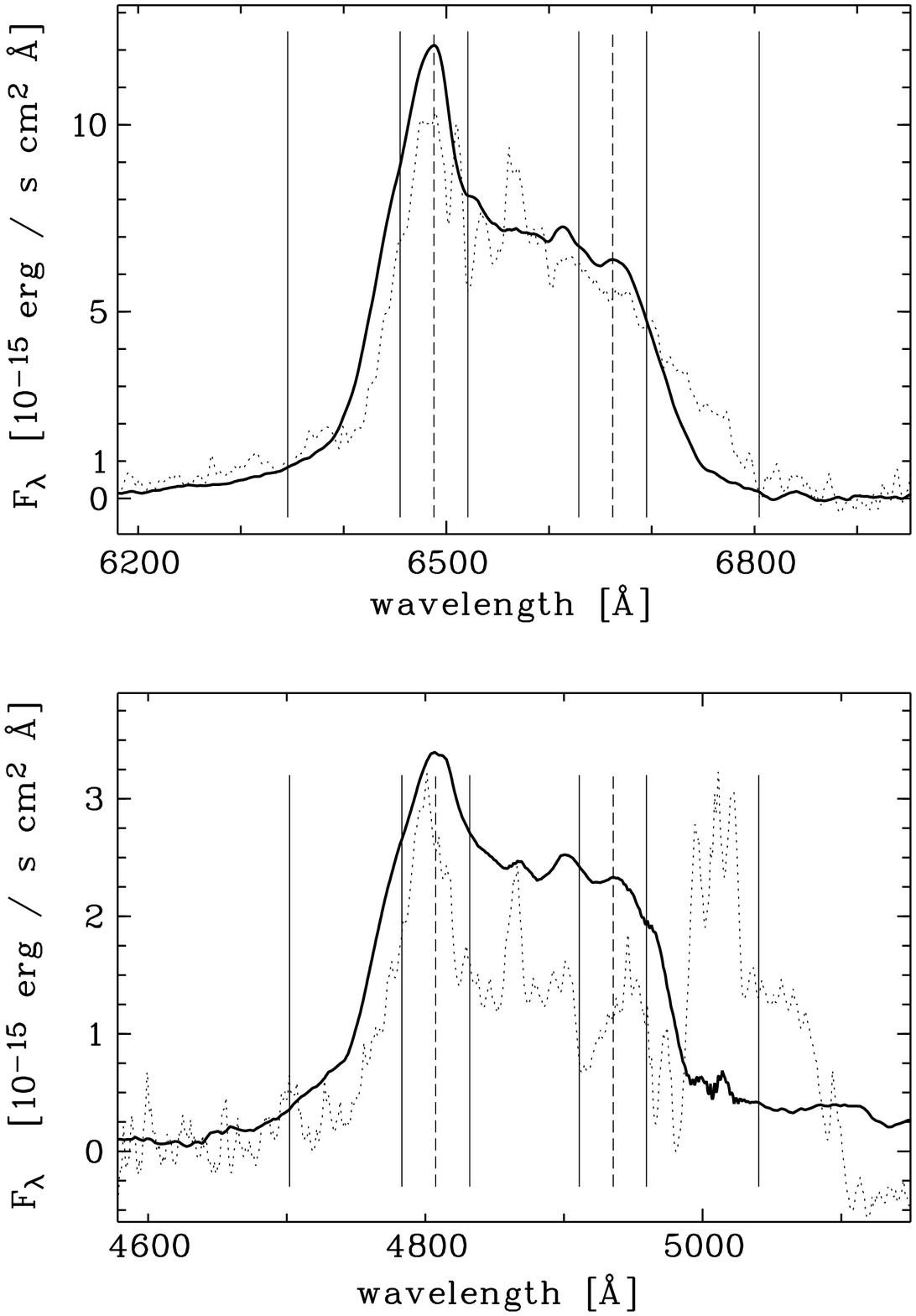}
\label{fig14}

\noindent
Figure 14 -- The location of the extraction windows for the blue and red wing,
             the blue and red peak, and the center of the broad line profiles 
             of the H$\alpha$ (top panel) and H$\beta$ (bottom panel) emission
             lines using the mean spectrum (solid line) and the rms spectrum
             (dotted line, scaled by a factor of 20).
             The long dashed line indicates the measured location of the blue 
             and red peak.
\end{figure}

%\clearpage

\begin{figure}
\epsscale{0.75}
\plotone{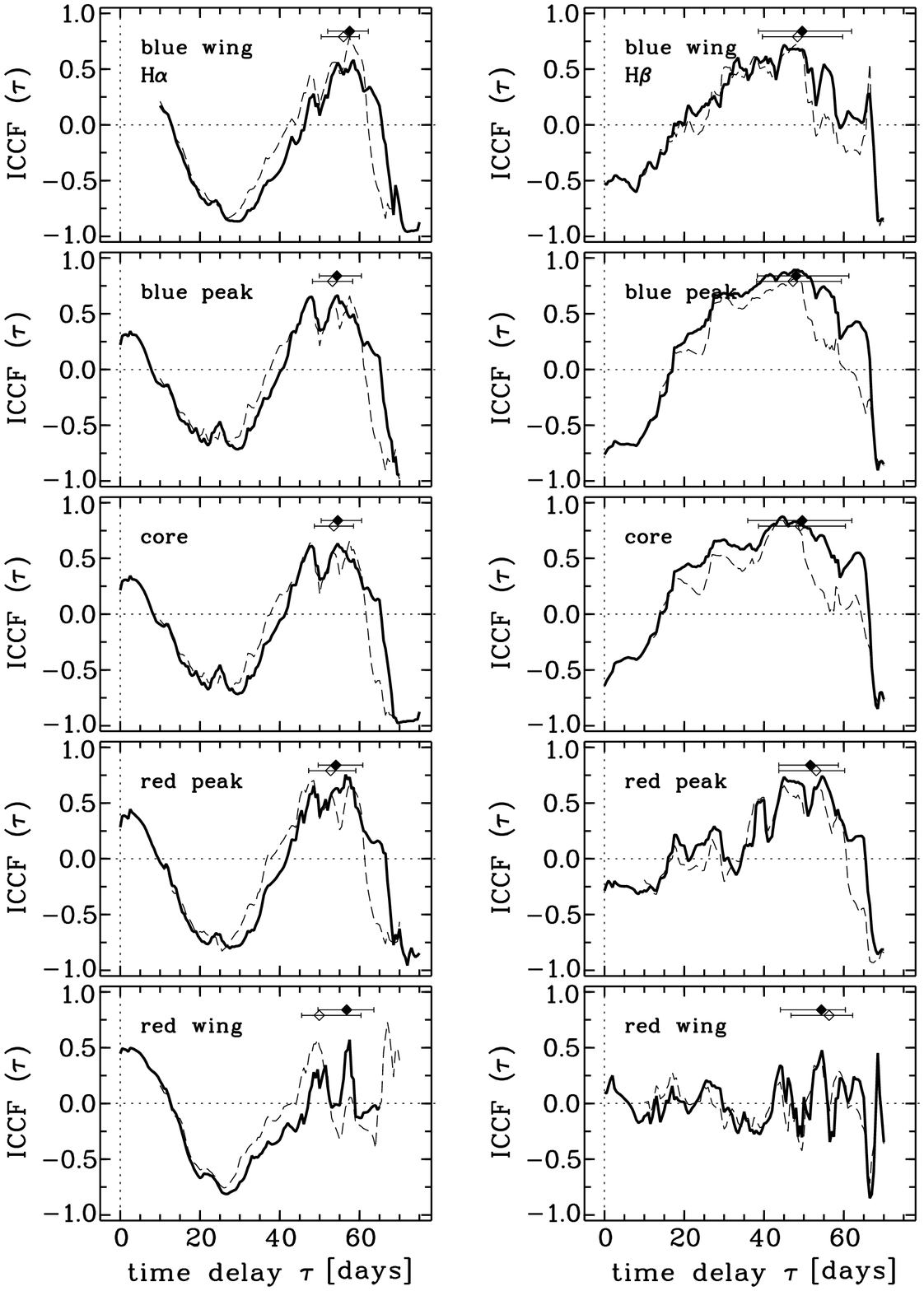}
\label{fig15}

\noindent
Figure 15 -- The cross-correlation functions using the AGN continuum 
             F$_\lambda$(5100\AA) as driver light curve (solid line) for the 
             blue wing, blue peak, line center, red peak, and red wing (top to 
             bottom) of the broad H$\alpha$ line profile (on the left) and
             of the broad H$\beta$ line profile (on the right). The ICCFs 
             obtained with the {\it g}-band variations as driving continuum are
             shown as dashed line. The location of the centroid, based on a 
             threshold of 80\,\%\ of ICCF$_{max}$ and the corresponding 
             uncertainty is shown as filled diamonds (F$_\lambda$(5100\AA)) and
             open diamonds ({\it g}-band), respectively.
\end{figure}

%\clearpage

\begin{figure}
\epsscale{0.75}
\plotone{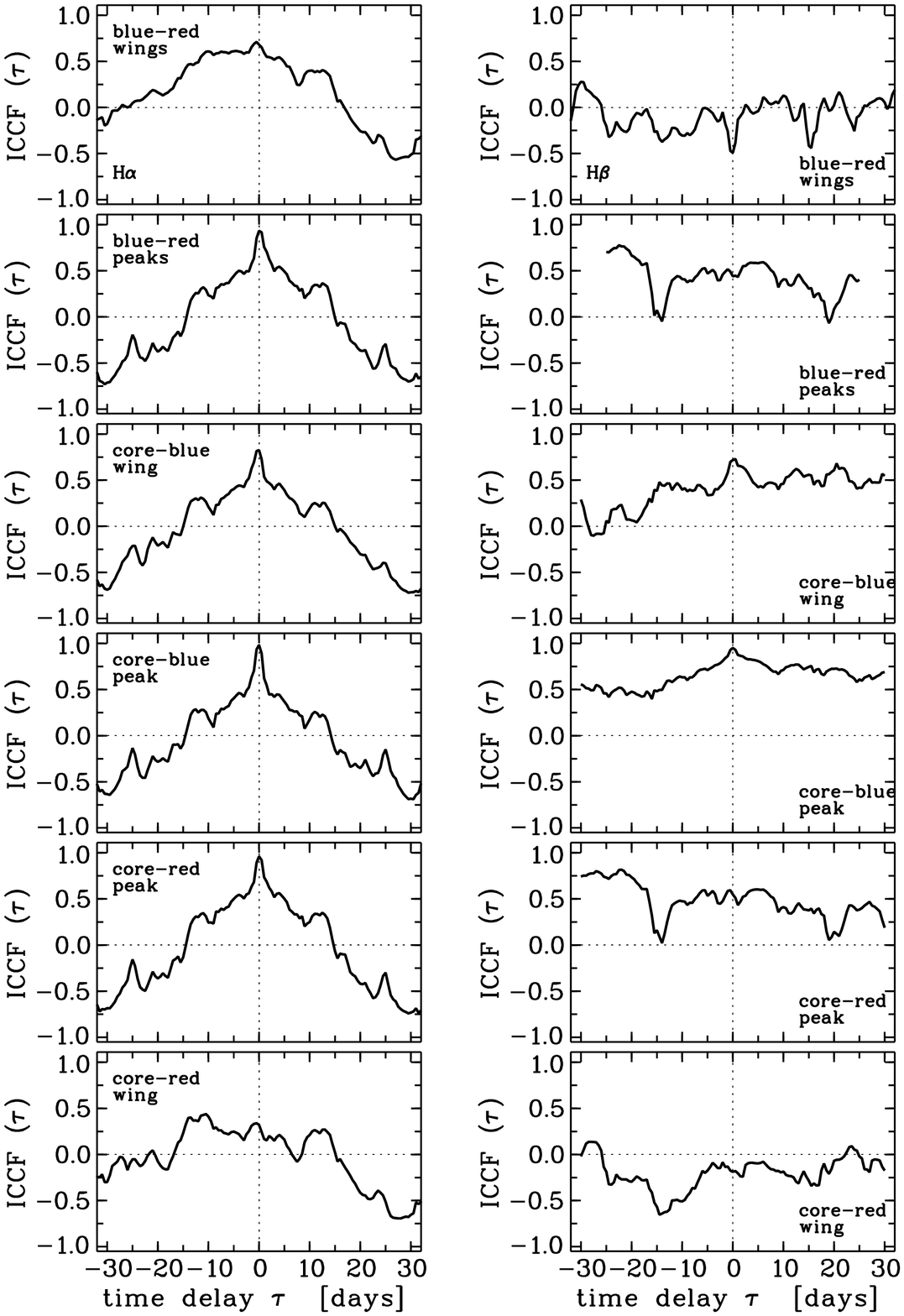}
\label{fig16}

\noindent
Figure 16 -- Cross-correlation functions for H$\alpha$ (left panels) and for
             H$\beta$ (right panels) correlating the blue and red wing,
             as well as the blue and red peak of the line profiles directly, 
             with the blue light curve as driver and the red light curve in 
             response (top two panels). The following panels show the ICCFs 
             using the variations of the line center as driving light curve 
             and the profile wings and peaks in response.
\end{figure}

%\clearpage

\begin{figure}
\epsscale{0.70}
\plotone{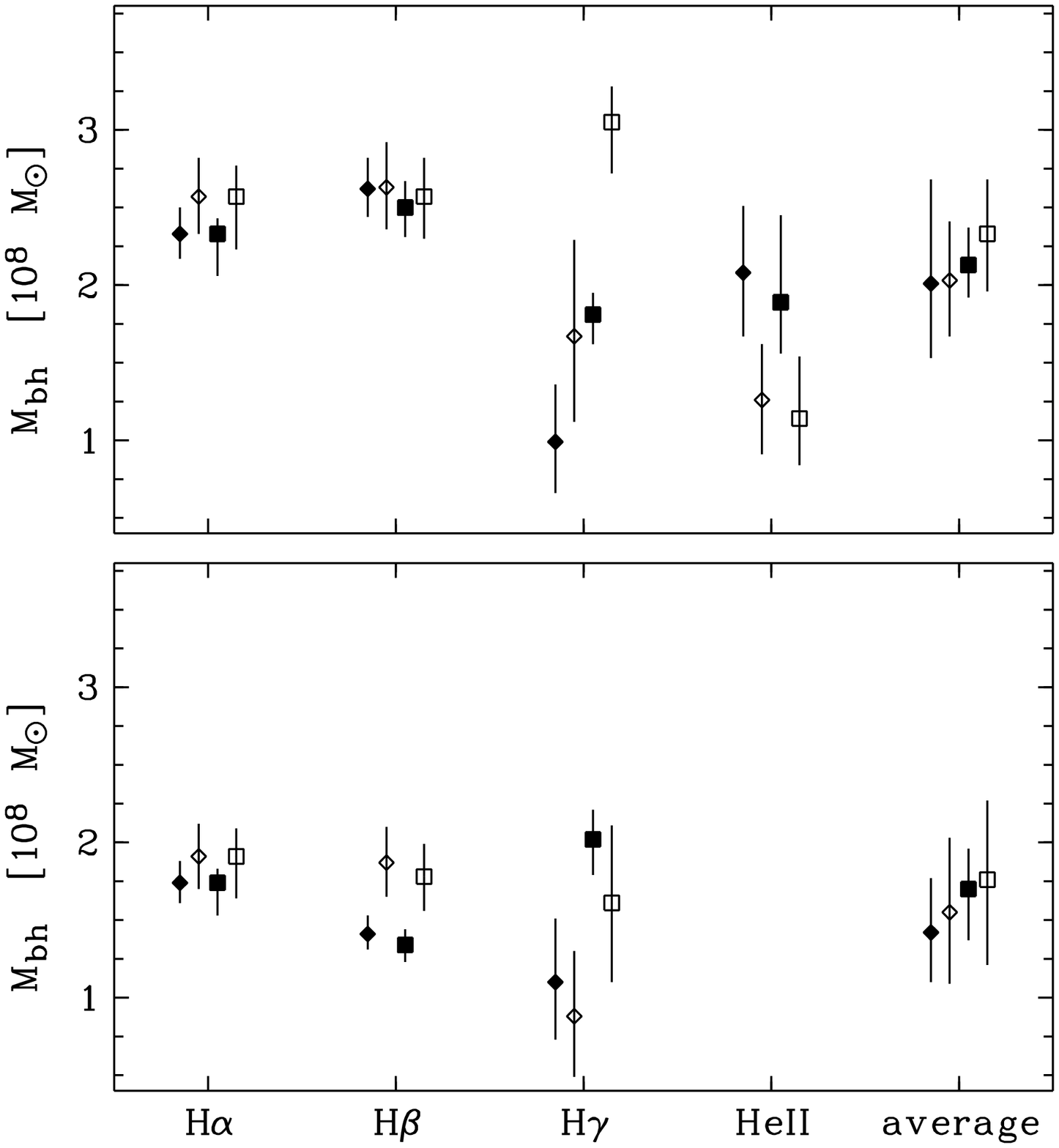}
\label{fig17}

\noindent
Figure 17 -- Comparison of the black hole mass estimates that are based on
             the measured time delay for the Balmer lines and the helium
             \heii\,$\lambda 4686$ line using the second moment 
             $\sigma_{line}$ of the line profiles (top panel) and the 
             blue -- red peak separation (bottom panel).
             The black hole masses based on the mean line profiles and 
             of the blue -- red peak separations are shown as filled symbols 
             while the rms spectra results are given by the open symbols. The 
             diamonds represent the black hole masses using the AGN continuum 
             F$_\lambda$(5100\AA) to determine the time delay while the boxes 
             display the results using the {\it g}-band variations.
\end{figure}

%\clearpage

\begin{figure}
\epsscale{0.70}
\plotone{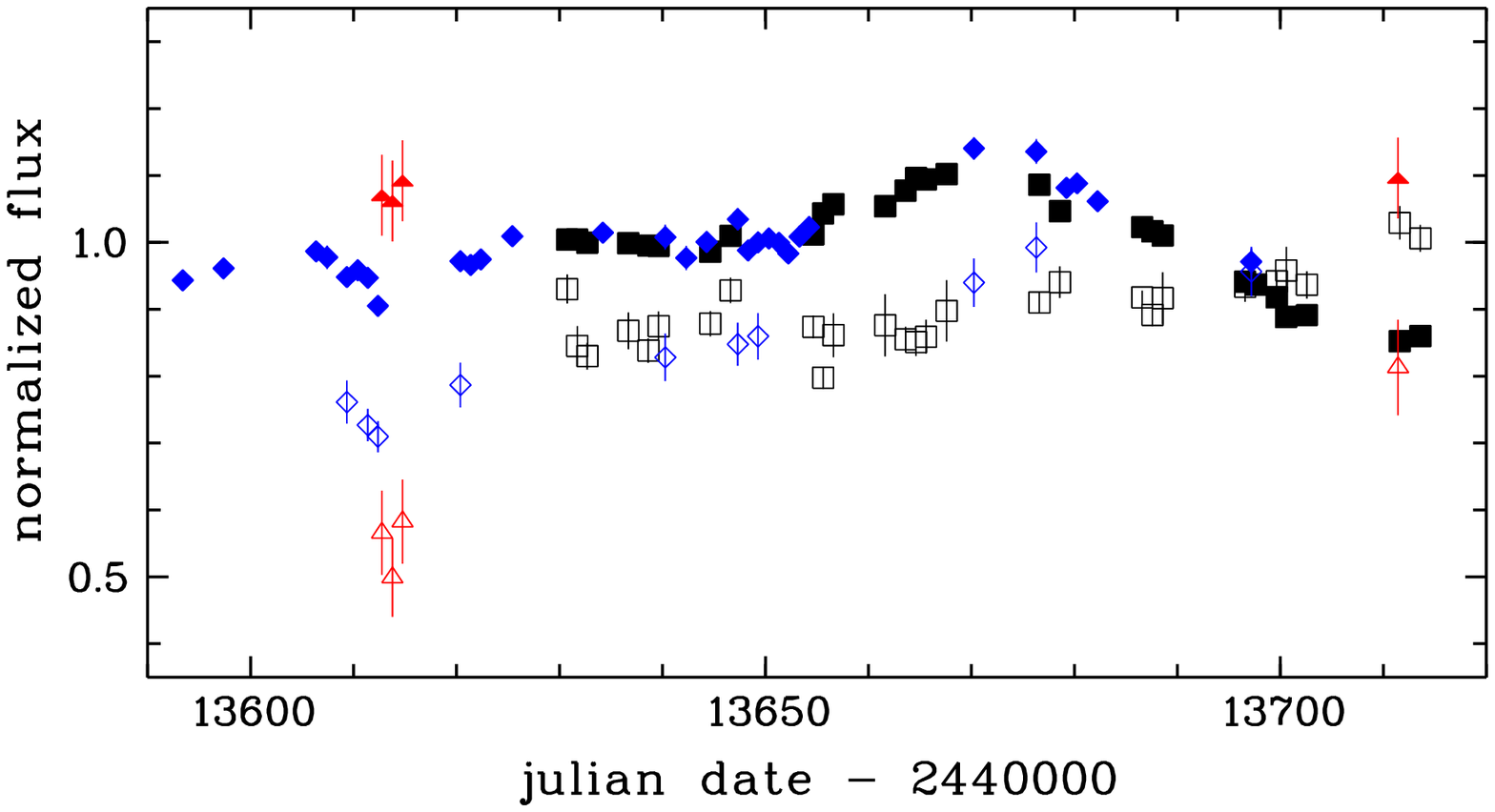}
\label{fig18}

\noindent
Figure 18 -- Comparison of the normalized light curves of the AGN continuum
             and of the broad H$\beta$ emission line flux of this study
             (black boxes), of Sergeev et al.\,(2002,2011, blue diamonds), and 
             of Shapovalova et al.\,(2010, red triangles). The continuum is 
             shown as filled symbols and the H$\beta$ emission-line flux is
             shown as open symbols. For better comparison the normalized light
             curves have been offset.
\end{figure}

\clearpage

%Table 1
%\setcounter{table}{1}
%\ptlandscape
\hspace*{-15mm}
\begin{deluxetable}{ccc}
%\tablewidth{215pt}
\tablewidth{0pt}
\tabletypesize{\scriptsize}
\tablecaption{Imaging and Spectroscopy Observing Log of 3C\,390.3.}
\tablehead{
\colhead{civil date$^a$}       &
\multicolumn{2}{c}{Julian Date - 245\,0000}\\
\colhead{mm-dd-yy} &
\colhead{photom.}&
\colhead{spectros.}\\
\colhead{(1)} &
\colhead{(2)} &
\colhead{(3)} 
}
\startdata
09 16 05 & 3630.71895 & 3630.74012 \\
09 17 05 & 3631.68545 & 3631.71968 \\
09 18 05 & 3632.67934 & 3632.69208 \\
09 22 05 &  \nodata   & 3636.67738 \\
09 24 05 & 3638.60241 & 3638.61469 \\ 
09 25 05 & 3639.60314 & 3639.61829 \\
09 30 05 & 3644.61169 & 3644.63453 \\
10 02 05 & 3646.58313 & 3646.60055 \\
10 10 05 & 3654.65719 & 3654.64235 \\
10 11 05 & 3655.60982 & 3655.59675 \\
10 12 05 & 3656.60524 & 3656.59131 \\
10 17 05 &  \nodata   & 3661.61921 \\
10 19 05 & 3663.61128 & 3663.59896 \\
10 20 05 & 3664.62103 & 3664.60890 \\
10 21 05 & 3665.60079 & 3665.58876 \\
10 23 05 & 3667.56875 & 3667.58457 \\
11 01 05 & 3676.57946 & 3676.59541 \\
11 03 05 & 3678.59495 & 3678.58313 \\
11 11 05 & 3686.58735 & 3686.57751 \\
11 12 05 & 3687.57943 & 3687.56689 \\
11 13 05 & 3688.57791 & 3688.56608 \\
11 21 05 & 3696.55431 & 3696.57008 \\
11 22 05 & \nodata    & 3697.63839 \\
11 24 05 & 3699.61123 & 3699.63440 \\
11 25 05 & 3700.55593 & 3700.56501 \\
11 27 05 & 3702.57947 & 3702.56630 \\
12 06 05 & 3711.59450 & 3711.58072 \\
12 08 05 & 3713.56451 & 3713.58304 \\
\enddata
\tablenotetext{a}{beginning of the night}
\end{deluxetable}

%\clearpage

%Table 2
%\setcounter{table}{2}
%\ptlandscape
\hspace*{-15mm}
\begin{deluxetable}{lcc}
%\tablewidth{215pt}
\tablewidth{0pt}
\tabletypesize{\scriptsize}
\tablecaption{Mean NLR emission-line rest-frame fluxes, relative to the 
              flux of H$\beta$.}
\tablehead{
\colhead{line}             &
\colhead{flux\,$^a _{obs}$}&
\colhead{flux\,$^{a,b} _{cor}$}\\
\colhead{(1)} &
\colhead{(2)} &
\colhead{(3)} 
}
\startdata
{[}\oii ]$\lambda 3727$  &$1.18\pm0.08$&$1.41\pm0.10$\\
{[}\neiii ]$\lambda 3869$&$1.20\pm0.08$&$1.41\pm0.10$\\
{[}\neiii ]$\lambda 3968$&$0.73\pm0.02$&$0.84\pm0.02$\\
H$\gamma \,\lambda 4340$ &$0.44\pm0.02$&$0.48\pm0.02$\\
{[}\oiii ]$\lambda 4363$ &$0.85\pm0.02$&$0.92\pm0.03$\\
\heii $\lambda 4686$     &$0.15\pm0.01$&$0.15\pm0.01$\\
H$\beta \,\lambda 4861$  &$1.00\pm0.01$&$1.00\pm0.02$\\
{[}\oiii ]$\lambda 4959$ &$2.48\pm0.02$&$2.44\pm0.05$\\
{[}\oiii ]$\lambda 5007$ &$7.30\pm0.07$&$7.12\pm0.15$\\
{[}\fevii ]$\lambda 5721$&$0.15\pm0.01$&$0.13\pm0.01$\\
\hei $\lambda 5876$      &$0.15\pm0.01$&$0.13\pm0.01$\\
{[}\fevii ]$\lambda 6087$&$0.36\pm0.02$&$0.30\pm0.02$\\
{[}\oi ]$\lambda 6300$   &$0.99\pm0.01$&$0.82\pm0.02$\\
{[}\oi ]$\lambda 6364$   &$0.33\pm0.01$&$0.27\pm0.01$\\
{[}\fex ]$\lambda 6374$  &$0.06\pm0.01$&$0.05\pm0.01$\\
{[}\nii ]$\lambda 6548$  &$0.44\pm0.01$&$0.35\pm0.01$\\
H$\alpha \,\lambda 6563$ &$3.53\pm0.07$&$2.85\pm0.08$\\
{[}\nii ]$\lambda 6583$  &$1.34\pm0.04$&$1.08\pm0.04$\\
{[}\sii ]$\lambda 6716$  &$0.20\pm0.01$&$0.16\pm0.01$\\
{[}\sii ]$\lambda 6731$  &$0.22\pm0.01$&$0.18\pm0.01$\\
\enddata
\tablenotetext{a}{integrated H$\beta$-narrow line flux, 
                  F$(H\beta) = (26.50\pm0.19)\,\times\,10^{-15}$ erg\,s$^{-1}$\,cm$^{-2}$}
\tablenotetext{b}{ {\bf line} ratios are reddening corrected assuming a Balmer 
                  decrement of F(H$\alpha$)/F(H$\beta$) = 2.85.}
\end{deluxetable}

%\clearpage

%Table 3
%\setcounter{table}{3}
\ptlandscape
\hspace*{-15mm}
\begin{deluxetable}{cc|cccccc}
\rotate
%\tablewidth{-250pt}
\tablewidth{0pt}
\tabletypesize{\scriptsize}
\tablecaption{Observed {\it g}-band magnitudes, measured rest-frame continuum
              flux density at $\lambda = 5100$\,\AA , and integrated flux of
              the broad emission lines of H$\alpha$, H$\beta$, H$\gamma$, and
              \heii $\lambda 4686$.}
\tablehead{
\colhead{Julian Date}               &
\colhead{{\it g}-band}                    &
\colhead{Julian Date}               &
\colhead{F$_\lambda 5100$\,\AA $^a$}&
\colhead{H$\alpha \,\lambda 6563$\,$^b$}      &
\colhead{H$\beta \,\lambda 4861$\,$^b$}       &
\colhead{H$\gamma \,\lambda 4340$\,$^b$}      &
\colhead{\heii $\lambda 4686$\,$^b$}\\
\colhead{-245\,0000}&
\colhead{{[}mag]}   &
\colhead{-245\,0000}&
\colhead{ }       &
\colhead{ }       &
\colhead{ }       &
\colhead{ }       &
\colhead{ }       \\
\colhead{( 1)} &
\colhead{( 2)} &
\colhead{( 3)} &
\colhead{( 4)} &
\colhead{( 5)} &
\colhead{( 6)} &
\colhead{( 7)} &
\colhead{( 8)} \\
}
\startdata
3630.71895&$14.543\pm 0.001$&3630.74012&$6.66\pm 0.07$&$2123.4\pm 33.7$&$617.7\pm 13.0$&$114.2\pm 17.0$&$ 91.6\pm 15.7$\\
3631.68545&$14.559\pm 0.001$&3631.71968&$6.66\pm 0.07$&$2203.3\pm 40.3$&$566.6\pm 17.8$&$125.5\pm 17.1$&$ 85.0\pm 18.8$\\
3632.67934&$14.566\pm 0.001$&3632.69208&$6.63\pm 0.07$&$2457.7\pm 35.3$&$557.5\pm 12.0$&$135.8\pm 13.1$&$ 65.9\pm 13.7$\\
 \nodata  &   \nodata       &3636.67738&$6.63\pm 0.07$&$2433.4\pm 45.7$&$580.2\pm 16.4$&$135.0\pm 17.2$&$ 48.4\pm 17.7$\\
3638.60241&$14.589\pm 0.001$&3638.61469&$6.60\pm 0.07$&$2379.1\pm 43.4$&$562.5\pm 10.9$&$128.2\pm 17.1$&$ 59.1\pm 12.4$\\
3639.60314&$14.590\pm 0.001$&3639.61829&$6.60\pm 0.07$&$2320.6\pm 29.5$&$584.4\pm 13.1$&$130.1\pm 17.1$&$ 78.4\pm 14.5$\\
3644.61169&$14.598\pm 0.001$&3644.63453&$6.54\pm 0.07$&$2368.3\pm 29.3$&$586.6\pm 11.4$&$127.8\pm 14.0$&$ 80.0\pm 13.3$\\
3646.58313&$14.583\pm 0.001$&3646.60055&$6.70\pm 0.07$&$2642.3\pm 39.7$&$616.4\pm 11.3$&$125.4\pm 15.9$&$ 89.1\pm 14.0$\\
3654.65719&$14.566\pm 0.001$&3654.64235&$6.71\pm 0.07$&$2545.6\pm 38.1$&$583.7\pm 10.2$&$127.0\pm 17.2$&$ 77.6\pm 12.0$\\
3655.60982&$14.559\pm 0.001$&3655.59675&$6.92\pm 0.07$&$2410.9\pm 33.3$&$538.2\pm  9.8$&$133.5\pm 15.3$&$ 67.2\pm 10.5$\\
3656.60524&$14.549\pm 0.001$&3656.59131&$7.01\pm 0.07$&$2437.8\pm 47.5$&$576.2\pm 19.5$&$131.9\pm 17.3$&$ 66.3\pm 20.4$\\
 \nodata  &   \nodata       &3661.61921&$6.99\pm 0.07$&$2673.7\pm 49.6$&$585.2\pm 27.9$&$150.2\pm 35.8$&$ 51.0\pm 28.7$\\
3663.61128&$14.526\pm 0.001$&3663.59896&$7.14\pm 0.07$&$2627.9\pm 47.4$&$572.5\pm 11.3$&$159.4\pm 12.8$&$ 67.2\pm 12.1$\\
3664.62103&$14.525\pm 0.002$&3664.60890&$7.27\pm 0.07$&$2488.4\pm 33.6$&$570.2\pm 12.4$&$156.6\pm 16.6$&$ 73.9\pm 13.0$\\
3665.60079&$14.522\pm 0.002$&3665.58876&$7.26\pm 0.08$&$2534.5\pm 34.4$&$574.9\pm 15.2$&$150.2\pm 16.6$&$ 79.8\pm 15.7$\\
3667.56875&$14.495\pm 0.004$&3667.58457&$7.31\pm 0.08$&$2611.2\pm 52.2$&$598.1\pm 27.4$&$151.4\pm 15.5$&$ 84.0\pm 28.0$\\
3676.57946&$14.520\pm 0.002$&3676.59541&$7.20\pm 0.08$&$2460.7\pm 46.3$&$605.6\pm  9.2$&$160.6\pm 16.9$&$ 99.0\pm 10.7$\\
3678.59495&$14.529\pm 0.002$&3678.58313&$6.94\pm 0.07$&$2488.4\pm 37.3$&$623.8\pm 14.0$&$152.5\pm 17.5$&$116.7\pm 15.9$\\
3686.58735&$14.574\pm 0.001$&3686.57751&$6.78\pm 0.07$&$2334.8\pm 40.1$&$610.1\pm  6.2$&$169.2\pm 16.8$&$109.6\pm  7.9$\\
3687.57943&$14.591\pm 0.002$&3687.56689&$6.74\pm 0.07$&$2340.7\pm 36.4$&$594.4\pm 10.0$&$170.6\pm 16.7$&$107.7\pm 11.0$\\
3688.57791&$14.592\pm 0.002$&3688.56608&$6.70\pm 0.07$&$2523.8\pm 34.1$&$609.5\pm 22.9$&$170.2\pm 17.0$&$108.3\pm 23.7$\\
3696.55431&$14.633\pm 0.001$&3696.57008&$6.24\pm 0.07$&$2291.8\pm 30.0$&$619.6\pm 13.3$&$169.8\pm 16.9$&$110.2\pm 14.3$\\
\nodata   &   \nodata       &3697.63839&$6.21\pm 0.07$&$2162.3\pm 40.4$&$632.8\pm 15.5$&$145.3\pm 20.6$&$110.2\pm 15.6$\\
3699.61123&$14.651\pm 0.002$&3699.63440&$6.09\pm 0.06$&$2427.7\pm 33.6$&$624.6\pm  8.6$&$170.5\pm 17.1$&$123.3\pm 10.6$\\
3700.55593&$14.647\pm 0.002$&3700.56501&$5.89\pm 0.06$&$2399.6\pm 40.6$&$634.3\pm 21.1$&$165.6\pm 17.4$&$113.0\pm 22.3$\\
3702.57947&$14.659\pm 0.002$&3702.56630&$5.91\pm 0.06$&$2372.6\pm 36.2$&$621.5\pm 12.2$&$162.2\pm 17.2$&$111.2\pm 13.8$\\
3711.59450&$14.708\pm 0.002$&3711.58072&$5.65\pm 0.06$&$2441.6\pm 47.4$&$676.9\pm 15.0$&$171.0\pm 12.8$&$104.8\pm 16.2$\\
3713.56451&$14.694\pm 0.002$&3713.58304&$5.70\pm 0.06$&$2519.7\pm 31.8$&$663.0\pm 12.0$&$170.7\pm 17.1$&$103.9\pm 13.5$\\
\enddata
\tablenotetext{a}{continuum flux in [$10^{-15}$ erg\,s$^{-1}$\,cm$^{-2}$\,\AA $^{-1}$]}
\tablenotetext{b}{integrated emission-line flux in [$10^{-15}$ erg\,s$^{-1}$\,cm$^{-2}$]}
\end{deluxetable}

%\clearpage

%Table 4
%\setcounter{table}{4}
%\ptlandscape
\hspace*{-15mm}
\begin{deluxetable}{lcccc}
%\tablewidth{215pt}
\tablewidth{0pt}
\tabletypesize{\scriptsize}
\tablecaption{Measurements of the mean and rms line profile parameters of the 
              Balmer lines H$\alpha$, H$\beta$, H$\gamma$, and the helium 
              line \heii $\lambda 4686$.}
\tablehead{
\multicolumn{5}{c}{{\bf Mean Spectra}}\\[1mm]
\colhead{feature }&
\colhead{FWHM    }&
\colhead{$\sigma_{line}$}&
\colhead{FWHM    }&
\colhead{$\sigma_{line}$}\\
\colhead{ }      &
\colhead{{[}\AA ]}&
\colhead{{[}\AA ]}      &
\colhead{{[}km\,s$^{-1}$]}&
\colhead{{[}km\,s$^{-1}$]}\\
\colhead{(1)} &
\colhead{(2)} &
\colhead{(3)} &
\colhead{(4)} &
\colhead{(5)} 
}
\startdata
H$\alpha \,\lambda$\,6563      &$275\pm 1$&$101\pm 1$&$12563\pm  31$&$4607\pm  29$\\
H$\beta \,\lambda$\,4861       &$214\pm 1$&$ 87\pm 1$&$13211\pm  28$&$5377\pm  37$\\    
H$\gamma \,\lambda$\,4340      &$172\pm 2$&$ 58\pm 1$&$11875\pm 103$&$3991\pm  29$\\
\heii\,$\lambda 4686$          &$242\pm 8$&$103\pm 3$&$15480\pm 512$&$6590\pm 192$\\ 
 & & & & \\[-3mm]
\hline
 & & & & \\[-2mm]
\multicolumn{5}{c}{{\bf rms Spectra}}\\[1mm]
\colhead{        }&
\colhead{FWHM    }&
\colhead{$\sigma_{line}$}&
\colhead{FWHM    }&
\colhead{$\sigma_{line}$}\\
\colhead{ }      &
\colhead{{[}\AA ]}&
\colhead{{[}\AA ]}&
\colhead{{[}km\,s$^{-1}$]}&
\colhead{{[}km\,s$^{-1}$]}\\
 & & & & \\[-3mm]
\hline
 & & & & \\[-3mm]
H$\alpha \,\lambda$\,6563      &$278\pm 33$&$106\pm  6$&$12679\pm 1312$&$4839\pm 215$\\
H$\beta \,\lambda$\,4861       &$176\pm 26$&$ 88\pm  5$&$10872\pm 1670$&$5455\pm 278$\\
H$\gamma \,\lambda$\,4340      &$165\pm 19$&$ 75\pm  2$&$11422\pm 1458$&$5191\pm  82$\\
\heii\,$\lambda 4686$          &$207\pm 15$&$ 80\pm 11$&$13244\pm  960$&$5125\pm 704$\\ 
\enddata
\end{deluxetable}

%\clearpage

%Table 5
%\setcounter{table}{5}
%\ptlandscape
\hspace*{-15mm}
\begin{deluxetable}{lccccc}
%\tablewidth{240pt}
\tablewidth{0pt}
\tabletypesize{\scriptsize}
\tablecaption{Statistical properties of the continuum and broad 
              emission-line variations}
\tablehead{
\colhead{feature}&
\colhead{mean flux\,$^c$}&
\colhead{$\sigma _F^c$}&
\colhead{F$_{var}$}&
\colhead{min}&
\colhead{max}\\
\colhead{(1)} &
\colhead{(2)} &
\colhead{(3)} &
\colhead{(4)} & 
\colhead{(5)} &
\colhead{(6)} 
}
\startdata
{\it g}-band$^a$               &14.58&0.28&0.050&14.71&14.50\\
V-band$^b$               & 7.34&0.37&0.051& 6.53& 7.94\\
F$_\lambda$(5100\AA )$^b$& 6.63&0.47&0.070& 5.65& 7.31\\
H$\alpha \,\lambda$\,6563          & 2439& 128&0.041& 2123& 2674\\
H$\beta \,\lambda$\,4861           &  598&  32&0.047&  538&  677\\
H$\gamma \,\lambda$\,4340          &  149&  18&0.031&  114&  171\\
\heii \,$\lambda 4686$   &   88&  21&0.155&   48&  123\\
\enddata
\tablenotetext{a}{continuum flux in [mag]}
\tablenotetext{b}{continuum flux in [$10^{-15}$ erg\,s$^{-1}$\,cm$^{-2}$\,\AA$^{-1}$]}
\tablenotetext{c}{integrated line flux in [$10^{-15}$ erg\,s$^{-1}$\,cm$^{-2}$]}
\end{deluxetable}

%\clearpage

%Table 6
%\setcounter{table}{6}
%\ptlandscape
\hspace*{-15mm}
\begin{deluxetable}{lccccccc}
%\tablewidth{215pt}
\tablewidth{0pt}
\tabletypesize{\scriptsize}
\tablecaption{Results of the cross-correlation analysis of the broad line
              emission of the Balmer lines H$\alpha$, H$\beta$, H$\gamma$,
              the helium line \heii $\lambda 4686$, using AGN continuum 
              variations at F$_\lambda$(5100\AA) and the {\it g}-band 
              variations 
              as driving continuum. The centroid of the ICCF has been 
              computed for 80\,\%\ of the corresponding maximum of the 
              ICCF. The delays calculated with SPEAR are given in column 
              (8).}
\tablehead{
\colhead{feature}&
\multicolumn{3}{c}{F$_\lambda$(5100\AA)}&
\multicolumn{3}{c}{{\it g}-band}&
\colhead{SPEAR}\\
\colhead{ }&
\colhead{$\tau _{cent}$}&
\colhead{$\tau _{peak}$}&
\colhead{CCF$_{max}$   }&
\colhead{$\tau _{cent}$}&
\colhead{$\tau _{peak}$}&
\colhead{CCF$_{max}$   }&
\colhead{ }\\
\colhead{(1)} &
\colhead{(2)} &
\colhead{(3)} &
\colhead{(4)} &
\colhead{(5)} &
\colhead{(6)} &
\colhead{(7)} &
\colhead{(8)}
}
\tablecolumns{7}
\startdata
{\it g}-band              &$ 0.2^{ +1.1}_{ -1.1}$&$ 0.6^{ +1.0}_{ -1.5}$&$0.97\pm 0.01$&         ---          &      ---             &    ---       &$0.55^{+1.56}_{-0.45}$\\[1mm]
H$\alpha \,\lambda$\,6563     &$56.3^{ +4.2}_{ -4.0}$&$56.5^{ +3.5}_{ -4.0}$&$0.83\pm 0.39$&$56.3^{ +2.4}_{ -6.6}$&$56.5^{ +2.0}_{ -6.5}$&$0.87\pm 0.30$&$52.5^{+0.7}_{-0.6}$\\[1mm]
H$\beta \,\lambda$\,4861      &$46.4^{ +3.6}_{ -3.2}$&$46.5^{ +5.0}_{ -3.5}$&$0.91\pm 0.07$&$44.3^{ +3.0}_{ -3.3}$&$45.0^{ +3.0}_{ -4.0}$&$0.87\pm 0.10$&$47.9^{+2.4}_{-4.2}$\\[1mm]
H$\gamma \,\lambda$\,4340     &$31.7^{+11.8}_{-10.5}$&$32.0^{+11.5}_{-10.5}$&$0.75\pm 0.11$&$58.1^{ +4.3}_{ -6.1}$&$58.0^{ +4.5}_{ -6.0}$&$0.86\pm 0.34$&$32.1\pm17.3$\\[1mm]
\heii $\lambda 4686$&$24.5^{ +5.0}_{ -4.7}$&$24.5^{ +6.0}_{ -5.0}$&$0.75\pm 0.11$&$22.3^{ +6.5}_{ -3.8}$&$23.0^{ +7.0}_{ -4.5}$&$0.67\pm 0.13$&$36.0\pm 5.2$\\[1mm]
\hline
\enddata
\end{deluxetable}

%\clearpage

%Table 7
%\setcounter{table}{7}
%\ptlandscape
\hspace*{-15mm}
\begin{deluxetable}{lcccccc}
%\tablewidth{215pt}
\tablewidth{0pt}
\tabletypesize{\scriptsize}
\tablecaption{Location of the blue and red peak, as well as of a red
              feature in the broad line profiles of the H$\alpha$,
              H$\beta$, and H$\gamma$ emission, if detectable.}
\tablehead{
\multicolumn{7}{c}{{\bf Mean Spectra}}\\[1mm]
\colhead{feature       }     &
\multicolumn{2}{c}{blue peak}&
\multicolumn{2}{c}{red peak }&
\multicolumn{2}{c}{red feature}\\
%\colhead{ }&
%\multicolumn{2}{c}{$\lambda_c$}&
%\multicolumn{2}{c}{$\lambda_c$}&
%\multicolumn{2}{c}{$\lambda_c$}\\
\colhead{ } &
\colhead{{[}\AA ]} &
\colhead{{[}km\,s$^{-1}$]} &
\colhead{{[}\AA ]} &
\colhead{{[}km\,s$^{-1}$]} &
\colhead{{[}\AA ]} &
\colhead{{[}km\,s$^{-1}$]} \\
\colhead{(1)} &
\colhead{(2)} &
\colhead{(3)} & 
\colhead{(4)} &
\colhead{(5)} & 
\colhead{(6)} &
\colhead{(7)} 
}
\startdata
H$\alpha \,\lambda$\,6563&$6488.3\pm 1.0$&$-3413\pm  41$&$6662.1\pm 0.0$&$+4538\pm  54$&$6613.0\pm 1.0$&$+2292\pm  38$\\
H$\beta \,\lambda$\,4861 &$4807.4\pm 1.0$&$-3324\pm  49$&$4935.3\pm 0.0$&$+4563\pm  68$&$4901.6\pm 1.0$&$+2485\pm  42$\\
H$\gamma \,\lambda$\,4340&$4287.6\pm 1.0$&$-3650\pm  69$&$4409.7\pm 1.5$&$+4783\pm 104$& \nodata       & \nodata      \\
 & & & & & & \\[-3mm]
\hline
 & & & & & & \\[-2mm]
\multicolumn{7}{c}{{\bf rms Spectra}}\\[1mm]
\colhead{feature       }     &
\multicolumn{2}{c}{blue peak}&
\multicolumn{2}{c}{red peak }&
\multicolumn{2}{c}{red feature}\\
%\colhead{ }&
%\multicolumn{2}{c}{$\lambda_c$}&
%\multicolumn{2}{c}{$\lambda_c$}&
%\multicolumn{2}{c}{$\lambda_c$}\\
\colhead{ } &
\colhead{{[}\AA ]} &
\colhead{{[}km\,s$^{-1}$]} &
\colhead{{[}\AA ]} &
\colhead{{[}km\,s$^{-1}$]} &
\colhead{{[}\AA ]} &
\colhead{{[}km\,s$^{-1}$]} \\
%\colhead{(1)} &
%\colhead{(2)} &
%\colhead{(3)} & 
%\colhead{(4)} &
%\colhead{(5)} & 
%\colhead{(6)} &
%\colhead{(7)} 
 & & & & & & \\[-3mm]
\hline
 & & & & & & \\[-2mm]
H$\alpha \,\lambda$\,6563&$6483.5\pm  2.0$&$-3622\pm  91$&$6665.9\pm 2.0$&$+4711\pm 228$&$6619.3\pm 4.0$&$+2582\pm 183$\\
H$\beta \,\lambda$\,4861 &$4801.5\pm  3.0$&$-3689\pm 185$&$4945.8\pm 4.0$&$+5210\pm 247$&$4897.1\pm 6.0$&$+2207\pm 370$\\
H$\gamma \,\lambda$\,4340&$4293.0\pm 10.0$&$-3277\pm 690$&$4402.0\pm 6.0$&$+4251\pm 415$& \nodata       & \nodata      \\
\enddata
\end{deluxetable}

%\clearpage

%Table 8
%\setcounter{table}{8}
%\ptlandscape
\hspace*{-15mm}
\begin{deluxetable}{lcc}
%\tablewidth{215pt}
\tablewidth{0pt}
\tabletypesize{\scriptsize}
\tablecaption{Range of the line profile regions to extract light curves for
              the profile wings, the line center, and the blue and read
              peak in the broad emission-line profiles of the H$\alpha$
              and H$\beta$ lines. In brackets the velocity-ranges are given
              which are used by Gezari et al.\,(2007).
}
\tablehead{
\colhead{feature       }     &
\colhead{H$\alpha \,\lambda$\,6563}&
\colhead{H$\beta \,\lambda$\,4861 }\\
\colhead{(1)} &
\colhead{(2)} &
\colhead{(3)}
}
\startdata
blue wing& 6345.5\,\AA --- 6455.0\,\AA &4702.0\,\AA $-$ 4783.0\,\AA \\
         &$-$9916\,km\,s$^{-1}$ --- $-$4916\,km\,s$^{-1}$&$-$9819\,km\,s$^{-1}$ --- $-$4819\,km\,s$^{-1}$\\
         &($-$12300\,km\,s$^{-1}$ --- $-$4500\,km\,s$^{-1}$)& \\
blue peak& 6455.0\,\AA --- 6521.0\,\AA &4783.0\,\AA $-$ 4832.0\,\AA \\
         &$-$4916\,km\,s$^{-1}$ --- $-$1916\,km\,s$^{-1}$&$-$4819\,km\,s$^{-1}$ --- $-$1819\,km\,s$^{-1}$\\
         &($-$4500\,km\,s$^{-1}$ --- $-$1700\,km\,s$^{-1}$)& \\
center   & 6521.0\,\AA --- 6629.0\,\AA &4832.0\,\AA $-$ 4911.0\,\AA \\
         &$-$1916\,km\,s$^{-1}$ --- $+$3032\,km\,s$^{-1}$&$-$1819\,km\,s$^{-1}$ --- $+$3063\,km\,s$^{-1}$\\
         &($-$1700\,km\,s$^{-1}$ --- $+$3300\,km\,s$^{-1}$)& \\
red peak & 6629.0\,\AA --- 6695.0\,\AA &4911.0\,\AA $-$ 4959.5\,\AA \\
         &$+$3032\,km\,s$^{-1}$ --- $+$6032\,km\,s$^{-1}$&$+$3063\,km\,s$^{-1}$ --- $+$6063\,km\,s$^{-1}$\\
         &($+$3300\,km\,s$^{-1}$ --- $+$5900\,km\,s$^{-1}$)& \\
red wing & 6695.0\,\AA --- 6804.5\,\AA &4959.5\,\AA $-$ 5040.5\,\AA \\
         &$+$6032\,km\,s$^{-1}$ --- $+$11032\,km\,s$^{-1}$&$+$6063\,km\,s$^{-1}$ --- $+$11063\,km\,s$^{-1}$\\
         &($+$5900\,km\,s$^{-1}$ --- $+$15000\,km\,s$^{-1}$)& \\
\enddata
\end{deluxetable}

%\clearpage

%Table 9
%\setcounter{table}{9}
%\ptlandscape
\hspace*{-15mm}
\begin{deluxetable}{lcccccc}
%\tablewidth{215pt}
\tablewidth{0pt}
\tabletypesize{\scriptsize}
\tablecaption{Results of the cross-correlation analysis of the wings, the
              center, and the blue and red peak of the broad Balmer lines 
              H$\alpha$ and H$\beta$ using AGN continuum variations at 
              F$_\lambda$(5100\AA) and the {\it g}-band variations as driving 
              continuum. The centroid of the ICCF has been computed for 
              80\,\%\ of the corresponding maximum of the ICCF.}
\tablehead{
\colhead{feature}&
\multicolumn{3}{c}{F$_\lambda$(5100\AA)}&
\multicolumn{3}{c}{{\it g}-band}\\
\colhead{ }&
\colhead{$\tau _{cent}$}&
\colhead{$\tau _{peak}$}&
\colhead{CCF$_{max}$   }&
\colhead{$\tau _{cent}$}&
\colhead{$\tau _{peak}$}&
\colhead{CCF$_{max}$   }\\
\colhead{(1)} &
\colhead{(2)} &
\colhead{(3)} &
\colhead{(4)} &
\colhead{(5)} &
\colhead{(6)} &
\colhead{(7)} 
}
\tablecolumns{7}
\startdata
H$\alpha$ blue wing&$57.5^{ +4.7}_{ -5.5}$&$58.0^{ +5.0}_{ -4.5}$&$0.88\pm 0.47$&$55.9^{ +4.0}_{ -5.6}$&$57.0^{ +3.0}_{ -5.0}$&$0.92\pm 0.59$\\[1mm]
H$\alpha$ blue peak&$54.3^{ +6.2}_{ -4.4}$&$56.5^{ +4.5}_{ -5.0}$&$0.84\pm 0.49$&$53.2^{ +5.1}_{ -5.0}$&$55.5^{ +3.0}_{ -7.0}$&$0.85\pm 0.56$\\[1mm]
H$\alpha$ center   &$54.5^{ +6.0}_{ -4.2}$&$56.5^{ +4.5}_{ -4.5}$&$0.82\pm 0.41$&$53.5^{ +4.9}_{ -4.9}$&$56.0^{ +2.5}_{ -7.0}$&$0.85\pm 0.73$\\[1mm]
H$\alpha$ red peak &$54.1^{ +6.7}_{ -4.4}$&$56.5^{ +4.5}_{ -6.5}$&$0.85\pm 0.26$&$52.7^{ +6.3}_{ -5.5}$&$55.0^{ +4.0}_{ -7.5}$&$0.86\pm 0.20$\\[1mm]
H$\alpha$ red wing &$56.8^{ +6.8}_{ -7.2}$&$57.5^{ +6.0}_{ -8.0}$&$0.89\pm 0.40$&$49.9^{+10.5}_{ -4.4}$&$51.0^{+10.5}_{ -4.0}$&$0.95\pm 1.28$\\[1mm]
H$\alpha$ blue vs. red wing&$-3.4^{ +7.4}_{ -6.8}$&$-2.5^{ +7.5}_{ -8.0}$&$0.64\pm 0.10$& & & \\[1mm]
H$\alpha$ blue vs. red peak&$ 0.2^{ +3.9}_{ -3.6}$&$ 0.0^{ +1.0}_{ -0.5}$&$0.67\pm 0.10$& & & \\[1mm]
              & & & & & & \\
H$\beta$ blue wing &$49.6^{+12.4}_{-11.0}$&$52.0^{+10.0}_{-16.0}$&$0.84\pm 0.34$&$48.4^{+11.3}_{ -8.8}$&$49.5^{+11.0}_{-10.0}$&$0.84\pm 0.16$\\[1mm]
H$\beta$ blue peak &$48.0^{+13.3}_{ -9.7}$&$51.0^{+10.5}_{-14.5}$&$0.86\pm 0.71$&$47.2^{+12.1}_{ -9.1}$&$49.0^{+11.0}_{-11.5}$&$0.85\pm 0.27$\\[1mm]
H$\beta$ center    &$49.6^{+12.4}_{-13.7}$&$53.0^{ +9.0}_{-19.0}$&$0.85\pm 0.56$&$49.0^{+11.4}_{-10.4}$&$50.5^{+10.5}_{-12.0}$&$0.84\pm 0.19$\\[1mm]
H$\beta$ red peak  &$51.6^{ +7.0}_{ -7.9}$&$54.5^{ +4.5}_{-11.0}$&$0.80\pm 0.19$&$53.1^{ +7.1}_{ -9.4}$&$54.5^{ +6.0}_{-10.0}$&$0.86\pm 0.74$\\[1mm]
H$\beta$ red wing  &$54.4^{ +6.0}_{-10.3}$&$56.0^{ +4.5}_{-11.5}$&$0.81\pm 0.35$&$56.3^{ +5.9}_{ -9.6}$&$57.0^{ +5.5}_{ -8.0}$&$0.91\pm 0.71$\\[1mm]
H$\beta$ blue vs. red wing &$ 2.9^{+13.4}_{-18.5}$&$ 3.0^{+18.0}_{-22.0}$&$0.47\pm 0.10$& & & \\[1mm] 
H$\beta$ blue vs. red peak &$ 1.0^{ +8.8}_{ -9.8}$&$ 0.0^{+10.0}_{-10.0}$&$0.50\pm 0.11$& & & \\[1mm]
\enddata
\end{deluxetable}

%\clearpage

%Table 10
%\setcounter{table}{10}
%\ptlandscape
\hspace*{-15mm}
\begin{deluxetable}{lcccc}
%\tablewidth{215pt}
\tablewidth{0pt}
\tabletypesize{\scriptsize}
\tablecaption{Estimates of the black hole virial product M$_{bh}^{vir}$, based 
              on the Balmer and helium line profile properties ($\sigma_{line}$
              and separation of the blue and red peak in the double-peaked 
              profiles) and the time delays $\tau _{cent}$ (80\,\%\ threshold) 
              of the broad emission-line flux variations.}
\tablehead{
\multicolumn{5}{c}{{\bf Virial Black Hole Product $M_{bh}^{vir}$ in $10^8 M_\odot$}}\\[1mm]
\hline
\multicolumn{5}{c}{ }\\
\multicolumn{5}{c}{using $\sigma_{line}$ of the line profiles}\\[1mm]
\colhead{feature       }&
\multicolumn{2}{c}{F$_\lambda$(5100\AA)}&
\multicolumn{2}{c}{{\it g}-band        }\\
\colhead{ }             &
\colhead{M$_{bh}^{vir}$(mean)}&
\colhead{M$_{bh}^{vir}$(rms) }&
\colhead{M$_{bh}^{vir}$(mean)}&
\colhead{M$_{bh}^{vir}$(rms) }\\
\colhead{(1)} &
\colhead{(2)} &
\colhead{(3)} & 
\colhead{(4)} &
\colhead{(5)} 
}
\startdata
H$\alpha \,\lambda$\,6563      &$2.33^{+0.17}_{-0.16}$&$2.57^{+0.25}_{-0.24}$&$2.33^{+0.10}_{-0.27}$&$2.57^{+0.20}_{-0.34}$\\[1mm]
H$\beta \,\lambda$\,4861       &$2.62^{+0.20}_{-0.18}$&$2.69^{+0.29}_{-0.27}$&$2.50^{+0.17}_{-0.19}$&$2.57^{+0.25}_{-0.27}$\\[1mm]
H$\gamma \,\lambda$\,4340      &$0.99^{+0.37}_{-0.33}$&$1.67^{+0.62}_{-0.55}$&$1.81^{+0.14}_{-0.19}$&$3.05^{+0.23}_{-0.33}$\\[1mm]
\heii\,$\lambda 4686$&$2.08^{+0.43}_{-0.41}$&$1.26^{+0.36}_{-0.35}$&$1.89^{+0.56}_{-0.33}$&$1.14^{+0.40}_{-0.30}$\\[3mm]
average              &$2.01^{+0.67}_{-0.48}$&$2.05^{+0.38}_{-0.36}$&$2.13^{+0.24}_{-0.21}$&$2.33^{+0.35}_{-0.37}$\\[2mm]
\hline
                     &                      &                      &                      &                      \\
\multicolumn{5}{c}{using the separation of the blue and red peak}\\[1mm]
\colhead{feature       }&
\multicolumn{2}{c}{F$_\lambda$(5100\AA)}&
\multicolumn{2}{c}{{\it g}-band        }\\
\colhead{ }             &
\colhead{M$_{bh}^{vir}$(mean)}&
\colhead{M$_{bh}^{vir}$(rms) }&
\colhead{M$_{bh}^{vir}$(mean)}&
\colhead{M$_{bh}^{vir}$(rms) }\\
\colhead{(1)} &
\colhead{(2)} &
\colhead{(3)} & 
\colhead{(4)} &
\colhead{(5)} \\[1mm]
\hline
                     &                      &                      &                      &                      \\[-2mm]
H$\alpha \,\lambda$\,6563      &$1.74^{+0.14}_{-0.13}$&$1.91^{+0.21}_{-0.21}$&$1.74^{+0.09}_{-0.21}$&$1.91^{+0.18}_{-0.27}$\\[1mm]
H$\beta \,\lambda$\,4861       &$1.41^{+0.12}_{-0.10}$&$1.87^{+0.23}_{-0.22}$&$1.34^{+0.10}_{-0.11}$&$1.78^{+0.21}_{-0.22}$\\[1mm]
H$\gamma \,\lambda$\,4340      &$1.10^{+0.41}_{-0.37}$&$0.88^{+0.42}_{-0.39}$&$2.02^{+0.19}_{-0.23}$&$1.61^{+0.50}_{-0.51}$\\[2mm]
average              &$1.42^{+0.35}_{-0.32}$&$1.55^{+0.48}_{-0.46}$&$1.70^{+0.26}_{-0.33}$&$1.77^{+0.29}_{-0.31}$\\[2mm]
\enddata
\end{deluxetable}

\end{document}